\documentclass[12pt,twoside, a4paper]{article}
\def\pd{\partial}
\def\mc{\mathcal}

\usepackage[dvips]{graphicx}
\usepackage{amssymb}
\usepackage{amssymb,amsmath}
\usepackage{graphicx}
\usepackage{dsfont}
\usepackage{caption}
\usepackage{subcaption}
\usepackage{mathtools}
\usepackage{verbatim}
\usepackage{graphicx}
\usepackage{multirow}
\usepackage[outline]{contour}
\usepackage{xcolor,colortbl}
\usepackage{pdflscape}
\input{epsf.sty}  
\begin{document}
\begin{center}
\Large{\textbf{Holographic RG flows and symplectic deformations of $N=4$ gauged supergravity}}
\end{center}
\begin{center}
\large{\textbf{Parinya Karndumri}}
\end{center}
\begin{center}
String Theory and Supergravity Group, Department
of Physics, Faculty of Science, Chulalongkorn University, 254 Phayathai Road, Pathumwan, Bangkok 10330, Thailand
\end{center}
E-mail: parinya.ka@hotmail.com \vspace{1 cm}\\
\begin{abstract}
We study four-dimensional $N=4$ gauged supergravity with $SO(4)\times SO(4)\sim SO(3)\times SO(3)\times SO(3)\times SO(3)$ gauge group in the presence of symplectic deformations. There are in general four electric-magnetic phases corresponding to each $SO(3)$ factor, but two phases of the $SO(3)$ factors embedded in the $SO(6)$ R-symmetry are fixed. One phase can be set to zero by $SL(2,\mathbb{R})$ transformations. The second one gives equivalent theories for any non-vanishing values and can be set to $\frac{\pi}{2}$ resulting in gauged supergravities that admit $N=4$ supersymmetric $AdS_4$ vacua. The remaining two phases are truely deformation parameters leading to different $SO(4)\times SO(4)$ gauged supergravities. As in the $\omega$-deformed $SO(8)$ maximal gauged supergravity, the cosmological constant and scalar masses of the $AdS_4$ vacuum at the origin of the scalar manifold with $SO(4)\times SO(4)$ symmetry do not depend on the electric-magnetic phases. We find $N=1$ holographic RG flow solutions between $N=4$ critical points with $SO(4)\times SO(4)$ and $SO(3)_{\textrm{diag}}\times SO(3)\times SO(3)$ or $SO(3)\times SO(3)_{\textrm{diag}}\times SO(3)$ symmetries. We also give $N=2$ and $N=1$ RG flows from these critical points to various non-conformal phases. However, contrary to the $\omega$-deformed $SO(8)$ gauged supergravity, there exist non-trivial supersymmetric $AdS_4$ critical points only for particular values of the deformation parameters within the scalar sectors under consideration.    
\end{abstract}
\newpage

\section{Introduction}
The discovery of a new family of maximal gauged supergravities in \cite{omega_N8_1}, see also \cite{omega_N8_2,omega_Range1,omega_deWit}, called $\omega$-deformed gauged supergravity has led to various interesting consequences. The new $\omega$-deformed $SO(8)$ gauged supergravity is obtained from a symplectic deformation of the original $SO(8)$ gauged supergravity constructed in \cite{SO8_deWit}. In the context of the AdS/CFT correspondence \cite{maldacena,Gubser_AdS_CFT,Witten_AdS_CFT}, the $\omega$-deformed $SO(8)$ gauged supergravity admits a much richer structure of supersymmetric $AdS_4$ vacua and other holographic solutions such as holographic RG flows and Janus solutions \cite{omega_N8_1,omega_vacua1,omega_vacua2,omega_vacua,Guarino_BPS_DW,Elec_mag_flows,Yi_4D_flow,N8_omega_Janus}. 
\\
\indent An extension of symplectic deformations to $N\geq 2$ gauged supergravities has been considered in \cite{Inverso_symplectic} in which some examples of symplectically deformed $N=2$ and $N=4$ gauged supergravities have been given. In the present paper, we are interested in symplectic deformations of $N=4$ gauged supergravity with $SO(4)\times SO(4)\sim SO(3)\times SO(3)\times SO(3)\times SO(3)$ gauge group. Generally, there can be four deformation parameters or electric-magnetic phases for the four $SO(3)$ factors, see also \cite{de_Roo_N4_4D,N4_Wagemans} for an eariler construction of $N=4$ gauged supergravity with these phases. In the notation of \cite{Inverso_symplectic}, these phases are denoted by $\alpha_0$, $\alpha$, $\beta_1$ and $\beta_2$. $\alpha_0$ can be set to zero by $SL(2,\mathbb{R})$ transformations. In addition, all values of $\alpha>0$ lead to equivalent gauged supergravities and can be set to $\frac{\pi}{2}$. The remaining phases $\beta_1$ and $\beta_2$ constitute free deformation parameters of the $SO(4)\times SO(4)$ gauged supergravity. With these two phases, we expect to find a rich structure of vacua and other interesting holographic solutions as in the $\omega$-deformed $SO(8)$ gauged supergravity.
\\
\indent The $SO(4)\times SO(4)$ gauged supergravity with particular values of $\alpha_0=\beta_1=0$ and $\alpha=\beta_2=\frac{\pi}{2}$ has been considered previously in \cite{dS_Roest,4D_N4_flows,N4_Janus,N3_4_AdS4_BH}. In particular, a number of supersymmetric $AdS_4$ vacua, holographic RG flows, Janus solutions and $AdS_4$ black holes have been found in \cite{4D_N4_flows,N4_Janus,N3_4_AdS4_BH}. In the present paper, we will consider $SO(4)\times SO(4)$ gauge group with arbitrary values of $\beta_1$ and $\beta_2$. We will mainly look for supersymmetric $AdS_4$ vacua and holographic RG flows interpolating between these vacua or from $AdS_4$ critical points to singular geometries in the IR. The former describe RG flows between conformal fixed points in the dual $N=4$ Chern-Simons-matter (CSM) theories in three dimensions, see for example \cite{ABJM,Bena,BL1,BL2,BL3,Gustavsson,Basu_Harvey,Schwarz_3D_SCFT,ABJ}, while the latter correspond to RG flows from a conformal fixed point to non-conformal phases. We will consider $SO(3)_{\textrm{diag}}\times SO(3)_{\textrm{diag}}$, $SO(2)\times SO(2)\times SO(2)\times SO(2)$ and $SO(3)_{\textrm{diag}}\times SO(3)$ scalar sectors. 
\\
\indent It turns out that, unlike the $\omega$-deformed $SO(8)$ gauged supergravity, there do not exist any supersymmetric $AdS_4$ critical points apart from those identified previously in \cite{4D_N4_flows} or critical points related to these at least within the aforementioned scalar sectors. However, we do find new classes of holographic RG flows with $N=1$ and $N=2$ supersymmetries. In particular, some of the $N=1$ solutions describe RG flows between $N=4$ critical points with $SO(4)\times SO(4)$ and $SO(3)_{\textrm{diag}}\times SO(3)\times SO(3)$ or $SO(3)\times SO(3)_{\textrm{diag}} \times SO(3)$ symmetries. To the best of the author's knowledge, these are the first examples of holographic RG flows between conformal fixed points preserving $N=1$ supersymmetry in the framework of $N=4$ gauged supergravity, see \cite{tri-sasakian-flow} for examples of $N=1$ RG flows to non-conformal phases, and should further extend the list of known $N=4$ and $N=2$ solutions given in \cite{4D_N4_flows} and \cite{N4_Janus}, see \cite{Guarino_BPS_DW,Elec_mag_flows,Yi_4D_flow,tri-sasakian-flow,Warner_membrane_flow,Warner_M_F_theory_flow,Warner_higher_Dflow,Flow_in_N8_4D,4D_G2_flow,
Warner_M2_flow,N3_SU2_SU3,N3_4D_gauging,orbifold_flow,N4_from_ISO7,ISO7_N3_flow,N5_flow,N6_flow} for an incomplete list of similar solutions in other four-dimensional gauged supergravities. 
\\
\indent It should be pointed out that the $N=4$ gauged supergravity under consideration here has currently no known higher dimensional origins as in the case of $\omega$-deformed $SO(8)$ gauged supergravity. Accordingly, the complete holographic interpretation in string/M-theory framework is unavailable. However, it is still useful to have holographic solutions in lower dimensional gauged supergravities, and with recent developments in double field theory formalism, particularly the result of \cite{Dibitetto_SL2_angle}, the embedding of $SO(4)\times SO(4)$ gauged supergravity in higher dimensions could be achieved.      
\\
\indent The paper is organized as follows. In section \ref{N4_SUGRA}, we review the general structure of $N=4$ gauged supergravity in the embedding tensor formalism together with symplectic deformations of $SO(4)\times SO(4)$ gauge group. The truncations to $SO(3)_{\textrm{diag}}\times SO(3)_{\textrm{diag}}$, $SO(2)\times SO(2)\times SO(2)\times SO(2)$ and $SO(3)_{\textrm{diag}}\times SO(3)$ singlet scalars are considered in sections \ref{SO3_SO3_sector}, \ref{SO2_4_sector} and \ref{SO3d_SO3_sector}, respectively. In these scalar sectors, we will focus on $AdS_4$ vacua and possible RG flow solutions between these vacua and RG flows to singular geometries. We end the paper by giving some conclusions and comments on the results in section \ref{conclusion}. Useful formulae and details on relevant BPS equations can be found in the appendix. 

\section{Matter-coupled $N=4$ gauged supergravity}\label{N4_SUGRA} 
We first review the general structure of $N=4$ gauged supergravity coupled to vector multiplets in the embedding tensor formalism \cite{N4_gauged_SUGRA}, see also \cite{Eric_N4_4D} for an eariler construction. In four dimensions, there are two types of $N=4$ supermultiplets, the gravity and vector multiplets, with the following field content
\begin{equation}
(e^{\hat{\mu}}_\mu,\psi^i_\mu,A_\mu^m,\chi^i,\tau)
\end{equation}
and
\begin{equation}
(A^a_\mu,\lambda^{ia},\phi^{ma}).
\end{equation} 
The component fields in the gravity multiplet are given by the graviton $e^{\hat{\mu}}_\mu$, four gravitini $\psi^i_\mu$, six vectors
$A_\mu^m$, four spin-$\frac{1}{2}$ fields $\chi^i$ and one complex scalar $\tau$ while those in a vector multiplet are given by a vector field $A_\mu$, four gaugini $\lambda^i$ and six scalars $\phi^m$. 

It is useful to note the convention for various types of indices here. Indices $\mu,\nu,\ldots =0,1,2,3$ and $\hat{\mu},\hat{\nu},\ldots=0,1,2,3$ are respectively space-time and tangent space (flat) indices while $m,n=1,\ldots, 6$ and $i,j=1,2,3,4$ indices describe fundamental representations of $SO(6)_R$ and $SU(4)_R$ R-symmetry. The vector
multiplets are labeled by indices $a,b=1,\ldots, n$. From both the gravity and vector multiplets, there are $6+n$ vector fields $A^{+M}=(A^m_\mu,A^a_\mu)$. These are called electric gauge fields and appear in the ungauged Lagrangian with the usual Yang-Mills kinetic term. Indices $M,N,\ldots =1,2,\ldots, 6+n$ denote fundamental representation of $SO(6,n)$. Together with the magnetic dual $A^{-M}$, the resulting $2(6+n)$ vector fields form a doublet under $SL(2,\mathbb{R})$ and will be denoted by $A^{\alpha M}$ with $\alpha=(+,-)$ being an index of $SL(2,\mathbb{R})$ fundamental representation. 
\\
\indent All fermionic fields and supersymmetry parameters transform in fundamental representation of $SU(4)_R\sim SO(6)_R$ and are subject to the chirality projections
\begin{equation}
\gamma_5\psi^i_\mu=\psi^i_\mu,\qquad \gamma_5\chi^i=-\chi^i,\qquad \gamma_5\lambda^i=\lambda^i
\end{equation}
while those transforming in anti-fundamental representation of $SU(4)_R$ satisfy
\begin{equation}
\gamma_5\psi_{\mu i}=-\psi_{\mu i},\qquad \gamma_5\chi_i=\chi_i,\qquad \gamma_5\lambda_i=-\lambda_i\, .
\end{equation}
\indent The complex scalar $\tau$ consists of the dilaton $\phi$ and the axion $\chi$ which parametrize $SL(2,\mathbb{R})/SO(2)$ coset manifold. This $SL(2,\mathbb{R})/SO(2)$ can be described by the coset representative $\mc{V}_\alpha$ of the form
\begin{equation}
\mc{V}_\alpha=e^{\frac{\phi}{2}}\left(
                                         \begin{array}{c}
                                           \chi+ie^{\phi} \\
                                           1 \\
                                         \end{array}
                                       \right).
\end{equation}
with  
\begin{equation}
\tau=\chi+ie^\phi\, .
\end{equation}
Similarly, the $6n$ scalars $\phi^{ma}$ parametrize $SO(6,n)/SO(6)\times SO(n)$ coset manifold with the coset representative denoted by $\mc{V}_M^{\phantom{M}A}$. Under the global $SO(6,n)$ and local $SO(6)\times SO(n)$ symmetries, $\mc{V}_M^{\phantom{M}A}$ transforms by left and right multiplications, respectively. Accordingly, the $SO(6)\times SO(n)$ index $A$ can be split as $A=(m,a)$ resulting in the following components of the coset representative
\begin{equation}
\mc{V}_M^{\phantom{M}A}=(\mc{V}_M^{\phantom{M}m},\mc{V}_M^{\phantom{M}a}).
\end{equation}
The matrix $\mc{V}_M^{\phantom{M}A}$ satisfies the relation
\begin{equation}
\eta_{MN}=-\mc{V}_M^{\phantom{M}m}\mc{V}_N^{\phantom{M}m}+\mc{V}_M^{\phantom{M}a}\mc{V}_N^{\phantom{M}a}
\end{equation}
with $\eta_{MN}=\textrm{diag}(-1,-1,-1,-1,-1,-1,1,\ldots,1)$ being the $SO(6,n)$ invariant tensor. The inverse of $\mc{V}_M^{\phantom{M}A}$ will be denoted by ${\mc{V}_A}^M=({\mc{V}_m}^M,{\mc{V}_a}^M)$.
\\
\indent All possible gaugings of the aforementioned matter-coupled $N=4$ supergravity are encoded in the embedding tensor \cite{N4_gauged_SUGRA}. $N=4$ supersymmetry allows only two non-vanishing components of the embedding tensor denoted by $\xi^{\alpha M}$ and $f_{\alpha MNP}$. A given gauge group $G_0\subset SL(2,\mathbb{R})\times SO(6,n)$ can be embedded in both $SL(2,\mathbb{R})$ and
$SO(6,n)$ and can be gauged by either electric or magnetic vector fields or combinations thereof. We also note that each magnetic vector field must be accompanied by an auxiliary two-form field in order to remove the extra degrees of freedom. The embedding tensor also needs to satisfy the quadratic constraint in order for the resulting gauge generators to form a closed subalgebra of $SL(2,\mathbb{R})\times SO(6,n)$. 
\\
\indent In this paper, we are mainly interested in gauge groups admitting supersymmetric $AdS_4$ vacua. As shown in \cite{AdS4_N4_Jan}, see also \cite{de_Roo_N4_4D,N4_Wagemans} for an earlier result, this requires the gauge groups to be embedded solely in $SO(6,n)$ and gauged by both electric and magnetic vector fields. This implies that both electric and magnetic components of $f_{\alpha MNP}$ must be non-vanishing and $\xi^{\alpha M}=0$. Accordingly, we will set $\xi^{\alpha M}$ to zero from now on. Furthermore, since we will study supersymmetric $AdS_4$ vacua and domain wall solutions that involve only the metric and scalar fields, we will also set all vector and fermionic fields to zero. 
\\
\indent With all these, the bosonic Lagrangian can be written as
\begin{equation}
e^{-1}\mc{L}=\frac{1}{2}R+\frac{1}{16}\pd_\mu M_{MN}\pd^\mu
M^{MN}-\frac{1}{4(\textrm{Im}\tau)^2}\pd_\mu \tau \pd^\mu \tau^*-V
\end{equation}
where $e=\sqrt{-g}$ is the vielbein determinant. The scalar potential is given in terms of the scalar coset representative and the embedding tensor by
\begin{eqnarray}
V&=&\frac{1}{16}\left[f_{\alpha MNP}f_{\beta
QRS}M^{\alpha\beta}\left[\frac{1}{3}M^{MQ}M^{NR}M^{PS}+\left(\frac{2}{3}\eta^{MQ}
-M^{MQ}\right)\eta^{NR}\eta^{PS}\right]\right.\nonumber \\
& &\left.-\frac{4}{9}f_{\alpha MNP}f_{\beta
QRS}\epsilon^{\alpha\beta}M^{MNPQRS}\right].
\end{eqnarray}
We also note that $f_{\alpha MNP}$ include the gauge coupling constants.
\\
\indent The symmetric matrix $M_{MN}$ is defined by 
\begin{equation}
M_{MN}=\mc{V}_M^{\phantom{M}m}\mc{V}_N^{\phantom{M}m}+\mc{V}_M^{\phantom{M}a}\mc{V}_N^{\phantom{M}a}
\end{equation}
with $M^{MN}$ denoting its inverse. The tensor $M^{MNPQRS}$ is obtained from
\begin{equation}
M_{MNPQRS}=\epsilon_{mnpqrs}\mc{V}_{M}^{\phantom{M}m}\mc{V}_{N}^{\phantom{M}n}
\mc{V}_{P}^{\phantom{M}p}\mc{V}_{Q}^{\phantom{M}q}\mc{V}_{R}^{\phantom{M}r}\mc{V}_{S}^{\phantom{M}s}\label{M_6}
\end{equation}
by raising indices with $\eta^{MN}$. Finally, $M^{\alpha\beta}$ is the inverse of the symmetric $2\times 2$ matrix $M_{\alpha\beta}$ defined by
\begin{equation}
M_{\alpha\beta}=\textrm{Re}(\mc{V}_\alpha\mc{V}^*_\beta).
\end{equation}
\indent Fermionic supersymmetry transformations are given by
\begin{eqnarray}
\delta\psi^i_\mu &=&2D_\mu \epsilon^i-\frac{2}{3}A^{ij}_1\gamma_\mu
\epsilon_j,\\
\delta \chi^i &=&-\epsilon^{\alpha\beta}\mc{V}_\alpha D_\mu
\mc{V}_\beta\gamma^\mu \epsilon^i-\frac{4}{3}iA_2^{ij}\epsilon_j,\\
\delta \lambda^i_a&=&2i\mc{V}_a^{\phantom{a}M}D_\mu
\mc{V}_M^{\phantom{M}ij}\gamma^\mu\epsilon_j-2iA_{2aj}^{\phantom{2aj}i}\epsilon^j
\end{eqnarray}
with the fermion shift matrices defined by
\begin{eqnarray}
A_1^{ij}&=&\epsilon^{\alpha\beta}(\mc{V}_\alpha)^*\mc{V}_{kl}^{\phantom{kl}M}\mc{V}_N^{\phantom{N}ik}
\mc{V}_P^{\phantom{P}jl}f_{\beta M}^{\phantom{\beta M}NP},\nonumber
\\
A_2^{ij}&=&\epsilon^{\alpha\beta}\mc{V}_\alpha\mc{V}_{kl}^{\phantom{kl}M}\mc{V}_N^{\phantom{N}ik}
\mc{V}_P^{\phantom{P}jl}f_{\beta M}^{\phantom{\beta M}NP},\nonumber
\\
A_{2ai}^{\phantom{2ai}j}&=&\epsilon^{\alpha\beta}\mc{V}_\alpha
{\mc{V}_a}^M{\mc{V}_{ik}}^N\mc{V}_P^{\phantom{P}jk}f_{\beta
MN}^{\phantom{\beta MN}P}\, .
\end{eqnarray}
$\mc{V}_M^{\phantom{M}ij}$ and ${\mc{V}_{ij}}^M$ are defined in terms of the 't Hooft
symbols $G^{ij}_m$ as
\begin{equation}
\mc{V}_M^{\phantom{M}ij}=\frac{1}{2}\mc{V}_M^{\phantom{M}m}G^{ij}_m
\end{equation}
and
\begin{equation}
{\mc{V}_{ij}}^M=-\frac{1}{2}{\mc{V}_{m}}^M(G^{ij}_m)^*\,
.
\end{equation}
The explicit representation of $G^{ij}_m$ used in this paper is given in the appendix. It is also useful to note that upper and lower $i,j,\ldots$ indices are related by complex conjugation.

\subsection{$SO(4)\times SO(4)$ gauge group and symplectic deformations}
In this work, we only consider $SO(4)\times SO(4)\sim SO(3)\times SO(3)\times SO(3)\times SO(3)$ gauge group. The embedding of this gauge group into the global symmetry $SO(6,n)$ requires at least $n=6$ vector multiplets. We will only consider the minimal case of $n=6$. All four factors of $SO(3)$'s are embedded in $SO(6,6)$ via the maximal compact subgroup $SO(6)_R\times SO(6)$ with $SO(6)_R\sim SU(4)_R$ being the R-symmetry. Components of the embedding tensor describing the full gauge group are given by the $SO(3)$ structure constant for each $SO(3)$ factor. The existence of supersymmetric $AdS_4$ vacua requires the first two $SO(3)$ factors embedded in $SO(6)_R$ to be gauged differently, one factor electrically gauged and the other magnetically gauged.  
\\
\indent In general, each $SO(3)$ factor can acquire a non-trivial $SL(2,\mathbb{R})$ phase resulting in symplectic deformations of a particular $SO(4)\times SO(4)$ gauging such as purely electric gauged $SO(4)\times SO(4)$ gauge group \cite{Inverso_symplectic}, see also \cite{de_Roo_N4_4D,N4_Wagemans} for an earlier consideration of this deformation. To give an explicit form of the embedding tensor, it is convenient to split the $SO(6,6)$ fundamental index as $M=(\hat{m},\tilde{m},\hat{a},\tilde{a})$ for $\hat{m},\tilde{m},\hat{a},\tilde{a}=1,2,3$. The embedding tensor for symplectically deformed $SO(4)\times SO(4)$ gauging as given in \cite{Dibitetto_SL2_angle} can be written, in the notation of \cite{Inverso_symplectic}, as
\begin{eqnarray}
& &f_{+\hat{m}\hat{n}\hat{p}}=-g_0\cos\alpha_0 \epsilon_{\hat{m}\hat{n}\hat{p}}, \qquad f_{-\hat{m}\hat{n}\hat{p}}=g_0\sin\alpha_0 \epsilon_{\hat{m}\hat{n}\hat{p}}, \nonumber \\
& & f_{+\tilde{m}\tilde{n}\tilde{p}}=-g\cos\alpha \epsilon_{\tilde{m}\tilde{n}\tilde{p}},\qquad f_{-\tilde{m}\tilde{n}\tilde{p}}=g\sin\alpha \epsilon_{\tilde{m}\tilde{n}\tilde{p}},\nonumber \\ 
& &f_{+\hat{a}\hat{b}\hat{c}}=h_1\cos\beta_1 \epsilon_{\hat{a}\hat{b}\hat{c}}, \qquad f_{-\hat{a}\hat{b}\hat{c}}=h_1\sin\beta_1 \epsilon_{\hat{a}\hat{b}\hat{c}}, \nonumber \\
& & f_{+\tilde{a}\tilde{b}\tilde{c}}=h_2\cos\beta_2 \epsilon_{\tilde{a}\tilde{b}\tilde{c}},\qquad f_{-\tilde{a}\tilde{b}\tilde{c}}=h_2\sin\beta_2 \epsilon_{\tilde{a}\tilde{b}\tilde{c}}\, .
\end{eqnarray}      
The constants $\alpha_0$, $\alpha$, $\beta_1$ and $\beta_2$ are the electric-magnetic phases while $g_0$, $g$, $h_1$ and $h_2$ are the corresponding gauge coupling constants for each $SO(3)$ factor. 

A particular case of $\alpha_0=\beta_1=0$ and $\alpha=\beta_2=\frac{\pi}{2}$, after a redefinition of gauge coupling constants, has been considered in \cite{dS_Roest,4D_N4_flows,N4_Janus}. For later convenience, we will call the $SO(4)\times SO(4)$ gauge group with this particular choice of phases ``undeformed'' $SO(4)\times SO(4)$ gauge group. It has been pointed out in \cite{Inverso_symplectic} that by gauge fixing the $SL(2,\mathbb{R})$ symmetry, we can set $\alpha_0=0$. In addition, all the gaugings with $\alpha>0$ are equivalent to the gauge group with $\alpha=\frac{\pi}{2}$ up to a shift of gauge invariant theta terms and a redefinition of the axion. We will set $\alpha_0=0$ but keep $\alpha$ generic to keep track of the effects of symplectic deformations. We also note that if all the phases are not $0$ or $\frac{\pi}{2}$, see \cite{Inverso_symplectic} for possible ranges of these phases, all four $SO(3)$ factors are dyonically gauged by both electric and magnetic vector fields since all $f_{\pm MNP}$ are non-vanishing. For convenience, we will introduce the notation $SO(3)_0$, $SO(3)_\alpha$, $SO(3)_1$ and $SO(3)_2$ for these four $SO(3)$ factors. $SO(3)_0\times SO(3)_\alpha$ and $SO(3)_1\times SO(3)_2$ are embedded in $SO(6)_R$ and $SO(6)$, respectively. 

We also note that for particular values of the electric-magnetic phases
\begin{equation}
\alpha_0=0,\qquad \alpha=\frac{\pi}{2},\qquad \beta=\frac{\pi}{2}-2\omega,\qquad \beta_2=-2\omega
\end{equation}
with $\omega\in [0,\frac{\pi}{8}]$ and $g=-g_0=h_1=h_2$, the resulting $N=4$ gauged supergravity is a truncation of the $\omega$-deformed $SO(8)$ maximal gauged supergravity constructed in \cite{omega_N8_1}.        

\subsection{Parametrization of scalar manifold and BPS equations}\label{BPS_ansatz}
Since we are mainly interested in holographic RG flow solutions in the form of supersymmetric domain walls, an explicit parametrization of the scalar manifold $SO(6,6)/SO(6)\times SO(6)$ is crucial. To give the $SO(6,6)/SO(6)\times SO(6)$ coset representative, we first define $SO(6,6)$ generators in the fundamental representation by
\begin{equation}
(t_{MN})_P^{\phantom{P}Q}=2\delta^Q_{[M}\eta_{N]P}\, .
\end{equation}
The $SO(6,6)$ non-compact generators are then given by
\begin{equation}
Y_{ma}=t_{m,a+6}\, .
\end{equation}
To make things more manageable, we will only consider particular truncations of the full $36$-dimensional coset to submanifolds with a few scalars non-vanishing. The truncations we will consider contain singlet scalars under $SO(4)_{\textrm{diag}}\sim SO(3)_{\textrm{diag}}\times SO(3)_{\textrm{diag}}$, $SO(2)\times SO(2)\times SO(2)\times SO(2)$ and $SO(3)_{\textrm{diag}}\times SO(3)$. 

To find supersymmetric domain wall solutions, we use the standard metric ansatz
\begin{equation}
ds^2=e^{2A(r)}dx^2_{1,2}+dr^2\label{DW_metric}
\end{equation}
with $dx^2_{1,2}$ being the metric on three-dimensional Minkowski space. The only remaining non-vanishing fields are given by scalars. To preserve the isometry of $dx_{1,2}^2$, scalar fields can depend only on the radial coordinate $r$. 
\\
\indent Supersymmetric solutions can be found by considering solutions to the BPS equations obtained by setting fermionic supersymmetry transformations to zero. With the metric ansatz \eqref{DW_metric}, the variations of gravitini along $\mu=0,1,2$ directions give
\begin{equation}
A'\gamma_{\hat{r}}\epsilon^i-\frac{2}{3}A^{ij}_1\epsilon_j=0\, .
\end{equation}
To proceed, we will use Majorana representation for space-time gamma matrices with all $\gamma^{\hat{\mu}}$ real and $\gamma_5$ purely imaginary. Left and right chiralities of fermions are then related to each other by complex conjugation. The symmetric matrix $A_1^{ij}$ can be diagonalized with eiganvalues denoted by $\mc{A}_i$. In general, in the presence of unbroken supersymmetry, some or all of these eigenvalues will give rise to the superpotential in terms of which the scalar potential can be written. Let $\hat{\alpha}$ be the eigenvalue of the Killing spinors $\epsilon^{\hat{i}}$ corresponding to the unbroken supersymmetry. We can rewrite the above equation as
\begin{equation}
A'\gamma^{\hat{r}}\epsilon^{\hat{i}}-\frac{2}{3}\hat{\alpha}\epsilon_{\hat{i}}=0\, .\label{con1}
\end{equation}
\indent To proceed further, we impose the following projector
\begin{equation}
\gamma^{\hat{r}}\epsilon^{\hat{i}}=e^{i\Lambda}\epsilon_{\hat{i}}\label{gamma_r_pro}
\end{equation}
with an $r$-dependent phase $\Lambda$. This projector relates the two chiralities of $\epsilon^{\hat{i}}$ breaking half of the supersymmetry. A domain wall solution is then half-supersymmetric. By defining the superpotential
\begin{equation}
\mc{W}=\frac{2}{3}\hat{\alpha},
  \end{equation} 
we obtain the BPS condition
\begin{equation}
A'e^{i\Lambda}-\mc{W}=0
\end{equation}
which implies
\begin{equation}
A'=\pm W\label{Ap_general_eq}
\end{equation}
and 
\begin{equation}
e^{i\Lambda}=\pm\frac{\mc{W}}{W}\label{general_phase}
\end{equation}
for $W= |\mc{W}|$.
\\
\indent Repeating the same procedure for $\delta \psi^i_{\hat{r}}$, we find a differential condition on the Killing spinors
\begin{equation}
2\pd_r\epsilon^{\hat{i}}-\mc{W}\gamma_{\hat{r}}\epsilon_{\hat{i}}=0\, .
\end{equation}
With the condition \eqref{con1}, we find 
\begin{equation}
\epsilon^{\hat{i}}=e^{\frac{A}{2}}\epsilon_0^{\hat{i}}
\end{equation}
for constant spinors $\epsilon_0^{\hat{i}}$. Finally, using the $\gamma_{\hat{r}}$ projector \eqref{gamma_r_pro} in the variations $\delta \chi^i$ and $\delta \lambda^i_a$, we can determine all the BPS equations for scalars. In subsequent sections, we will find explicit solutions for various residual symmetries and different numbers of unbroken supersymmetries.

\section{$SO(3)_{\textrm{diag}}\times SO(3)_{\textrm{diag}}$ sector}\label{SO3_SO3_sector}
We begin with a simple case of $SO(3)_{\textrm{diag}}\times SO(3)_{\textrm{diag}}$ singlet scalars. This sector has been considered in the undeformed $SO(4)\times SO(4)$ gauge group in \cite{4D_N4_flows}. In this work, we will consider effects of arbitrary electric-magnetic phases. There are two singlets from $SO(6,6)/SO(6)\times SO(6)$ coset corresponding to the non-compact generators
\begin{equation}
\hat{Y}_1=Y_{11}+Y_{22}+Y_{33}\qquad \textrm{and}\qquad  \hat{Y}_2=Y_{44}+Y_{55}+Y_{66}\label{SO4_inv_Y}\, .
\end{equation}
Accordingly, the coset representative can be written as
\begin{equation}
\mc{V}=e^{\phi_1\hat{Y}_1}e^{\phi_2\hat{Y}_2}\, .\label{L_SO4_inv}
\end{equation}
The dilaton and axion are also $SO(3)_{\textrm{diag}}\times SO(3)_{\textrm{diag}}$ singlets since these scalars are singlets under the full $SO(6,6)$ global symmetry. Therefore, the $SO(3)_{\textrm{diag}}\times SO(3)_{\textrm{diag}}$ sector consists of $4$ scalars.

\subsection{Supersymmetric $AdS_4$ vacua}
We first look at possible supersymmetric $AdS_4$ vacua within the $SO(3)_{\textrm{diag}}\times SO(3)_{\textrm{diag}}$ sector. The scalar potential is given by
\begin{eqnarray}
V&=&\frac{1}{4}\left[2g_0^2e^{-\phi}\cosh^2\phi_1(\cosh2\phi_1-2)+8gg_0\cosh^3\phi_1\cosh^3\phi_2\sin\alpha\right. \nonumber \\
& &+8gh_1\cosh^3\phi_2\sin(\alpha+\beta_1)\sinh^3\phi_1-8g_0h_2\sin\beta_2\cosh^3\phi_1\sinh^3\phi_2\nonumber \\
& & +8h_1h_2\sin(\beta_1-\beta_2)\sinh^3\phi_1\sinh^3\phi_2+e^{-\phi}g_0h_1\sinh^32\phi_1(\cos\beta_1-\sin\beta_1\chi)\nonumber \\
& &+2e^{-\phi}g^2\cosh^4\phi_2(\cosh2\phi_2-2)(\cos^2\alpha+\sin^2\alpha e^{2\phi}+\chi\sin2\alpha +\sin^2\alpha \chi^2)\nonumber \\
& &+2e^{-\phi}h_1^2\sinh^4\phi_1(\cos^2\beta_1+e^{2\phi}\sin^2\beta_1-\chi\sin2\beta_1+\sin^2\beta_1\chi^2)\times \nonumber \\
& &(\cosh2\phi_1+2)+e^{-\phi}gh_2\sinh^32\phi_2\left[\cos\alpha\cos\beta_2-e^{2\phi}\sin\alpha\sin\beta_2\right. \nonumber \\
& &\left.+\sin(\alpha-\beta_2)\chi-\sin\alpha\sin\beta_2\chi^2\right]+2e^{-\phi}h_2^2\sinh^4\phi_2(\cosh2\phi_2+2)\nonumber \\
& &\left. (\cos^2\beta_2+e^{2\phi}\sin^2\beta_2-\sin 2\beta_2+\sin^2\beta_2\chi^2)\right].
\end{eqnarray}
\indent As in the undeformed $SO(4)\times SO(4)$ gauge group, it turns out that $A_1^{ij}$ tensor is proportional to the identity matrix with the four equal eigenvalues given by 
\begin{eqnarray}
\mc{A}&=&-\frac{3}{4}e^{-\frac{\phi}{2}}\left[h_1\sinh^3\phi_1(\cos\beta_1+ie^{\phi}\sin\beta_1)-ih_2\sinh^3\phi_2(\cos\beta_2+ie^\phi\sin\beta_2) \right.\nonumber \\
& &-(h_1\sin\beta_1\sinh^3\phi_1+ig\sin\alpha\cosh^3\phi_2-ih_2\sin\beta_2\sinh^3\phi_2)\chi \nonumber \\
& &\left. +g_0\cosh^3\phi_1-g\cosh^3\phi_2(i\cos\alpha+e^{\phi}\sin\alpha)\right].
\end{eqnarray}
\indent We now look for possible supersymmetric $AdS_4$ vacua. We begin with a simple case of $\phi_1=\phi_2=0$. It can be straightforwardly verified that this choice satisfies all the BPS conditions provided that
\begin{equation}
\phi=\ln\left[-\frac{g_0}{g\sin\alpha}\right]\qquad \textrm{and} \qquad \chi=-\frac{\cos\alpha}{\sin\alpha}\, .
\end{equation} 
This leads to a supersymmetric $AdS_4$ vacuum preserving $N=4$ supersymmetry and the full $SO(4)\times SO(4)$ gauge symmetry with the corresponding cosmological constant given by
\begin{equation}
V_0=3gg_0\sin\alpha\, .
\end{equation}
The $AdS_4$ radius is given by
\begin{equation}
L=\sqrt{-\frac{3}{V_0}}=\sqrt{-\frac{1}{gg_0\sin\alpha}}\, .
\end{equation}
It should be noted that $V_0<0$ since the reality condition on $\phi$ implies $gg_0\sin\alpha<0$. We also note that for $\alpha=0$, the $AdS_4$ vacuum does not exist. This is in agreement with the fact that $\alpha=0$ together with the previous choice of $\alpha_0=0$ imply that the $SO(3)_0$ and $SO(3)_\alpha$ are both electrically gauged leading to no supersymmetric $AdS_4$ vacua \cite{AdS4_N4_Jan}.
\\
\indent We can bring this $SO(4)\times SO(4)$ vacuum to the origin of the scalar manifold by shifting the dilaton and axion or equivalently choosing 
\begin{equation}
\alpha=\frac{\pi}{2}\qquad \textrm{and}\qquad g_0=-g\, .
\end{equation}
This simply realizes the general result of \cite{Inverso_symplectic} that any values of $\alpha>0$ lead to physically equivalent theories up to a redefinition of the axion. Therefore, this $N=4$ $AdS_4$ vacuum is the same as that of the undeformed $SO(4)\times SO(4)$ gauged supergravity considered in \cite{4D_N4_flows}. Moreover, the scalar masses turn out to be independent of all electric-magnetic phases with all scalar masses equal $m^2L^2=-2$. This result is similar to the maximally supersymmetric $AdS_4$ vacuum at the origin of the scalar manifold in $\omega$-deformed $SO(8)$ $N=8$ gauged supergravity.
\\
\indent To look for other supersymmetric vacua, it is more convenient to first analyze the resulting BPS conditions. The conditions arising from $\delta\lambda^i_a$ reduce to the following two equations
\begin{eqnarray}
e^{-i\Lambda}\phi_1'+\frac{1}{2}e^{-\frac{\phi}{2}}\sinh2\phi_1\left[h_1\sinh\phi_1(ie^\phi\sin\beta_1-\cos\beta_1)\right.& &\nonumber \\
\left.+h_1\sin\beta_1\sinh\phi_1\chi-g_0\cosh\phi_1\right]&=&0,\qquad \\
e^{-i\Lambda}\phi_2'+\frac{i}{2}e^{-\frac{\phi}{2}}\sinh2\phi_2\left[h_2\sin\beta_2\sinh\phi_2(\chi+ie^\phi)-h_2\cos\beta_2\sinh\phi_2 \right.& &\nonumber \\
\left. +g\cosh\phi_2[\cos\alpha+\sin\alpha(\chi+ie^\phi)]\right]&=&0\, .\qquad 
\end{eqnarray}
In these equations and in the following analysis, we choose an upper sign choice in \eqref{Ap_general_eq} and \eqref{general_phase} for definiteness. This also identifies the trivial $SO(4)\times SO(4)$ critical point in the limit $r\rightarrow \infty$ in the RG flow solutions.
\\ 
\indent At the vacua with constant $\phi_1$ and $\phi_2$, we have $\phi_1'=\phi_2'=0$, and consistency of the above two equations imposes the following conditions
\begin{eqnarray}
h_1e^\phi\sin\beta_1\sinh\phi_1\sinh2\phi_1&=&0,\nonumber\\
e^{-\frac{\phi}{2}}\sinh2\phi_1\left[h_1\sin\beta_1\sinh\phi_1\chi-h_1\sinh\phi_1\cos\beta_1-g_0\cosh\phi_1\right]&=&0,\nonumber \\
e^{\frac{\phi}{2}}\sinh2\phi_2(h_2\sin\beta_2\sinh\phi_2-g\cosh\phi_2)&=&0,\nonumber \\
e^{\frac{\phi}{2}}\sinh2\phi_2\left[(h_2\sin\beta_2\sinh\phi_2-g\cosh\phi_2)\chi-h_2\cos\beta_2\sinh\phi_2\right]&=&0
\end{eqnarray}  
after setting $\alpha=\frac{\pi}{2}$. All these conditions imply that for $\phi_1\neq 0$ or $\phi_2\neq0$, $AdS_4$ vacua are possible only for 
\begin{equation}
\beta_1=0\qquad \textrm{and}\qquad  \beta_2=\frac{\pi}{2}\, .\label{beta_SO3_SO3}
\end{equation}
Accordingly, non-trivial $N=4$ supersymmetric $AdS_4$ vacua with at least $SO(3)_{\textrm{diag}}\times SO(3)_{\textrm{diag}}$ symmetry only exist in the undeformed $SO(4)\times SO(4)$ gauge group considered in \cite{4D_N4_flows}. We also note that for both $\phi_1$ and $\phi_2$ non-vanishing, the residual symmetry is give by $SO(3)\times SO(3)$ while for one of them vanishing, the vacua preserve larger $SO(4)\times SO(3)$ or $SO(3)\times SO(4)$ symmetries.

Further analysis also shows that consistency of the full BPS equations from $\delta\lambda^a_i=0$ conditions with $r$-dependent scalars also requires \eqref{beta_SO3_SO3} for any values of $\alpha=0,\frac{\pi}{2}$. This also implies that apart from the solutions found in \cite{4D_N4_flows} no supersymmetric domain walls or RG flows with at least $SO(3)_{\textrm{diag}}\times SO(3)_{\textrm{diag}}$ symmetry exist in the symplectically deformed $SO(4)\times SO(4)$ gauge group. Therefore, in $SO(3)_{\textrm{diag}}\times SO(3)_{\textrm{diag}}$ sector, no new $AdS_4$ vacua and holographic RG flows interpolating between them exist apart from those already given in \cite{4D_N4_flows}.

\subsection{Holographic RG flows and supersymmetric domain walls}
We end this section by giving supersymmetric domain wall solutions with $\phi_1=\phi_2=0$. The solutions can be considered to be solutions of pure $N=4$ gauged supergravity with $SO(3)_0\times SO(3)_\alpha$ gauge group. Although all values of $\alpha>0$ give physically equivalent gauged supergravities, we keep $\alpha$ to be arbitrary here for generality of the expressions. With $\phi_1=\phi_2=0$, the scalar potential reads
\begin{eqnarray}
V&=&4\left(\frac{\pd W}{\pd \phi}\right)^2+4e^{2\phi}\left(\frac{\pd W}{\pd \chi}\right)^2-3W^2 \nonumber \\
&=&2gg_0\sin\alpha-\frac{1}{2}g^2\sin^2\alpha e^\phi-\frac{1}{2}e^{-\phi}(g^2\cos^2\alpha+g_0^2)\nonumber \\
& &-\frac{1}{2}e^{-\phi}\chi(g^2\sin 2\alpha +g^2\sin^2\alpha\chi)
\end{eqnarray} 
with the superpotential given by
\begin{equation}
\mc{W}=\frac{1}{2}e^{-\frac{\phi}{2}}(ig\cos\alpha-g_0+e^\phi g\sin\alpha+ig\sin\alpha \chi).
\end{equation}
\indent With this superpotential, we find the following BPS equations 
\begin{eqnarray}
A'&=&W\nonumber \\
&=&\frac{1}{2}e^{-\frac{\phi}{2}}\sqrt{g^2(\cos\alpha+\sin\alpha \chi)^2+(g_0-ge^\phi\sin\alpha)^2},\\
\phi'&=&-4\frac{\pd W}{\pd \phi}\nonumber \\
&=&\frac{e^{-\frac{\phi}{2}}\left[g_0^2+(\cos^2\alpha-e^{2\phi\sin^2\alpha})g^2g^2\sin\alpha\chi (2\cos\alpha+\sin\alpha\chi)\right]}
{\sqrt{(g_0-e^\phi g\sin\alpha)^2+(g\cos\alpha+g\sin\alpha\chi)^2}},\\
\chi'&=&-4e^{2\phi}\frac{\pd W}{\pd \chi}\nonumber \\
&=&-\frac{2e^{\frac{3\phi}{2}}g^2\sin\alpha(\cos\alpha+\sin\alpha\chi)}
{\sqrt{(g_0-e^\phi g\sin\alpha)^2+(g\cos\alpha+g\sin\alpha\chi)^2}}\, .
\end{eqnarray}
Combining $A'$ and $\phi'$ equations with $\chi'$ equation, we obtain the solutions for $A$ and $\phi$ as functions of $\chi$ 
\begin{eqnarray}
A&=&\frac{1}{4}\ln\left[4g_0^2-C_0\chi-2g^2(1+\chi^2)+[C_0\chi+2g^2(\chi^2-1)]\cos2\alpha\right. \nonumber \\
& &\left.-(C_0+4g^2\chi)\sin2\alpha\right]\nonumber \\
& &-\frac{1}{2}\ln\left[4\sqrt{2}g_0\sqrt{4g_0^2-C_0\chi-2g^2(1+\chi^2)+[C_0\chi+2g^2(\chi^2-1)]\cos 2\alpha}\right.\nonumber \\
& &\left.\overline{-(C_0+4g^2\chi)\sin 2\alpha} +\sqrt{2}\left\{8g_0^2+C_0\chi(\cos2\alpha-1)-C_0\sin2\alpha\right\}\right],\\
\phi&=&\frac{1}{2}\ln\left[\frac{1}{2g^2}\textrm{csc}^2\alpha\left\{2g_0^2-g^2(1+\cos2\alpha)-C_0\sin\alpha\cos\alpha\right.\right.\nonumber \\
& &\left.\phantom{\frac{1}{2}}\left.-\sin\alpha \chi(4g^2\cos\alpha+C_0\sin\alpha)\right\} -\chi^2\right]
\end{eqnarray} 
with an integration constant $C_0$. We have neglected an additive integration constant for $A$ which can be removed by rescaling coordinates of $dx^2_{1,2}$. Finally, by changing to a new radial coordinate $\rho$ defined by 
\begin{equation}
\frac{dr}{d\rho}=e^{-\frac{\phi}{2}}\sqrt{(g_0-g\sin\alpha e^\phi)^2+(g\cos\alpha +g\sin\alpha \chi)^2},
\end{equation}
we obtain the solution for $\chi$ of the form
\begin{eqnarray}
2gg_0\sin\alpha(\rho-\rho_0)&=&\ln\left[8g_0^2+C_0\chi(\cos2\alpha-1)-C_0\sin 2\alpha\phantom{\sqrt{g_0^2}}\right.\nonumber \\
& &\left. +4g_0\sqrt{4g_0^2-2(\chi+\cot\alpha)\sin^2\alpha[C_0+2g^2(\chi+\cot\alpha)]}\right]\nonumber \\
& &+\ln\left[\frac{\sin\alpha}{2\sqrt{2}gg_0(\chi+\cot\alpha)}\right]
\end{eqnarray}
with another integration constant $\rho_0$ which can also be removed by shifting the coordinate $\rho$. For $\alpha=\frac{\pi}{2}$, this is the holographic RG flow from a three-dimensional $N=4$ SCFT to a non-conformal field theory in the IR given in \cite{4D_N4_flows}. However, the $\chi(\rho)$ solution has not been given. Accordingly, the present result should fill this gap.
\\
\indent For $\alpha=0$, the BPS equations simplify considerably to
\begin{equation}
A'=\frac{1}{2}e^{-\frac{\phi}{2}}\sqrt{g^2+g_0^2}\qquad \textrm{and}\qquad \phi'=e^{-\frac{\phi}{2}}\sqrt{g^2+g_0^2}
\end{equation}  
together with $\chi'=0$. The solution takes a simple form
\begin{equation}
\phi=2\ln \left[\frac{1}{2}\sqrt{g^2+g_0^2}(r-r_0)\right]\qquad \textrm{and}\qquad A=\frac{1}{2}\phi
\end{equation}
with constant $\chi=\chi_0$. This gives a half-supersymmetric domain wall vacuum of $SO(3)_0\times SO(3)_{\alpha=0}$ gauged supergravity. 

\section{$SO(2)\times SO(2)\times SO(2)\times SO(2)$ sector}\label{SO2_4_sector}
The $SO(2)\times SO(2)\times SO(2)\times SO(2)$ sector of the undeformed $SO(4)\times SO(4)$ gauged supergravity has been considered recently in \cite{N4_Janus} in which a number of holographic RG flows and Janus solutions have been found. In the present paper, we will consider the same sector with electric-magnetic phases. 

As shown in \cite{N4_Janus}, there are four $SO(2)\times SO(2)\times SO(2)\times SO(2)$ singlet scalars from $SO(6,6)/SO(6)\times SO(6)$ corresponding to non-compact generators $Y_{33}$, $Y_{36}$, $Y_{63}$ and $Y_{66}$ in terms of which the coset representative can be written as
\begin{equation}
\mc{V}=e^{\phi_1 Y_{33}}e^{\phi_2 Y_{36}}e^{\phi_3 Y_{63}}e^{\phi_4 Y_{66}}\, .\label{SO2_4_coset}
\end{equation}
Together with the dilaton and axion from the gravity multiplet, there are six scalars in the $SO(2)\times SO(2)\times SO(2)\times SO(2)$ sector. The kinetic term for these six scalars takes the form
\begin{eqnarray}
\mc{L}_{\textrm{kin}}&=&\frac{1}{2}G_{rs}{\Phi^r}'{\Phi^s}'\nonumber \\
&=&-\frac{1}{4}(\phi'^2+e^{-2\phi}\chi'^2)-\frac{1}{16}\left[6+\cosh2(\phi_2-\phi_3)\right. \nonumber \\
& &
\left.+\cosh2(\phi_2+\phi_3)+2\cosh2\phi_4(\cosh2\phi_2\cosh2\phi_3-1)\right]\phi'^2_1\nonumber \\
& &-\cosh\phi_2\cosh\phi_4\sinh\phi_3\sinh\phi_4\phi'_1\phi'_2-\cosh\phi_3\cosh\phi_4\sinh\phi_2\sinh\phi_4\phi'_1\phi'_3\nonumber \\
& &+\sinh\phi_2\sinh\phi_3\phi'_1\phi'_4-\frac{1}{2}\cosh^2\phi_4\phi'^2_2-\frac{1}{2}\cosh^2\phi_4\phi'^2_3-\frac{1}{2}\phi'^2_4\label{scalar_kin}
 \end{eqnarray} 
in which we have introduced a symmetric matrix $G_{rs}$ and a notation $\Phi^r=(\phi,\chi,\phi_1,\phi_2,\phi_3,\phi_4)$, with $r,s=1,2,\ldots, 6$ for later convenience.

The resulting scalar potential is given by
\begin{eqnarray}
V&=&-\frac{1}{4}e^{-\phi}[g^2(1+\cos2\alpha)+2g_0^2+2g^2\sin\alpha\chi(2\cos\alpha+\sin\alpha\chi)]-\frac{1}{2}e^\phi g^2\sin^2\alpha \nonumber \\
& &+2gg_0\sin\alpha \cosh\phi_1\cosh\phi_2\cosh\phi_3\cosh\phi_4\, .\label{Potential_SO2_4}
\end{eqnarray}
In this case, the phases $\beta_1$ and $\beta_2$ do not appear in the scalar potential. In addition, as in the undeformed $SO(4)\times SO(4)$ gauge group, the potential admits only a trivial $AdS_4$ critical point at $\phi=\chi=\phi_1=\phi_2=\phi_3=\phi_4=0$ for $\alpha=\frac{\pi}{2}$ and $g_0=-g$.

Since all values of $\alpha>0$ are equivalent to $\alpha=\frac{\pi}{2}$, we also see that in this sector, no new results arise from the symplectic deformation. However, the RG flow solutions considered in \cite{N4_Janus} are obtained only in a subtrucation with $SO(2)\times SO(2)\times SO(2)\times SO(3)$ or $SO(2)\times SO(2)\times SO(3)\times SO(2)$ symmetries corresponding to setting $\phi_2=\phi_4=0$ or $\phi_1=\phi_3=0$, respectively. In the present paper, we will consider the most general solutions in the full $SO(2)\times SO(2)\times SO(2)\times SO(2)$ sector with all scalars non-vanishing. 

In this case, the $A_1^{ij}$ tensor is diagonal and takes the form of
\begin{equation}
A_1^{ij}=\textrm{diag}(\mc{A}_-,\mc{A}_+,\mc{A}_+,\mc{A}_-)
\end{equation}
with the two eigenvalues given by
\begin{eqnarray}
\mc{A}_\pm&=&\frac{1}{4}e^{-\frac{\phi}{2}}\left[3\cosh\phi_4[g\cosh\phi_3(e^\phi\sin\alpha+i\cos\alpha)\pm i g_0\sinh\phi_1\sinh\phi_3] \right.\nonumber \\
& &\left. -3g_0\cosh\phi_1(\cosh\phi_2\mp i\sinh\phi_2\sinh\phi_4)+3ig\sin\alpha \cosh\phi_3\cosh\phi_4\chi\right].\nonumber \\
& &
\end{eqnarray}
$\mc{A}_-$ and $\mc{A}_+$ eigenvalues correspond to unbroken $N=2$ supersymmetry with the Killing spinors given by $\epsilon^{1,4}$ and $\epsilon^{2,3}$, respectively. The two choices are equivalent, and we will choose $\epsilon^{1,4}$ as Killing spinors for definiteness. 

\begin{figure}
  \centering
  \begin{subfigure}[b]{0.45\linewidth}
    \includegraphics[width=\linewidth]{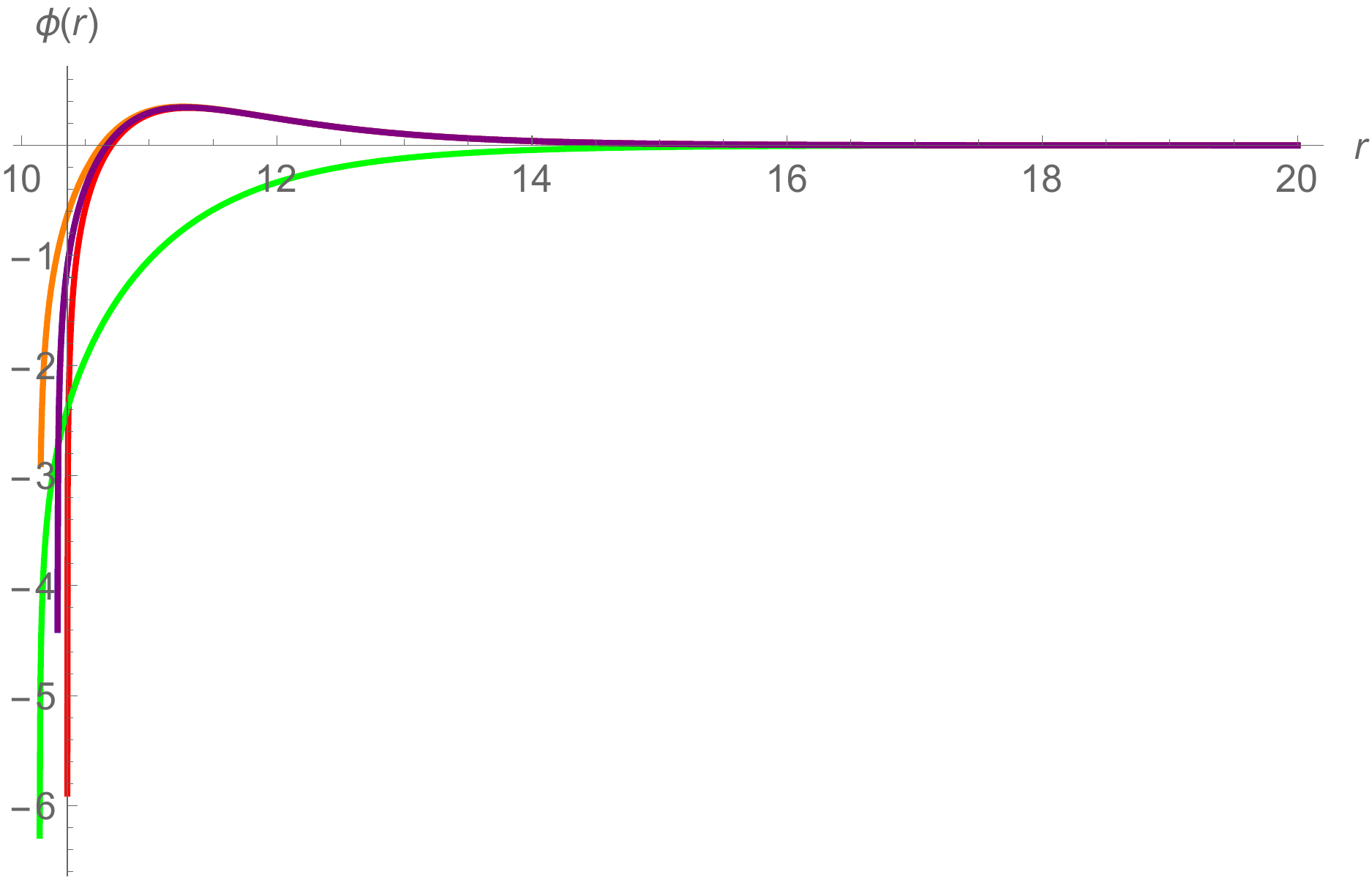}
  \caption{$\phi(r)$ solution}
  \end{subfigure}
  \begin{subfigure}[b]{0.45\linewidth}
    \includegraphics[width=\linewidth]{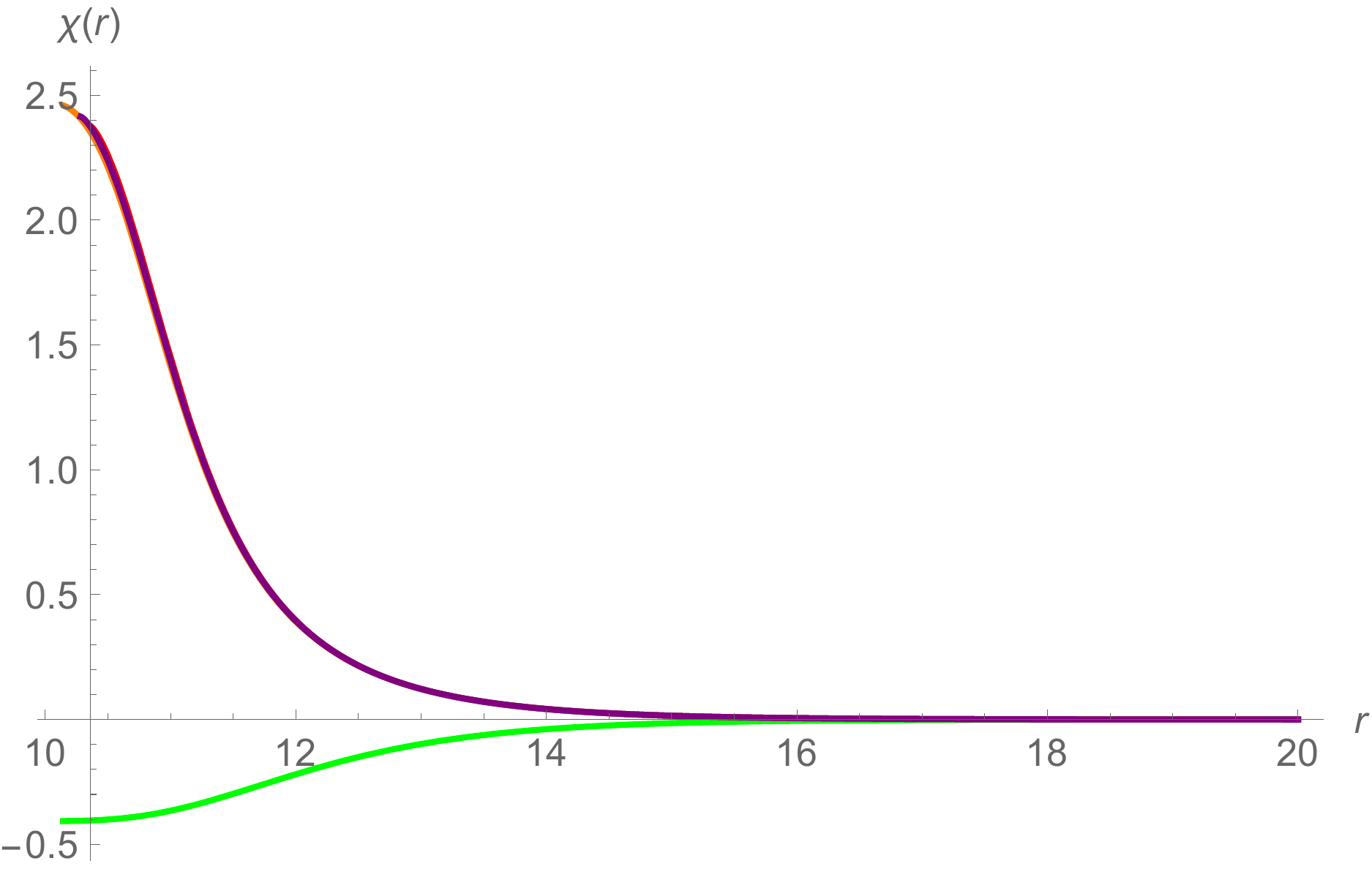}
  \caption{$\chi(r)$ solution}
  \end{subfigure}\\
    \begin{subfigure}[b]{0.45\linewidth}
    \includegraphics[width=\linewidth]{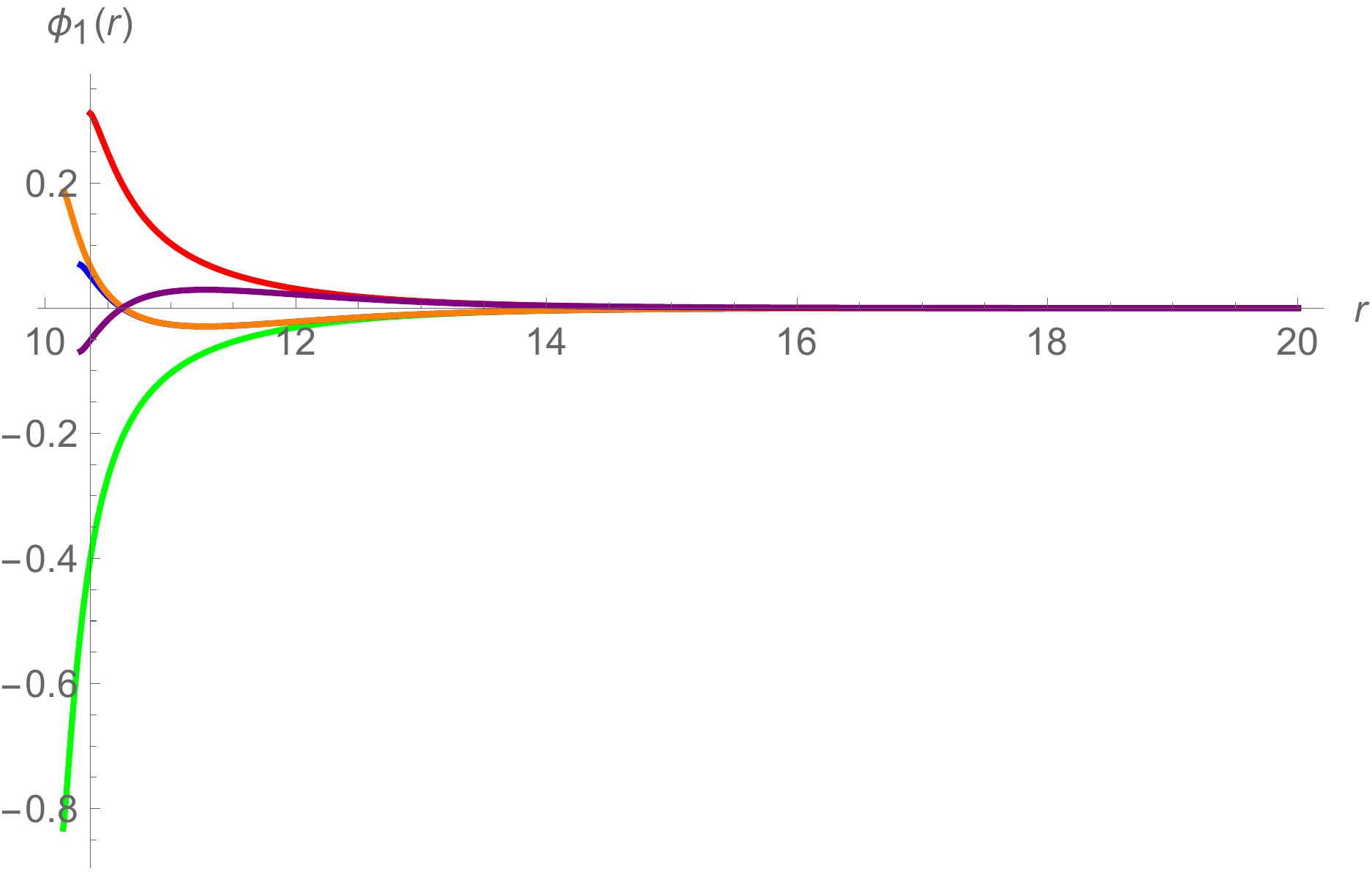}
  \caption{$\phi_1(r)$ solution}
  \end{subfigure}
  \begin{subfigure}[b]{0.45\linewidth}
    \includegraphics[width=\linewidth]{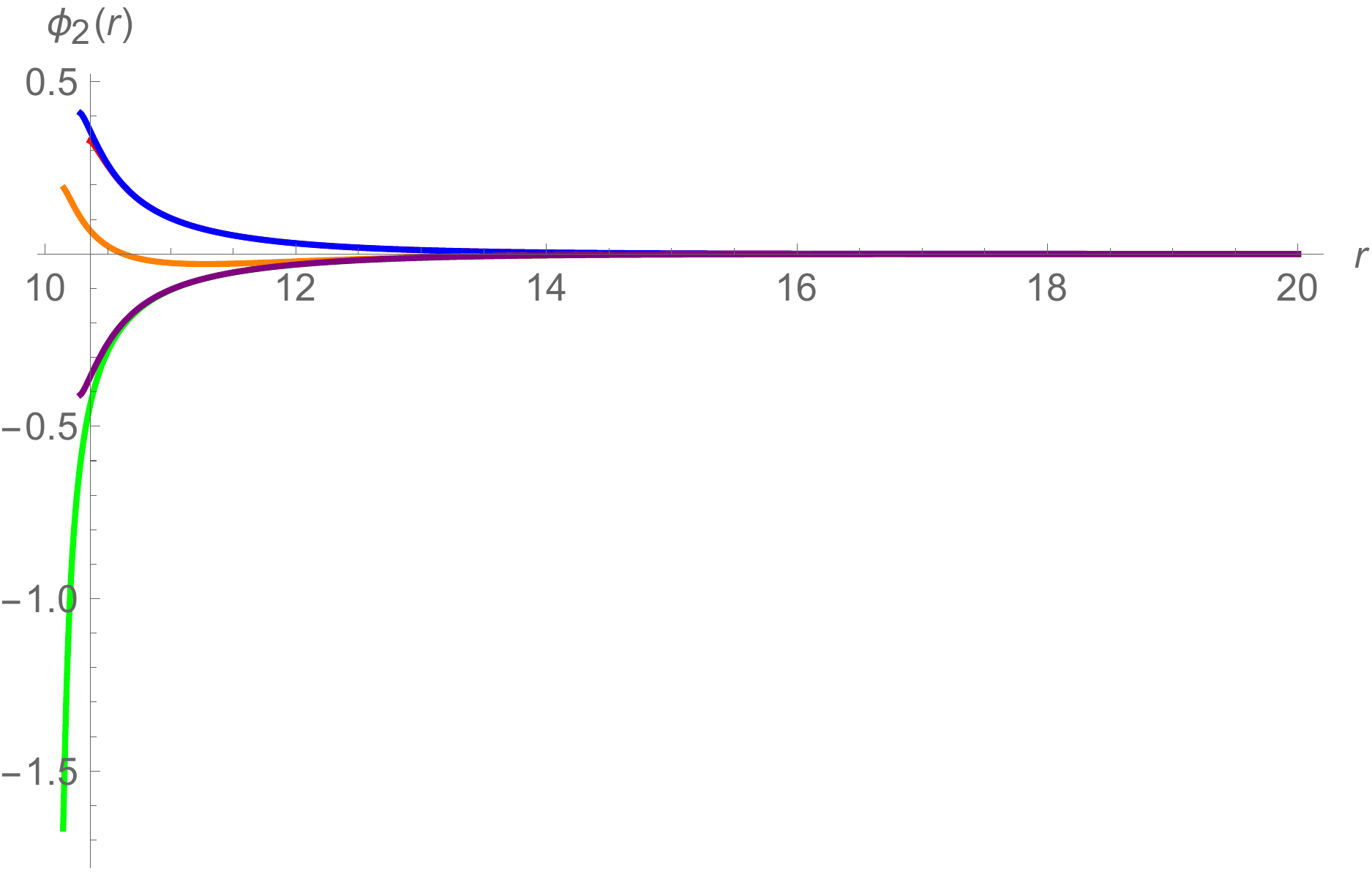}
  \caption{$\phi_2(r)$ solution}
  \end{subfigure}\\
  \begin{subfigure}[b]{0.45\linewidth}
    \includegraphics[width=\linewidth]{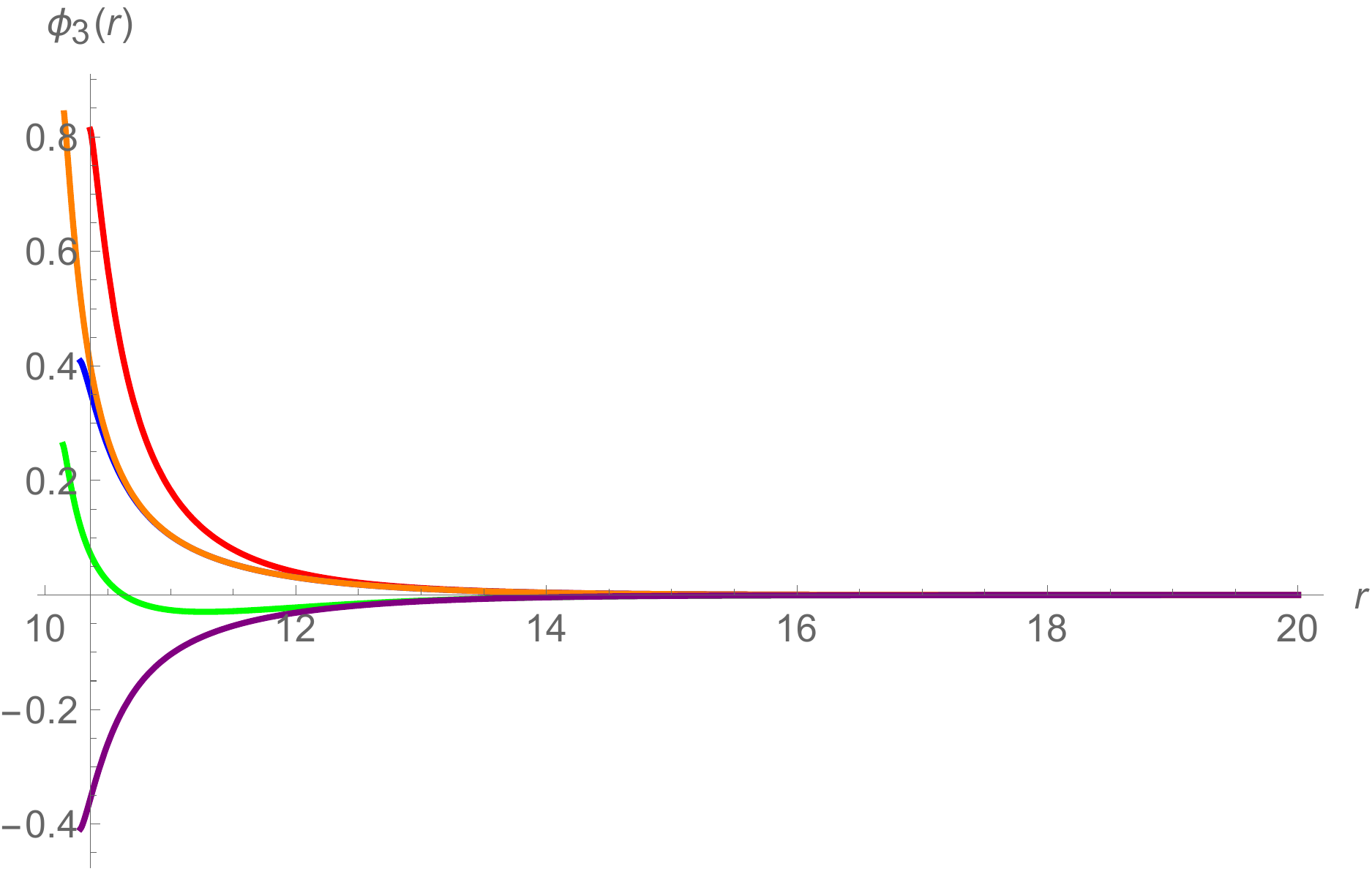}
  \caption{$\phi_3(r)$ solution}
  \end{subfigure}
  \begin{subfigure}[b]{0.45\linewidth}
    \includegraphics[width=\linewidth]{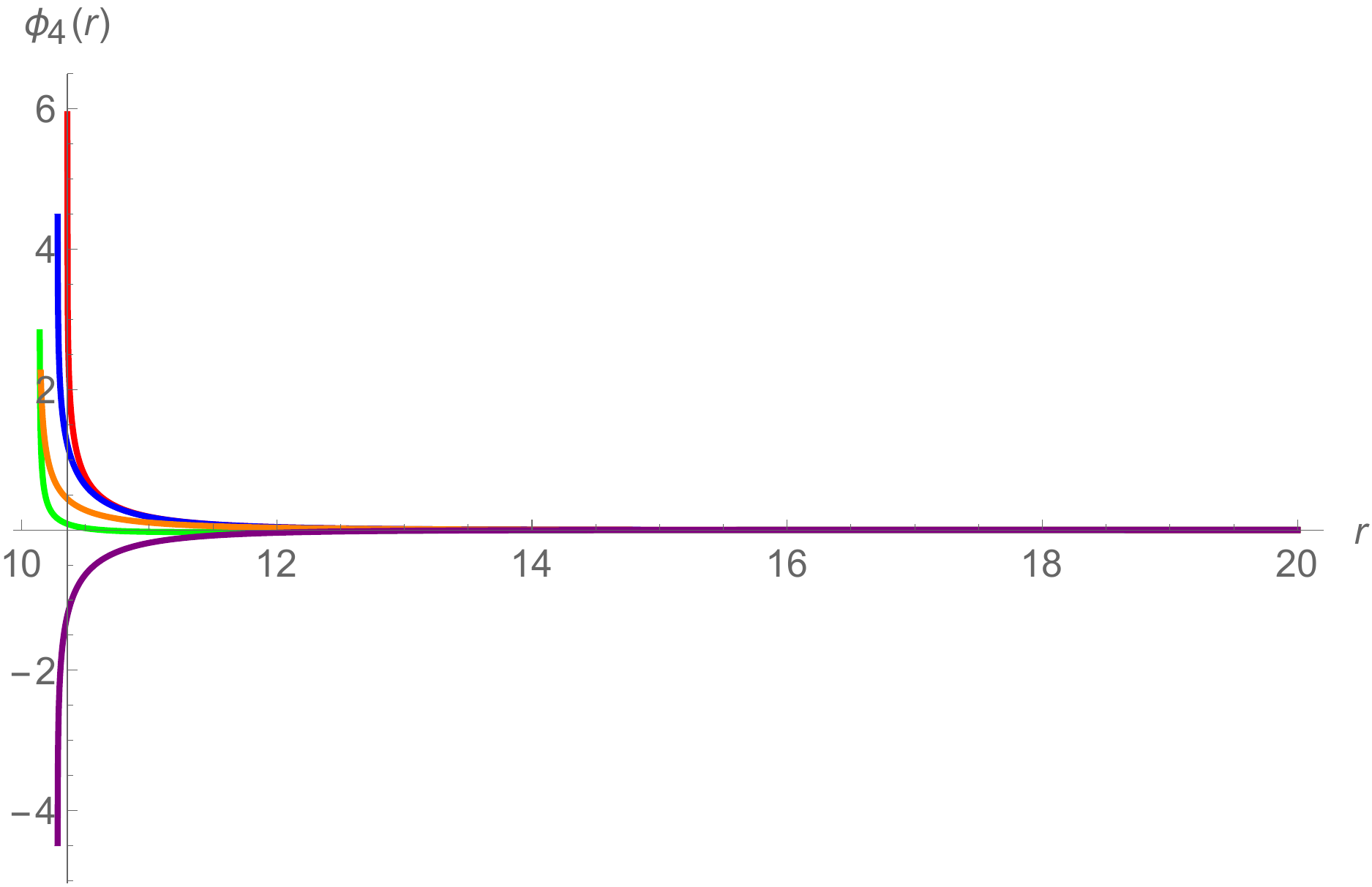}
  \caption{$\phi_4(r)$ solution}
  \end{subfigure}\\
   \begin{subfigure}[b]{0.45\linewidth}
    \includegraphics[width=\linewidth]{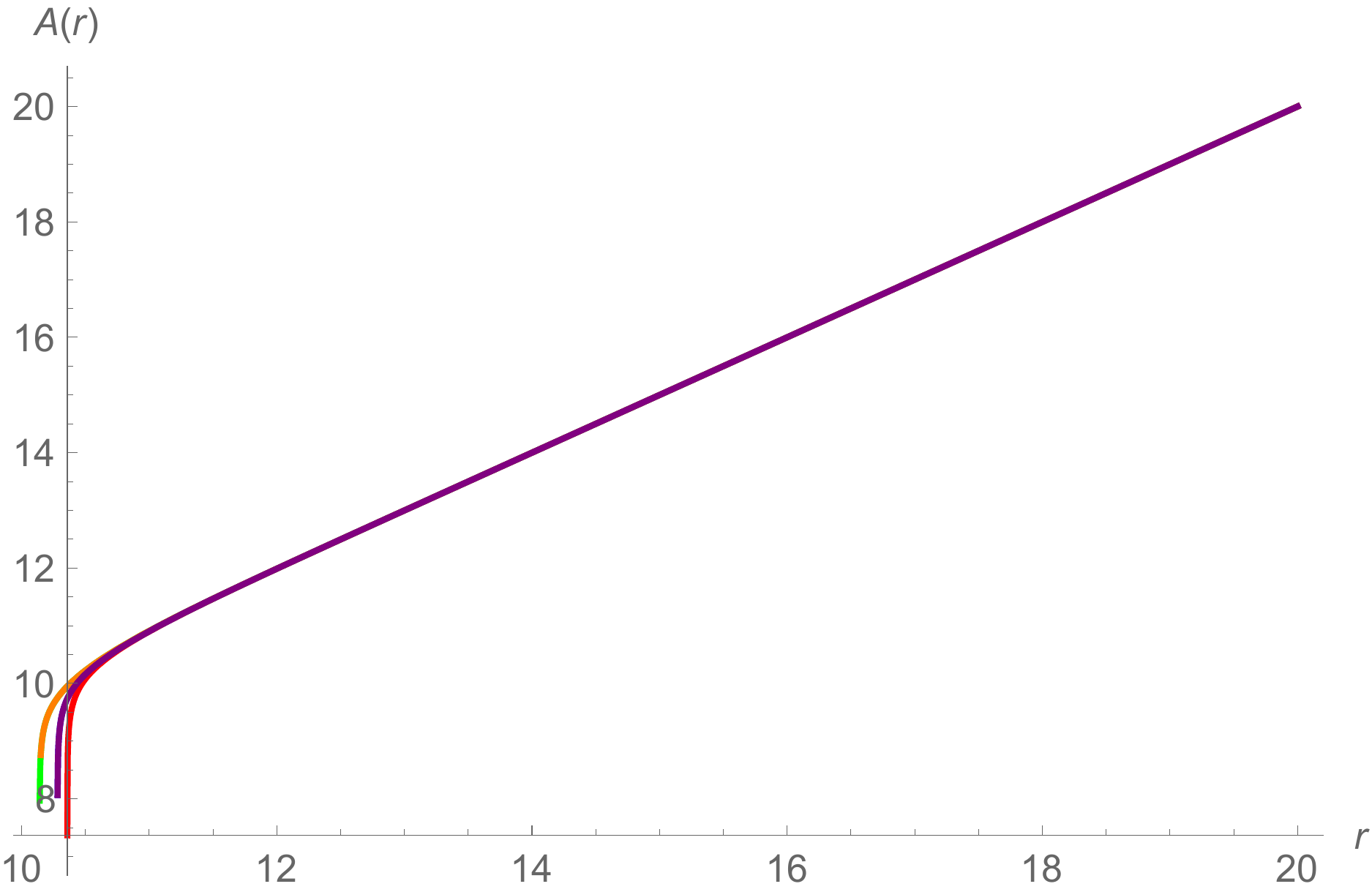}
  \caption{$A(r)$ solution}
   \end{subfigure} 
 \begin{subfigure}[b]{0.45\linewidth}
    \includegraphics[width=\linewidth]{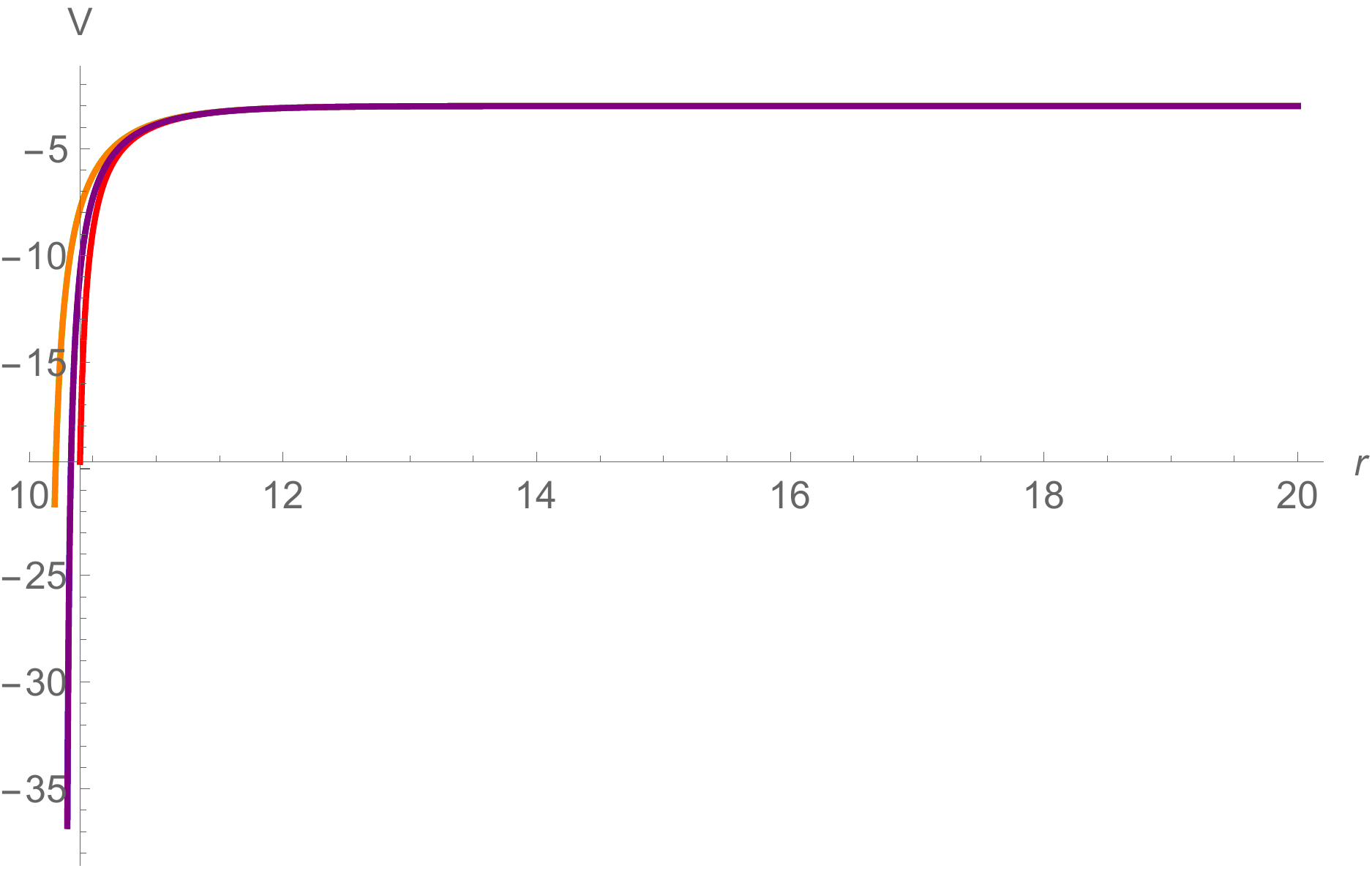}
  \caption{Scalar potential}
   \end{subfigure} 
  \caption{Examples of $N=2$ RG flows from the $N=4$ SCFT with $SO(4)\times SO(4)$ symmetry in the UV to non-conformal phases in the IR with $g=1$ and $\alpha=\frac{\pi}{2}$.}
  \label{Fig1}
\end{figure}

With $\epsilon^{2,3}=0$ and the superpotential of the form 
\begin{eqnarray}
\mc{W}&=&\frac{2}{3}\mc{A}_-\nonumber \\
&=&\frac{1}{2}e^{-\frac{\phi}{2}}\left[\cosh\phi_4[g\cosh\phi_3(e^\phi\sin\alpha+i\cos\alpha)-g_0\sinh\phi_1\sinh\phi_3] \right.\nonumber \\
& &\left. -g_0\cosh\phi_1(\cosh\phi_2+i\sinh\phi_2\sinh\phi_4)+ig\sin\alpha\cosh\phi_3\cosh\phi_4\chi\right],\quad
\end{eqnarray}
we find that all the BPS equations can be written collectively as
\begin{equation}
{\Phi^r}'=2G^{rs}\frac{\pd W}{\pd \Phi^s}\, .
\end{equation}
$G^{rs}$ is the inverse of the scalar matrix $G_{rs}$ which is in turn given by 
\begin{equation}
G_{rs}=\begin{pmatrix}	-\frac{1}{2} & 0 & 0 &0 &0 &0 \\
						0 &-\frac{1}{2}e^{-2\phi}  & 0 &0 &0 &0 \\
						  0&0  & \square & \Delta_1 & \Delta_2 & \Delta_3\\
						  0 &  0& \Delta_1 &-\cosh^2\phi_4 & 0&  0\\
						    0& 0 & \Delta_2 &0 &-\cosh^2\phi_4 & 0 \\
						    0 & 0 & \Delta_3 &0 &0 & -1
			\end{pmatrix}
\end{equation}
with
\begin{eqnarray}
\square&=&\frac{1}{8}\left[2\cosh2\phi_4(1-\cosh2\phi_2\cosh2\phi_3)-\cosh2(\phi_2-\phi_3)\right.\nonumber \\
& &\left.-\cosh2(\phi_2+\phi_3)-6\right],\nonumber \\
\Delta_1&=&-\cosh\phi_2\cosh\phi_4\sinh\phi_3\sinh\phi_4,\nonumber \\
\Delta_2&=&-\cosh\phi_3\cosh\phi_4\sinh\phi_2\sinh\phi_4,\nonumber \\
\Delta_3&=&\sinh\phi_2\sinh\phi_3\, .
\end{eqnarray}
We also note that the scalar potential can be written as
\begin{equation}
V=-2G^{rs}\frac{\pd W}{\pd \Phi^r}\frac{\pd W}{\pd \Phi^s}-3W^2\, .
\end{equation}
\indent For $\alpha=\frac{\pi}{2}$ and $g_0=-g$, the explicit form of the BPS equations reads
\begin{eqnarray}
W\phi'&=&-g^2\cosh\phi_3\cosh\phi_4(\cosh\phi_1\cosh\phi_2+e^\phi\cosh\phi_3\cosh\phi_4)\nonumber \\
& &+\frac{1}{2}e^{-\phi}g^2\left[\cosh\phi_4(\sinh\phi_1\sinh\phi_3+\cosh\phi_3 \chi)+\cosh\phi_1\sinh\phi_2\sinh\phi_4\right]^2\nonumber \\
& &+\frac{1}{2}e^{-\phi}g^2(\cosh\phi_1\cosh\phi_2+e^\phi\cosh\phi_3\cosh\phi_4)^2,\\
W\chi'&=&-e^{\phi}g^2\cosh\phi_3\cosh\phi_4\times \nonumber \\
& &(\cosh\phi_4\sinh\phi_1\sinh\phi_3+\cosh\phi_1\sinh\phi_2\sinh\phi_4+\cosh\phi_3\cosh\phi_4\chi),\\
W\phi_1'&=&-\frac{1}{2}e^{-\phi}g^2\left[e^\phi\cosh\phi_4\textrm{sech}\phi_2\textrm{sech}\phi_3\sinh\phi_1+\cosh\phi_1\tanh\phi_3\chi+\cosh\phi_1\times \right.\nonumber \\
& &\left. (\textrm{sech}^2\phi_3\sinh\phi_1-e^\phi\sinh\phi_4\tanh\phi_2\tanh\phi_3+\sinh\phi_1\tanh^2\phi_3)\right],\\
W\phi_2'&=&-\frac{1}{16}e^{-\phi}g^2\left[e^\phi\cosh\phi_1[6+2\cosh2\phi_3+\cosh2(\phi_3-\phi_4)-2\cosh2\phi_4 \right.\nonumber \\
& &+\cosh2(\phi_3+\phi_4)]\textrm{sech}\phi_3\textrm{sech}\phi_4\sinh\phi_2-8e^\phi\sinh\phi_1\sinh\phi_4\tanh\phi_3\nonumber \\
& &\left. +4\cosh^2\phi_1\sinh2\phi_2+8\cosh\phi_1\cosh\phi_2\textrm{sech}\phi_3\tanh\phi_4\chi \right],\\
W\phi_3'&=&\frac{1}{16}e^{-\phi}g^2\left[\sinh2(\phi_1-\phi_3)+(2-4e^{2\phi})\sinh2\phi_3-\sinh2(\phi_1+\phi_3) \right.\nonumber \\
& &-8e^\phi\cosh\phi_1\textrm{sech}\phi_4\sinh\phi_3(\textrm{sech}\phi_2+\cosh^2\phi_4\sinh\phi_2\tanh\phi_2)\nonumber \\
& &\left. +8e^\phi\sinh\phi_1\sinh\phi_4\tanh\phi_2-4\chi(2\cosh2\phi_3\sinh\phi_1+\sinh2\phi_3\chi) \right],\\
W\phi_4'&=&\frac{1}{8}e^{-\phi}g^2\left[\sin2\phi_4(\cosh2\phi_1-2e^{2\phi}\cosh^2\phi_3-\cosh2\phi_3\sinh^2\phi_1)\right.\nonumber \\
& &-2e^\phi\cosh\phi_1\textrm{sech}\phi_2\textrm{sech}\phi_3\sinh\phi_4(\cosh2\phi_2+\cosh2\phi_3)\nonumber \\
& &-4\cosh\phi_4\sinh\phi_1(2\cosh\phi_1\cosh\phi_4\sinh\phi_2\sinh\phi_3+e^\phi\tanh\phi_2\tanh\phi_3)\nonumber \\
& &-2\chi(\sinh\phi_1\sinh2\phi_3\sinh2\phi_4+\cosh^2\phi_3\sinh2\phi_4\chi)-\cosh^2\phi_1\cosh2\phi_2\times\nonumber \\
& &\left.\sinh2\phi_4 -4\chi\cosh\phi_1\sinh\phi_2(\cosh\phi_3\cosh2\phi_4+\sinh\phi_3\tanh\phi_3)\right].
\end{eqnarray}
We numerically solve these equations with some examples of possible solutions given in figure \ref{Fig1}.

The solutions interpolate between the supersymmetric $AdS_4$ vacuum with $SO(4)\times SO(4)$ symmetry and singular geometries with diverging scalars. In all solutions, we see that $V\rightarrow -\infty$ implying that all singularities are physical by the criterion given in \cite{Gubser_singularity}. Indeed, we find the scalar potential, for $\alpha=\frac{\pi}{2}$ and $g_0=-g$,  
\begin{equation}
V=-\frac{1}{2}e^{-\phi}g^2(1+e^{2\phi}+\chi^2+4e^\phi \cosh\phi_1\cosh\phi_2\cosh\phi_3\cosh\phi_4)
\end{equation}
which is always bounded from above. Therefore, for any diverging behaviors of scalar fields, all possible solutions in $SO(2)\times SO(2)\times SO(2)\times SO(2)$ truncation are physically acceptable and describe RG flows from the dual $N=4$ SCFT to various non-conformal phases in the IR. Using the scalar masses $m^2L^2=-2$ and asymptotic behaviors near the $AdS_4$ critical point 
\begin{equation}
\phi\sim \chi\sim \phi_i\sim e^{-gr}\sim e^{-\frac{r}{L}},\qquad i=1,2,3,4,
\end{equation}
we can determine that the flows are driven by relevant operators of dimensions $\Delta=1,2$ and preserve $N=2$ supersymmetry and $SO(2)\times SO(2)\times SO(2)\times SO(2)$ symmetry along the flows. This extends the results of \cite{N4_Janus}, in which only particular truncations to $SO(2)\times SO(2)\times SO(3)\times SO(2)$ or $SO(2)\times SO(2)\times SO(2)\times SO(3)$ singlet scalars have been considered, to the full $SO(2)\times SO(2)\times SO(2)\times SO(2)$ sector.  

\section{$SO(3)_{\textrm{diag}}\times SO(3)$ sector}\label{SO3d_SO3_sector}
In this section, we look at another scalar sector with residual symmetry $SO(3)_{\textrm{diag}}\times SO(3)$. Since this sector has not previously been studied, we will give the construction in more detail than the previous two cases. The $36$ scalars in $SO(6,6)/SO(6)\times SO(6)$ coset transform under the $SO(6)\times SO(6)$ compact subgroup as $(\mathbf{6},\mathbf{6})$. We further decompose each $SO(6)$ factor into the $SO(3)\times SO(3)$ subgroup under which the fundamental representation $\mathbf{6}$ decomposes as $(\mathbf{3},\mathbf{1})+(\mathbf{1},\mathbf{3})$. The $36$ scalar fields then transform under $SO(3)\times SO(3)\times SO(3)\times SO(3)$ as
\begin{equation}
(\mathbf{3},\mathbf{3},\mathbf{1},\mathbf{1})+(\mathbf{3},\mathbf{1},\mathbf{1},\mathbf{3})+(\mathbf{1},\mathbf{3},\mathbf{3},\mathbf{1})+(\mathbf{1},\mathbf{3},\mathbf{1},\mathbf{3}).
\end{equation}
By taking a diagonal subgroup of the first three $SO(3)$ factors, we find
\begin{eqnarray}
& &(\mathbf{3}\times \mathbf{3},\mathbf{1})+(\mathbf{3},\mathbf{3})+(\mathbf{3}\times \mathbf{3},\mathbf{1})+(\mathbf{3},\mathbf{3})\nonumber \\
\rightarrow& &2\times [(\mathbf{1},\mathbf{1})+(\mathbf{3},\mathbf{1})+(\mathbf{5},\mathbf{1})+(\mathbf{3},\mathbf{3})].\label{rep_decom}
\end{eqnarray}
Accordingly, there are two $SO(3)_{\textrm{diag}}\times SO(3)$ singlets corresponding to the two $(\mathbf{1},\mathbf{1})$ representations. These two singlets correspond to the following $SO(6,6)$ non-compact generators
\begin{equation}
\hat{Y}_1=Y_{11}+Y_{22}+Y_{33}+Y_{44}\qquad \textrm{and}\qquad \hat{Y}_3=Y_{51}+Y_{62}+Y_{73}+Y_{84}\, .
\end{equation}
If we consider an even smaller $SO(3)_{\textrm{diag}}$ residual symmetry, there are two additional singlets obtained from the last representation $(\mathbf{3},\mathbf{3})$ in \eqref{rep_decom}. These singlets correspond to the non-compact generators
\begin{equation}
\hat{Y}_2=Y_{15}+Y_{26}+Y_{37}+Y_{48}\qquad \textrm{and}\qquad \hat{Y}_4=Y_{55}+Y_{66}+Y_{77}+Y_{88}\, .
\end{equation}
We also note that $\hat{Y}_1$ and $\hat{Y}_4$ are $SO(3)_{\textrm{diag}}\times SO(3)_{\textrm{diag}}$ singlets considered in section \ref{SO3_SO3_sector}. The coset representative for $SO(3)_{\textrm{diag}}$ sector can then be written as
\begin{equation}
\mc{V}=e^{\phi_1\hat{Y}_1}e^{\phi_2\hat{Y}_2}e^{\phi_3\hat{Y}_3}e^{\phi_4\hat{Y}_4}\, .
\end{equation}
However, it turns out that the resulting scalar potential and BPS equations are highly complicated. Therefore, we refrain from giving the complete analysis of this sector but simply note that the $A^{ij}_1$ tensor takes the form
\begin{equation}
A^{ij}_1=\textrm{diag}(\mc{A},\mc{B},\mc{B},\mc{B}).\label{A1_SO3diag}
\end{equation}
\indent To make the analysis more traceable, we will perform further truncation to $SO(3)_{\textrm{diag}}\times SO(3)$ singlet scalars by setting $\phi_2=\phi_4=0$. Although this subtruncation leads to simpler expressions for the results, there are still some new interesting features. For simplicity of the results, in this section, we will set $\alpha=\frac{\pi}{2}$ and $g_0=-g$. The $A^{ij}_1$ tensor for the subtruncation still takes the form \eqref{A1_SO3diag} with the eigenvalues given by
\begin{eqnarray}
\mc{A}&=&\frac{3}{4}e^{-\frac{\phi}{2}}\left[g(\cosh\phi_1+i\sinh\phi_1\sinh\phi_3)^3-h_1\cos\beta_1(\sinh\phi_1+i\cosh\phi_1\sinh\phi_3)^3\right.\nonumber \\
& &+e^\phi[g\cosh^3\phi_3+h_1\sin\beta_1(i\sinh\phi_1-\cosh\phi_1\sinh\phi_3)^3]\nonumber \\
& &\left. +[h_1\sin\beta_1(\sinh\phi_1+i\cosh\phi_1\sinh\phi_3)^3+ig\cosh^3\phi_3]\chi\right],\\
\mc{B}&=&\frac{3}{4}e^{-\frac{\phi}{2}}\left[g(\cosh\phi_1\sinh^2\phi_1\sinh^2\phi_3-i\cosh^2\phi_1\sinh\phi_1\sinh\phi_3 \right.\nonumber\\ 
& &+\cosh^3\phi_1+e^\phi\cosh^3\phi_3-i\sinh^3\phi_1\sinh^3\phi_3+i\cosh^3\phi_3\chi)\nonumber \\
& &-h_1(\sinh\phi_1-i\cosh\phi_1\sinh\phi_3)^2(\sinh\phi_1+i\cosh\phi_1\sinh\phi_3)\times \nonumber\\
& &\left.(\cos\beta_1+ie^\phi\sin\beta_1-\sin\beta_1\chi) \right].
\end{eqnarray}
We find that the first eigenvalue $\mc{A}$ gives rise to the superpotential of the form
\begin{eqnarray}
\mc{W}&=&\frac{1}{2}e^{\frac{\phi}{2}}\left[g\cosh^3\phi_3+h_1\sin\beta_1(i\sinh\phi_1-\cosh\phi_1\sinh\phi_3)^3\right]\nonumber \\
& &+\frac{1}{2}e^{-\frac{\phi}{2}}\left[g(\cosh\phi_1+i\sinh\phi_1\sinh\phi_3)^3-(\sinh\phi_1+i\cosh\phi_1\sinh\phi_3)^3\times\right.\nonumber \\
& &\left. h_1\cos\beta_1\right]+\frac{1}{2}e^{-\frac{\phi}{2}}\left[ig\cosh^3\phi_3+h_1\sin\beta_1(\sinh\phi_1+i\cosh\phi_1\sinh\phi_3)^3\right]\chi\, .\nonumber \\
& &
\end{eqnarray}
Accordingly, RG flow solutions will preserve only $N=1$ supersymmetry. It should be noted that for $\phi_3=0$ or $\phi_1=0$, the two eigenvalues $\mc{A}$ and $\mc{B}$ are equal leading to an enhanced $N=4$ supersymmetry.

The scalar potential can be written in terms of the superpotential as follows
\begin{equation}
V=4\left(\frac{\pd W}{\pd \phi}\right)^2+4e^{2\phi}\left(\frac{\pd W}{\pd \chi}\right)^2+\frac{2}{3}\textrm{sech}^2\phi_3\left(\frac{\pd W}{\pd \phi_1}\right)^2+\frac{2}{3}\left(\frac{\pd W}{\pd \phi_3}\right)^2-3W^2\, .
\end{equation}
The explicit form of $V$ is given in the appendix. We note that only $\beta_1$ appears in the results. We can also make another subtruncation by setting $\phi_1=\phi_3=0$ in which only $\beta_2$ appears. This gives similar results with $(\phi_1,\phi_3)$ and $(\phi_2,\phi_4)$ together with $\beta_1$ and $\beta_2$ interchanged. On the other hand, if we consider the full $SO(3)_{\textrm{diag}}$ sector, both $\beta_1$ and $\beta_2$ appear in the scalar potential and the superpotential. 

From the superpotential given above, we have not found any non-trivial supersymmetric $AdS_4$ critical points for arbitrary values of $\beta_1$. However, there are two $AdS_4$ vacua for particular values of $\beta_1=0$ and $\beta_1=\frac{\pi}{2}$. These are given by
\begin{eqnarray}
i:\,\,\,\beta_1=0;\qquad & &\phi_3=\chi=0,\qquad \phi_1=\frac{1}{2}\ln\left[\frac{h_1+g}{h_1-g}\right],\nonumber \\
& &\phi=-\frac{1}{2}\ln\left[1-\frac{g^2}{h_1^2}\right],\qquad V_0=-\frac{3g^2h_1}{\sqrt{h_1^2-g^2}}
\end{eqnarray}
and
\begin{eqnarray}
ii:\,\,\,\beta_1=\frac{\pi}{2};\qquad & &\phi_1=\chi=0,\qquad \phi_3=\frac{1}{2}\ln\left[\frac{h_1+g}{h_1-g}\right],\nonumber \\
& &\phi=\frac{1}{2}\ln\left[1-\frac{g^2}{h_1^2}\right],\qquad V_0=-\frac{3g^2h_1}{\sqrt{h_1^2-g^2}}\, .
\end{eqnarray}
Both of these critical points preserve $N=4$ supersymmetry since the two eigenvalues of $A^{ij}_1$ are degenerate for $\phi_1=0$ or $\phi_3=0$ as previously mentioned. Critical point $i$ preserves $SO(3)_{\textrm{diag}}\times SO(3)_\alpha\times SO(3)_2$ with the $SO(3)_{\textrm{diag}}$ being a diagonal subgroup of $SO(3)_0\times SO(3)_1$. This is not a new $AdS_4$ critical point but one of the $N=4$ critical points identified in \cite{4D_N4_flows}. On the other hand, critical point $ii$ preserves $SO(3)_0\times SO(3)_{\textrm{diag}}\times SO(3)_2$ with the $SO(3)_{\textrm{diag}}$ being a diagonal subgroup of $SO(3)_\alpha\times SO(3)_1$. These two critical points appear to be related to each other by interchanging scalars and $SO(3)$ factors within the unbroken symmetry as well as a sign flip in the dilaton. However, the two vacua correspond to different values of electric-magnetic phase $\beta_1$. 

By the same procedure as in the previous sections, we find the BPS equations of the form
\begin{eqnarray}
\phi'&=&-4\frac{\pd W}{\pd \phi},\qquad \chi'=-4e^{2\phi}\frac{\pd W}{\pd \chi},\nonumber \\
\phi_1'&=&-\frac{2}{3}\textrm{sech}^2\phi_3\frac{\pd W}{\pd \phi_1},\qquad \phi_3'=-\frac{2}{3}\frac{\pd W}{\pd \phi_3},\qquad A'=W\, .\label{BPS_eq_N1_general}
\end{eqnarray}
The explicit form of these equations is rather long and given in the appendix.

\subsection{$N=4$ holographic RG flows}
We begin with holographic RG flow solutions preserving $N=4$ supersymmetry obtained by truncating out the axion $\chi$ together with one of the two scalars $\phi_1$ and $\phi_3$. Although the solution interpolating between the trivial critical point and critical point $i$ has already been given in \cite{4D_N4_flows}, it is useful to repeat it here in the present convention for the sake of comparison with the $N=1$ solutions given later on.  

With $\phi_3=\chi=0$, the BPS equations reduce to 
\begin{eqnarray}
\phi'&=&-\frac{1}{4}e^{-\frac{\phi}{2}}\left[4ge^\phi+h_1\sinh3\phi_1-3h_1\sinh\phi_1-g(3\cosh\phi_1+\cosh3\phi_1)\right],\qquad\\
\phi_1'&=&-e^{-\frac{\phi}{2}}\cosh\phi_1\sinh\phi_1(g\cosh\phi_1-h_1\sinh\phi_1),\\
A'&=&\frac{1}{2}e^{-\frac{\phi}{2}}(ge^\phi+g\cosh^3\phi_1-h_1\sinh^3\phi_1).
\end{eqnarray}
The solution can be found by first combining $\phi'$ and $\phi'_1$ equation and finding the solution for $\phi$ as a function of $\phi_1$. The result is given by
\begin{equation}
\phi=\ln\left[\frac{g\cosh\phi_1-h_1\sinh\phi_1}{g\cosh2\phi_1+C_1\sinh2\phi_1}\right].
 \end{equation}
We readily see that $\phi\rightarrow 0$ as $\phi_1\rightarrow 0$. To make the solution end at the IR fixed point given by critical point $i$, we choose the integration constant to be
\begin{equation}
C_1=-\frac{g^2+h_1^2}{2}
\end{equation}
resulting in 
\begin{equation}
\phi=-\ln\left[\cosh\phi_1-\frac{g}{h_1}\sinh\phi_1\right].
\end{equation}
Similarly, we find the solution for $A$ as follows
\begin{equation}
A=\frac{1}{2}\ln(g\sinh\phi_1-h_1\cosh\phi_1)+\ln(g\cosh\phi_1-h_1\sinh\phi_1)-\ln\sinh2\phi_1\, .
\end{equation}  
Finally, using the previous results and changing to a new radial coordinate $\rho$ given by $\frac{d\rho}{dr}=e^{-\frac{\phi}{2}}$, we find 
\begin{equation} 
gh_1(\rho-\rho_0)=h_1\ln\coth\frac{\phi_1}{2}+2\sqrt{h_1^2-g^2}\tanh^{-1}\left[\frac{g\tanh\frac{\phi_1}{2}-h_1}{\sqrt{h_1^2-g^2}}\right]-2\tan^{-1}\tanh\frac{\phi_1}{2}\label{N4_phi1_sol}
\end{equation}
with $\rho_0$ being an integration constant. Near the UV and IR fixed points, the asymptotic behaviors of the scalar fields are given respectively by
\begin{equation}      
\phi\sim\phi_1\sim e^{-gr}\sim e^{-\frac{r}{L}},\qquad L=\frac{1}{g}
\end{equation}
and
\begin{equation}
\phi\sim e^{-\frac{r}{L_i}},\qquad \phi_1\sim e^{\frac{r}{L_i}},\qquad L_i=\frac{1}{g}\left(1-\frac{g^2}{h_1^2}\right)^{\frac{1}{4}}\, .
\end{equation}
Accordingly, the flow is driven by relevant operators of dimensions $\Delta=1,2$. In the IR, the operator dual to $\phi_1$ becomes irrelevant with dimension $\Delta=4$ while that dual to $\phi$ is still relevant. 

Similarly, by the same procedure, we can find a flow solution interpolating between the trivial $AdS_4$ critical point and ciritcal point $ii$. In this case, the BPS equations read
\begin{eqnarray}
\phi_3'&=&-e^{\frac{\phi}{2}}\cosh\phi_3\sinh\phi_3(g\cosh\phi_3-h_1\sinh\phi_3),\quad\\
\phi'&=&-\frac{1}{4}e^{\frac{\phi}{2}}[g(2\cosh\phi_3+\cosh3\phi_3)+h_1(3\sinh\phi_3-\sinh3\phi_3)-4ge^{-\phi}],\\
A'&=&\frac{1}{2}e^{-\frac{\phi}{2}}[g+e^\phi(g\cosh^3\phi_3-h_1\sinh^3\phi_3)]
\end{eqnarray}
with the solution given by
\begin{eqnarray}
\phi&=&\ln\left[\cosh\phi_3-\frac{g}{h_1}\sinh\phi_3\right],\\
A&=&\frac{1}{2}\ln(g\sinh\phi_3-h_1\cosh\phi_3)+\ln(g\cosh\phi_3-h_1\sinh\phi_3)\nonumber \\
& &-\ln\sinh2\phi_3,\\
g(\rho-\rho_0)&=&\ln(g^2+h_1^2-2gh_1\coth2\phi_3)
\end{eqnarray}
with $\rho$ defined as in the previous case. The asymptotic behaviors and holographic interpretations are also similar. Furthermore, the solution for $\phi_3$ can be rewritten in a similar form as \eqref{N4_phi1_sol} by changing to another radial coordinate $\eta$ given by $\frac{d\eta}{d\rho}=e^\phi$ resulting in
\begin{eqnarray}
gh_1(\eta-\eta_0)&=&h_1\ln\coth\frac{\phi_3}{2}+2\sqrt{h_1^2-g^2}\tanh^{-1}\left[\frac{g\tanh\frac{\phi_3}{2}-h_1}{\sqrt{h_1^2-g^2}}\right]\nonumber \\
& &-2\tan^{-1}\tanh\frac{\phi_3}{2}\, .
\end{eqnarray}        
\indent In these two solutions, we see that the operators dual to $\phi$ and $\phi_1$ break conformal symmetry but preserve $N=4$ Poincare supersymmetry in three dimensions.           

\subsection{$N=1$ holographic RG flows}
We now consider holographic RG flows in the full $SO(3)_{\textrm{diag}}\times SO(3)$ sector. We first point out that setting $\chi=0$ still gives $\mc{A}\neq \mc{B}$ resulting in $N=1$ supersymmetry. However, truncating out only $\chi$ is not consistent with the BPS equations given in \eqref{BPS_eq_N1_general} unless $\phi_1=0$ or $\phi_3=0$. Accordingly, $N=1$ RG flow solutions to critical points $i$ or $ii$ involve all scalars in the $SO(3)_{\textrm{diag}}\times SO(3)$ sector. This makes finding the solutions more difficult, so we will numerically give some examples of possible solutions.

\begin{figure}
  \centering
  \begin{subfigure}[b]{0.45\linewidth}
    \includegraphics[width=\linewidth]{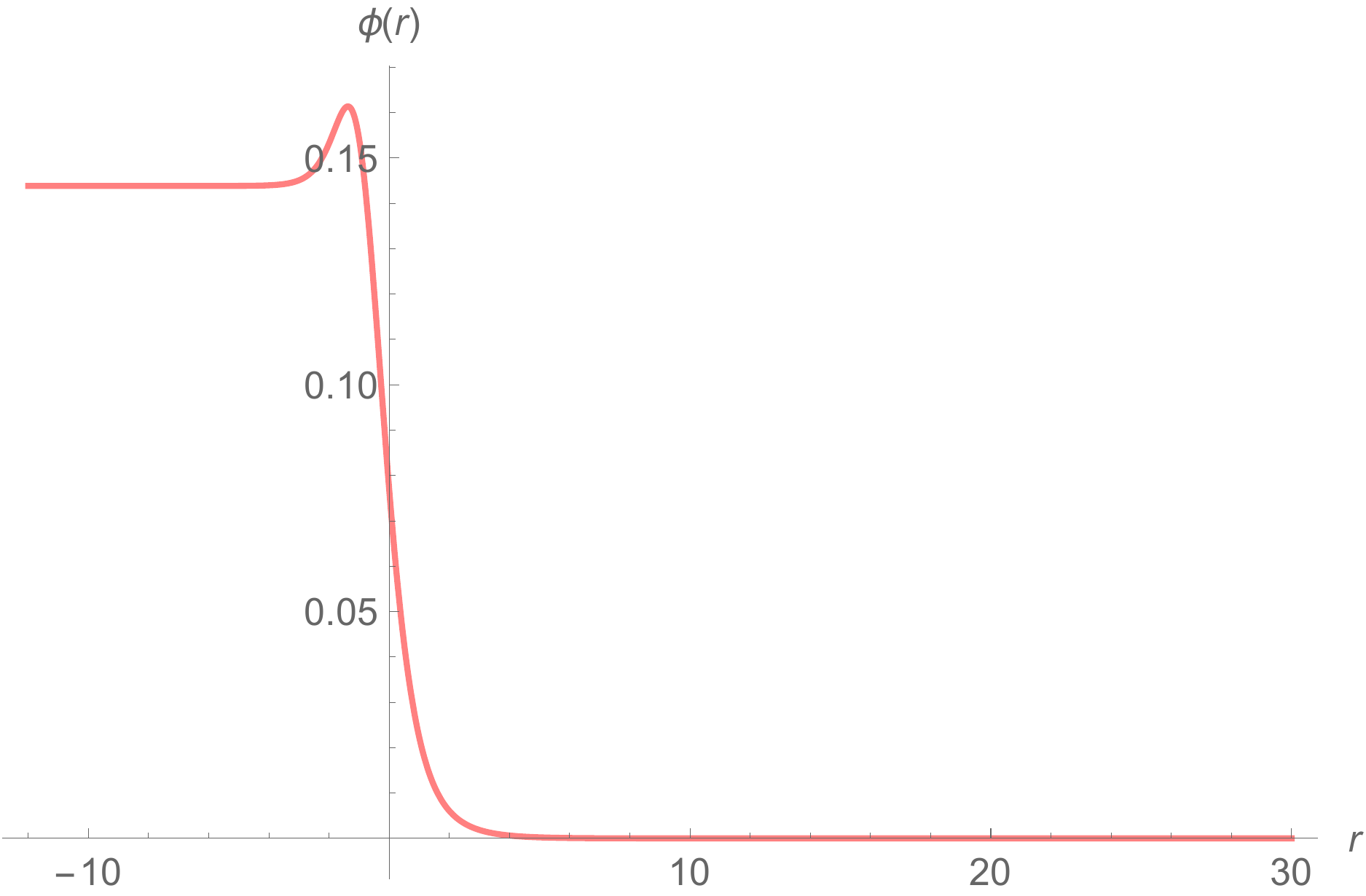}
  \caption{$\phi(r)$ solution}
  \end{subfigure}
  \begin{subfigure}[b]{0.45\linewidth}
    \includegraphics[width=\linewidth]{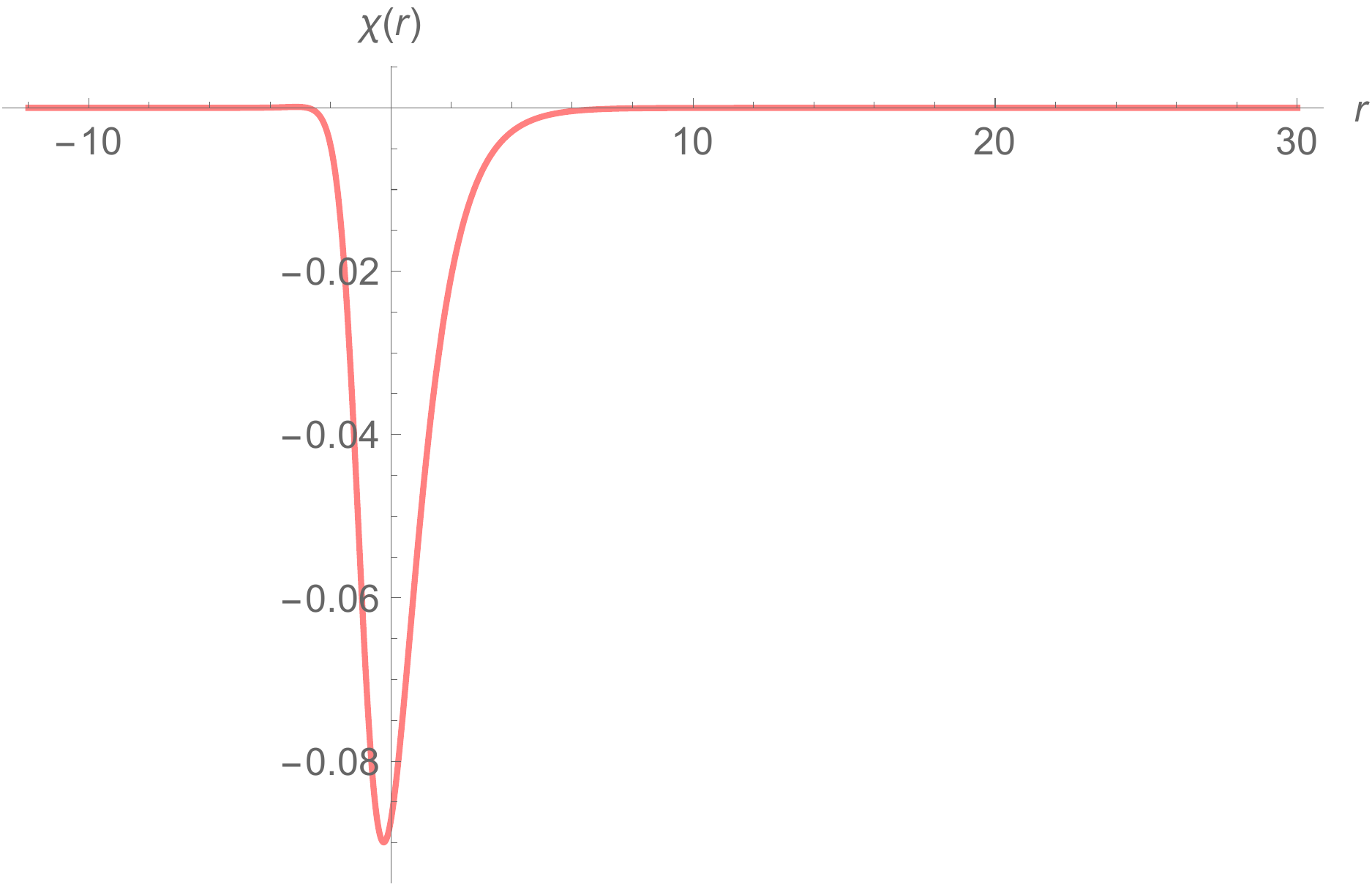}
  \caption{$\chi(r)$ solution}
  \end{subfigure}\\
    \begin{subfigure}[b]{0.45\linewidth}
    \includegraphics[width=\linewidth]{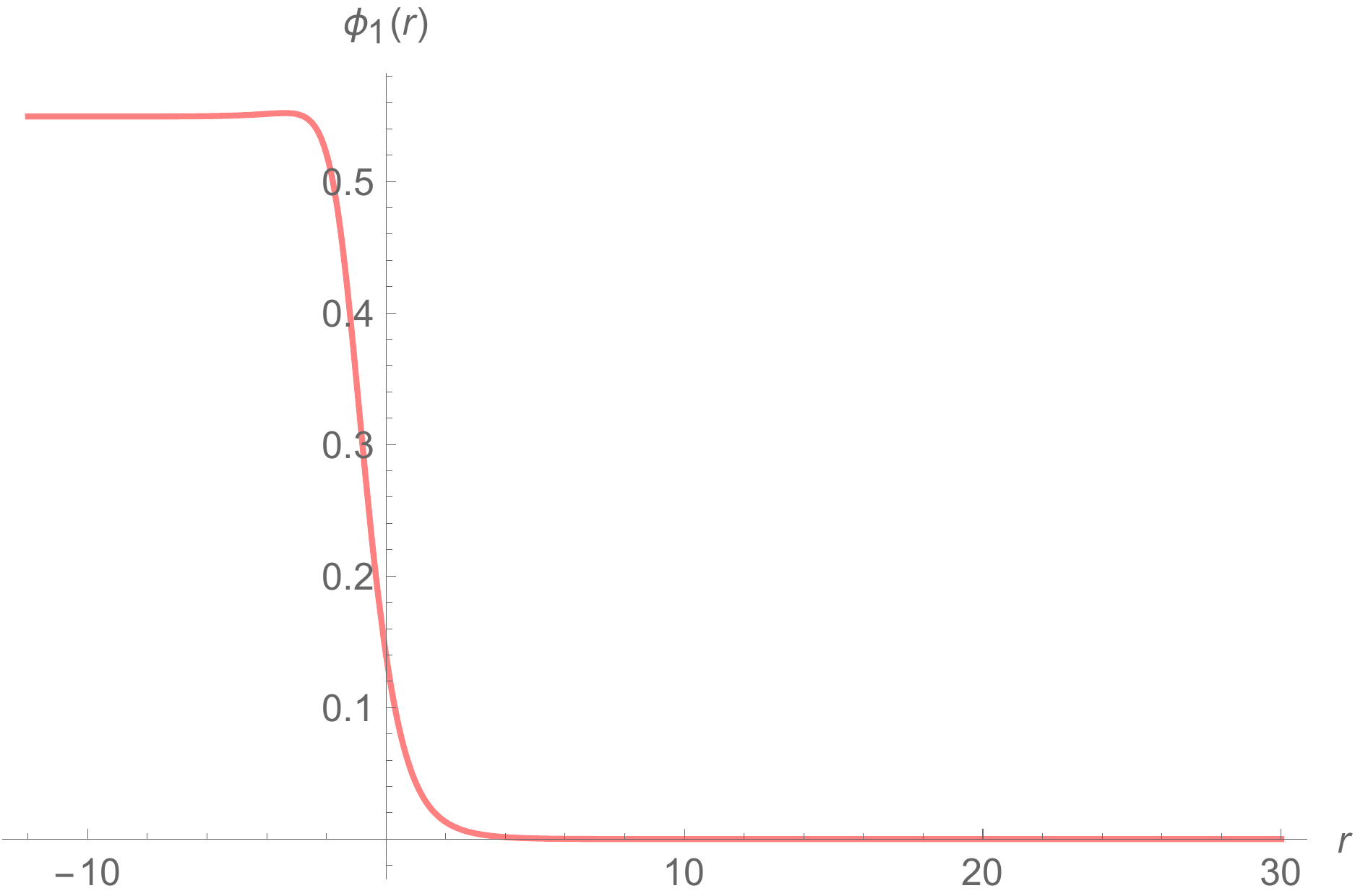}
  \caption{$\phi_1(r)$ solution}
  \end{subfigure}
  \begin{subfigure}[b]{0.45\linewidth}
    \includegraphics[width=\linewidth]{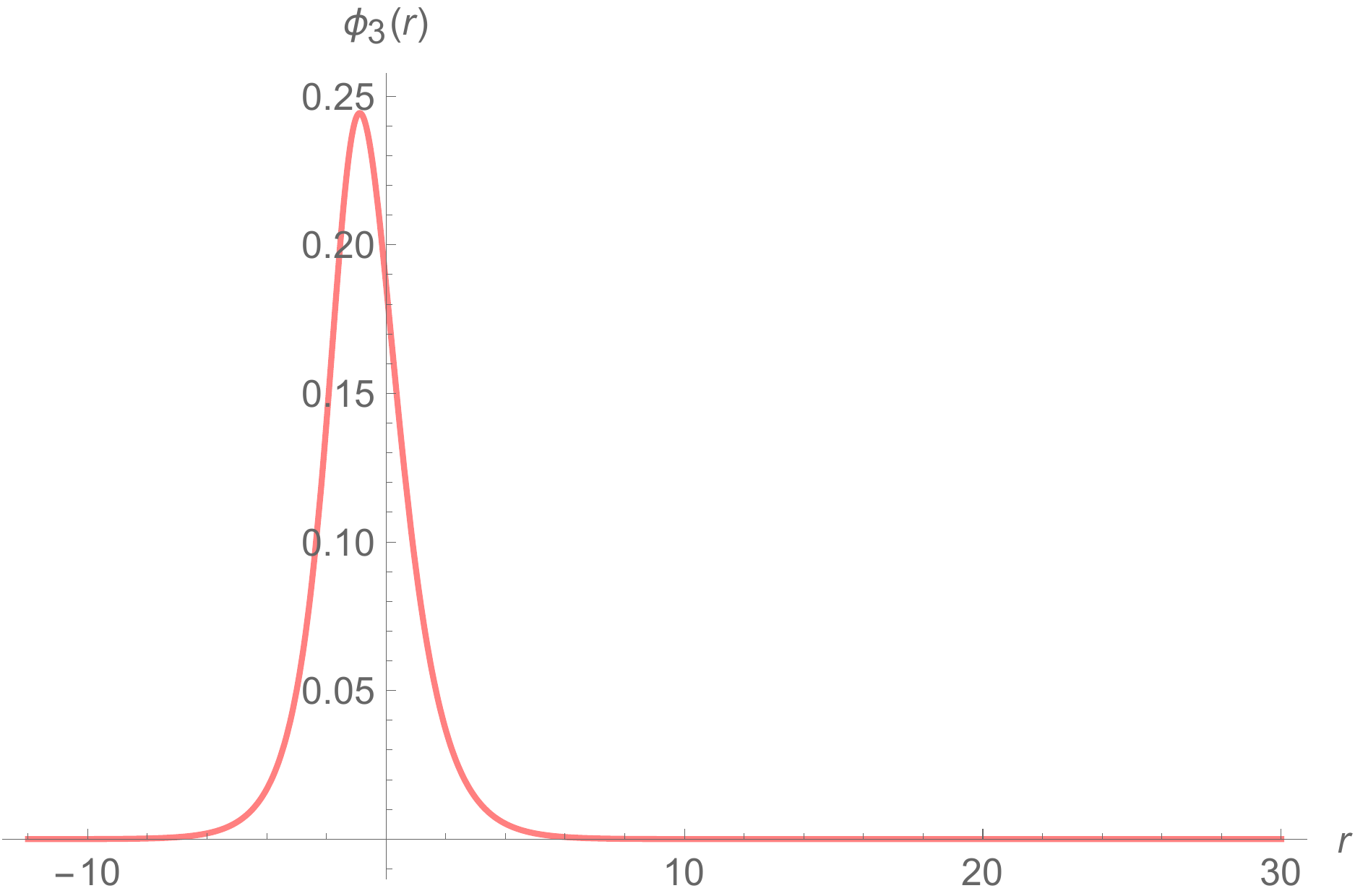}
  \caption{$\phi_3(r)$ solution}
  \end{subfigure}\\
   \begin{subfigure}[b]{0.45\linewidth}
    \includegraphics[width=\linewidth]{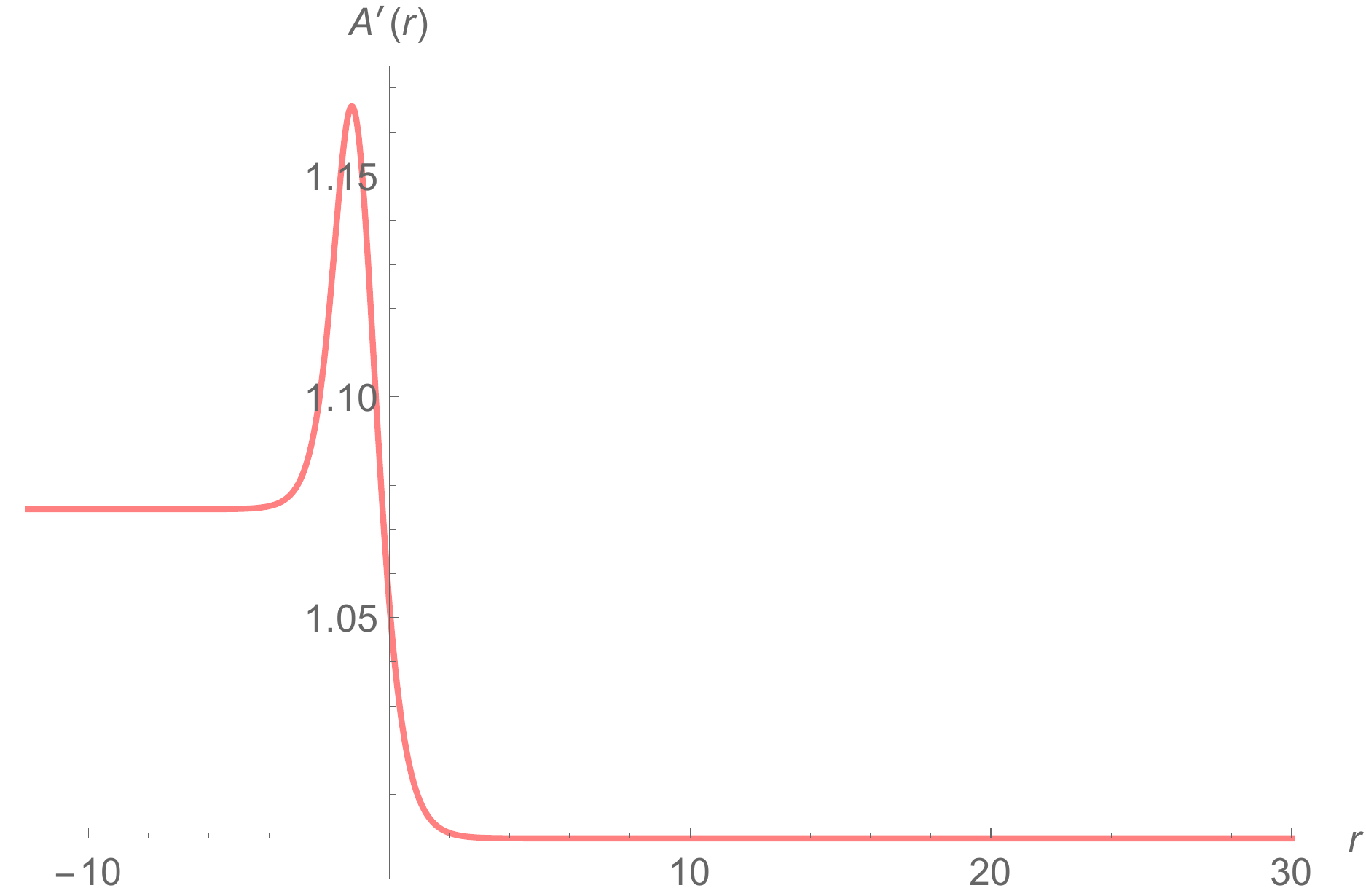}
  \caption{$A'(r)$ solution}
   \end{subfigure} 
  \caption{An $N=1$ RG flow from the $N=4$ SCFT with $SO(4)\times SO(4)$ symmetry to an $N=4$ conformal fixed point in the IR with $SO(3)_{\textrm{diag}}\times SO(3)\times SO(3)$ symmetry for $\beta_1=0$, $g=1$ and $h_1=2$.}
  \label{Fig2}
\end{figure}

Using the BPS equations given in the appendix, we find an RG flow solution interpolating between the trivial $N=4$ fixed point with $SO(4)\times SO(4)$ symmetry and critcal point $i$ as shown in figure \ref{Fig2}. Since all scalars have the same mass $m^2L^2=-2$ at the $SO(4)\times SO(4)$ critical point, the flow is again driven by relevant operators of dimensions $\Delta=1,2$. In the IR, using the scalar masses given in \cite{4D_N4_flows}, we find that $\phi_1$ is dual to an irrelevant operator of dimension $\Delta=4$, but $\phi$, $\chi$ and $\phi_3$ are dual to relevant operators of dimensions $\Delta=1,2$. Unlike the $N=4$ solutions given above, in addition to breaking conformal symmetry, turning on the operators dual to $\chi$ and $\phi_3$ along the flow further breaks the $N=4$ Poincare supersymmetry to $N=1$. However, at the IR fixed point, the conformal symmetry is restored, and the supersymmetry is enhanced to $N=4$ due to the vanishing of $\chi$ and $\phi_3$.  

\begin{figure}
  \centering
  \begin{subfigure}[b]{0.45\linewidth}
    \includegraphics[width=\linewidth]{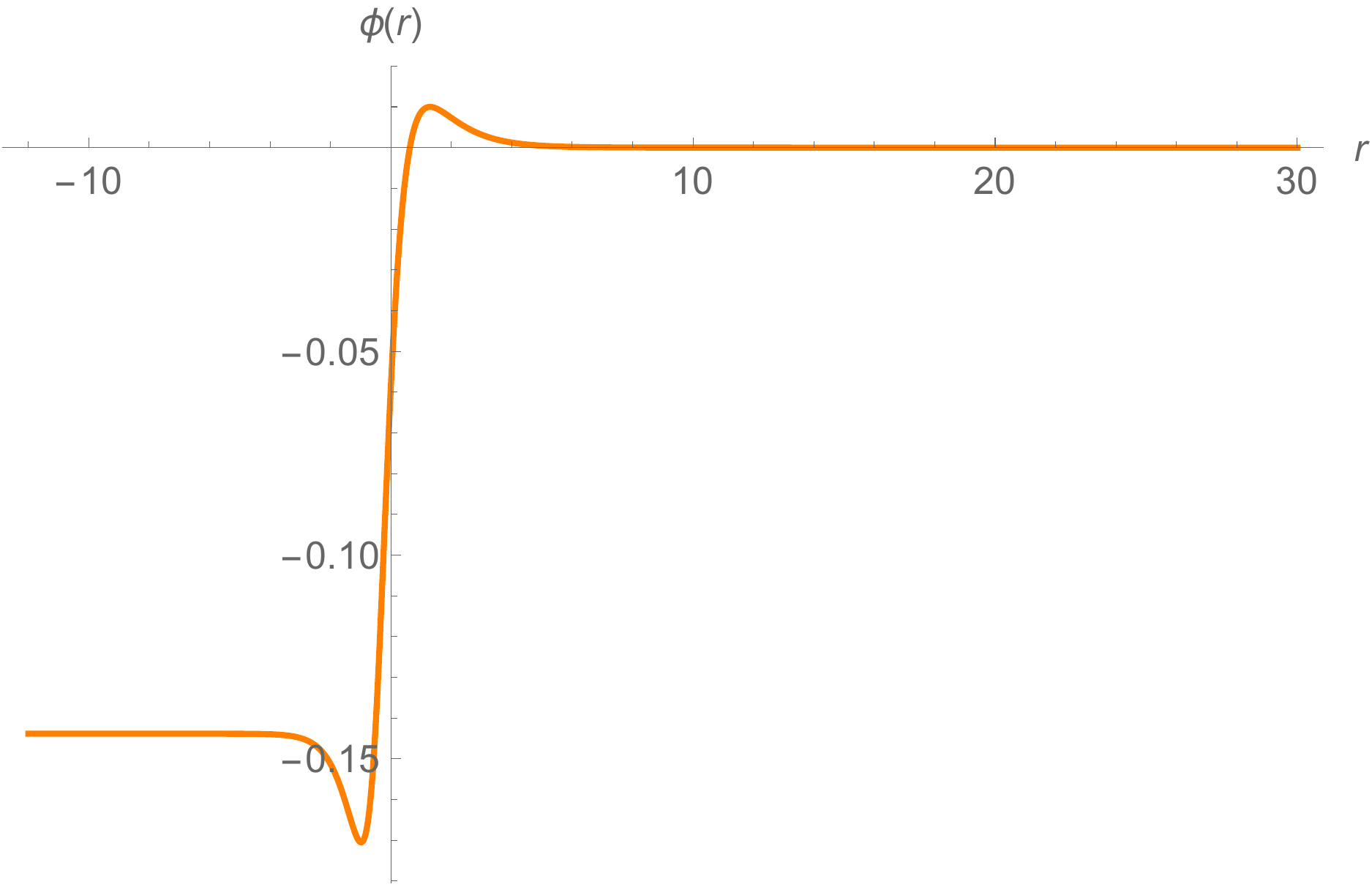}
  \caption{$\phi(r)$ solution}
  \end{subfigure}
  \begin{subfigure}[b]{0.45\linewidth}
    \includegraphics[width=\linewidth]{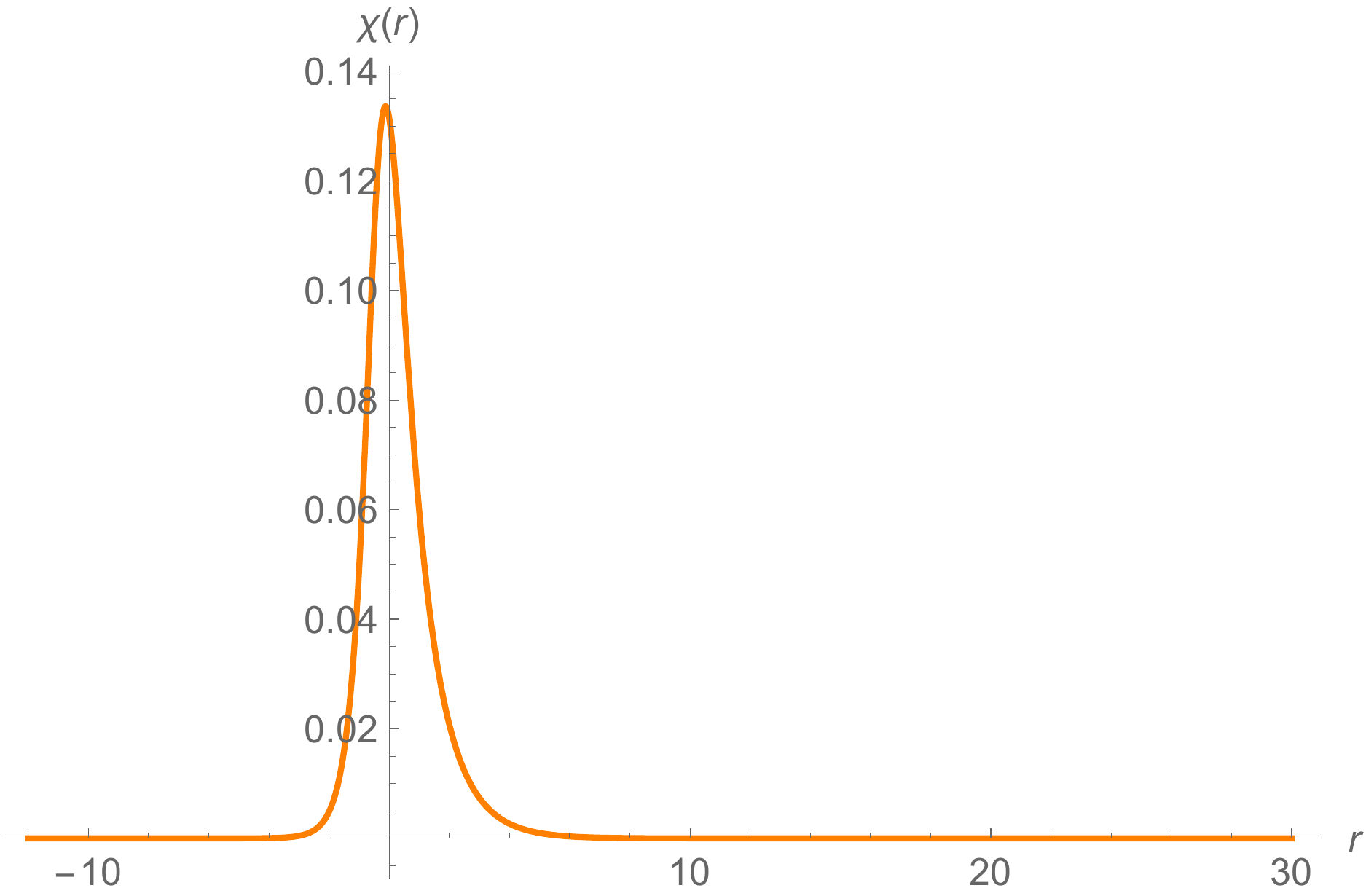}
  \caption{$\chi(r)$ solution}
  \end{subfigure}\\
    \begin{subfigure}[b]{0.45\linewidth}
    \includegraphics[width=\linewidth]{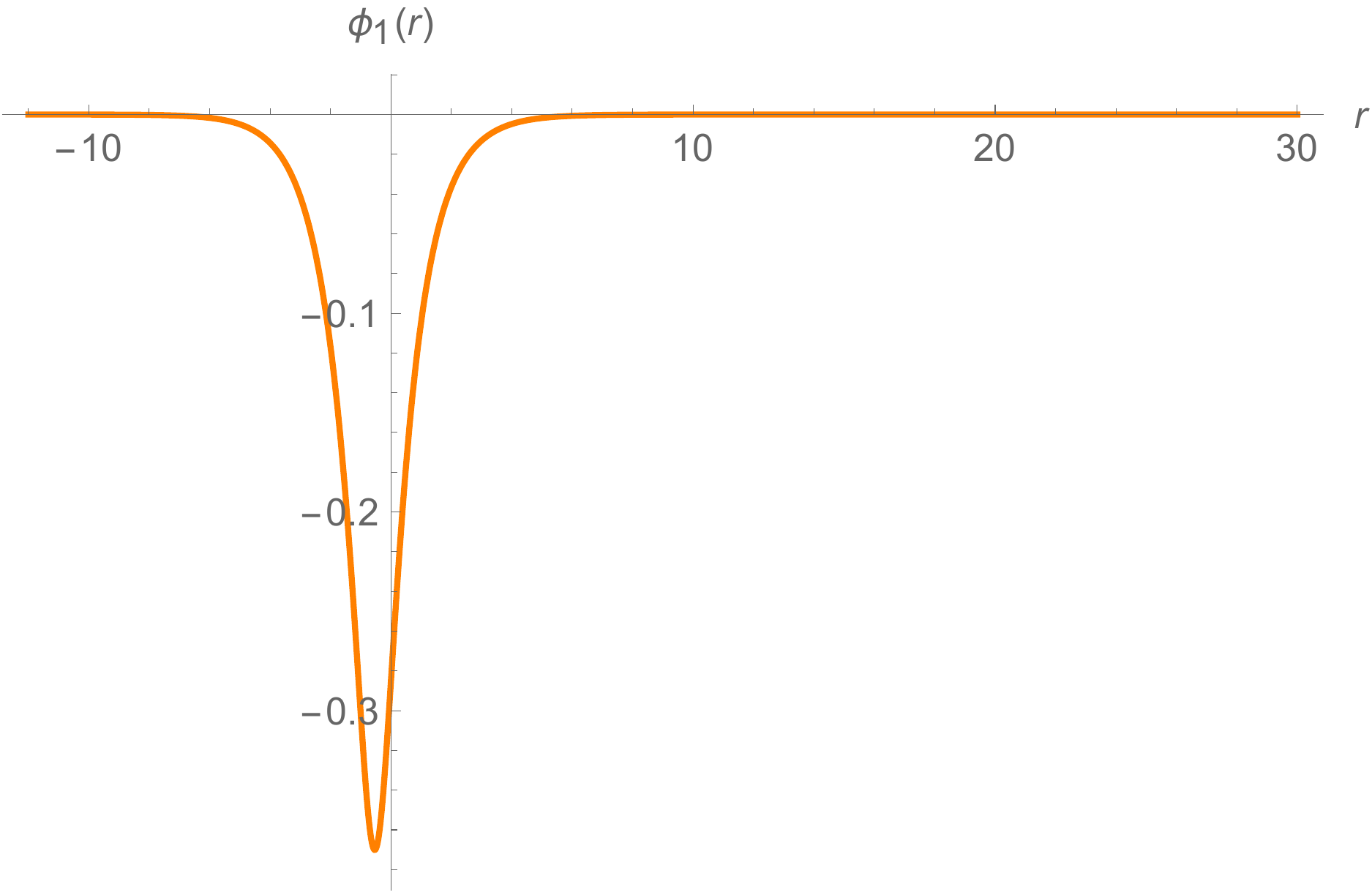}
  \caption{$\phi_1(r)$ solution}
  \end{subfigure}
  \begin{subfigure}[b]{0.45\linewidth}
    \includegraphics[width=\linewidth]{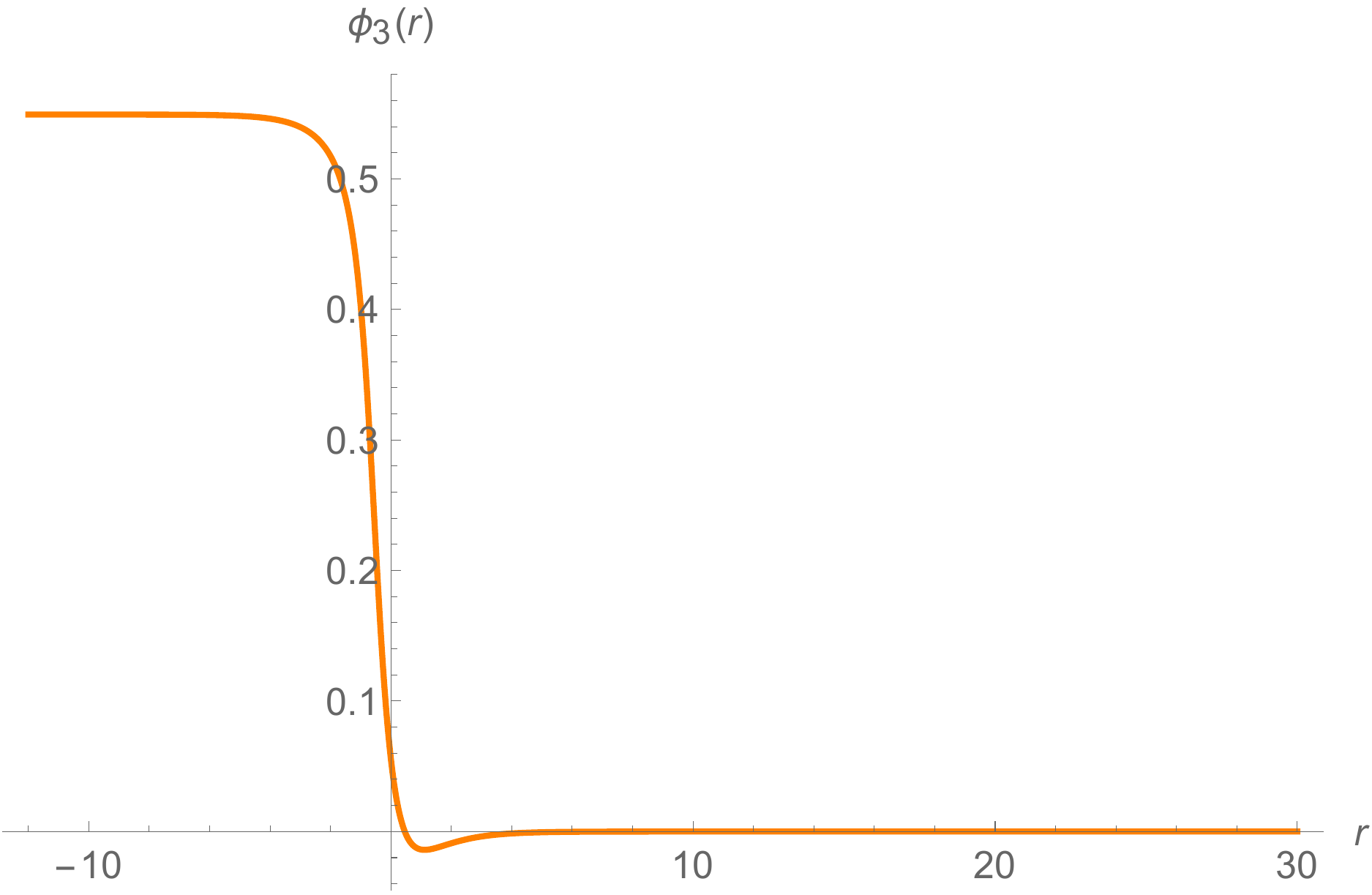}
  \caption{$\phi_3(r)$ solution}
  \end{subfigure}\\
   \begin{subfigure}[b]{0.45\linewidth}
    \includegraphics[width=\linewidth]{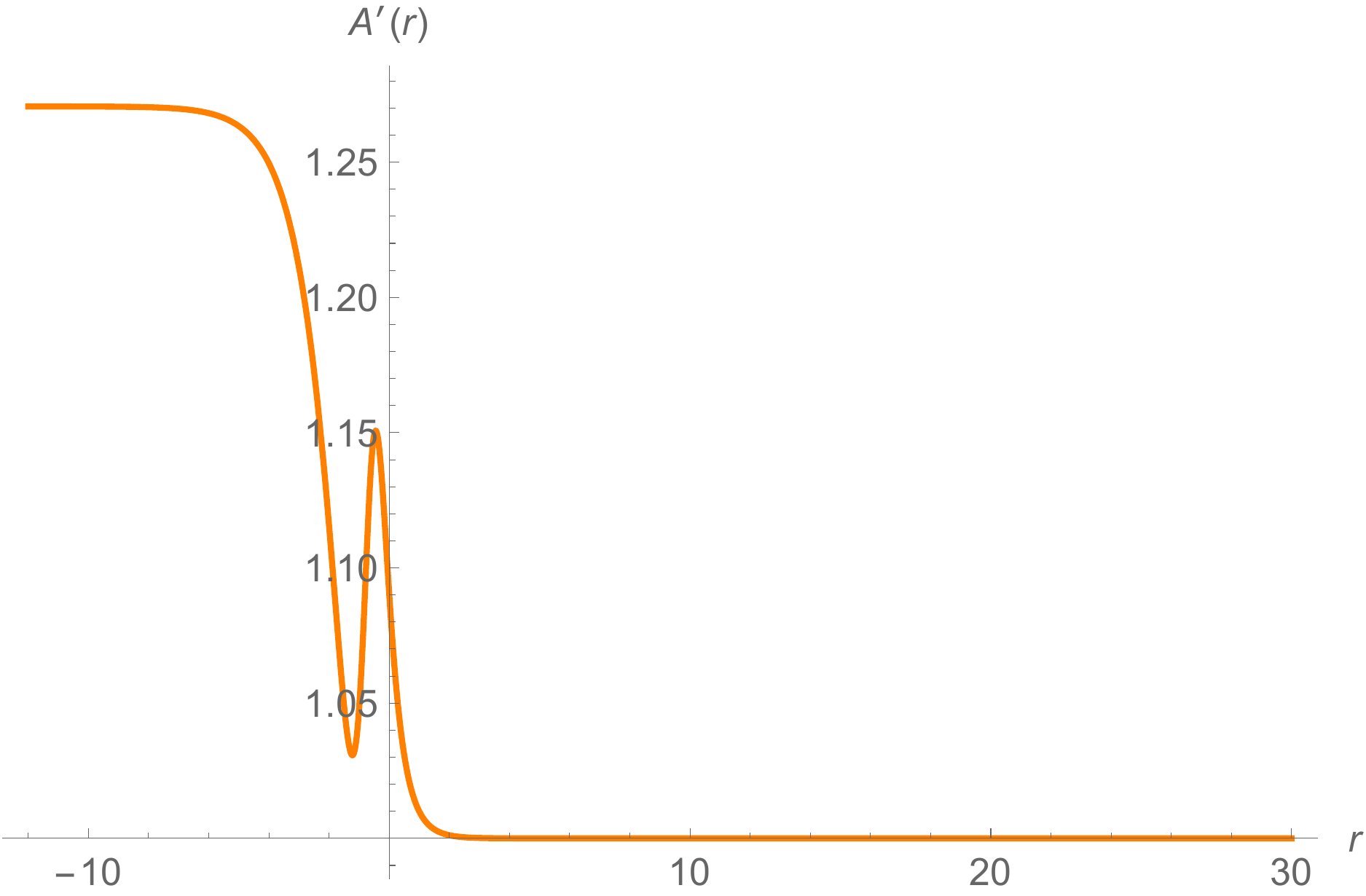}
  \caption{$A'(r)$ solution}
   \end{subfigure} 
  \caption{An $N=1$ RG flow from the $N=4$ SCFT with $SO(4)\times SO(4)$ symmetry to an $N=4$ conformal fixed point in the IR with $SO(3)\times SO(3)_{\textrm{diag}}\times SO(3)$ symmetry for $\beta_1=\frac{\pi}{2}$, $g=1$ and $h_1=2$.}
  \label{Fig3}
\end{figure}

\begin{figure}
  \centering
  \begin{subfigure}[b]{0.45\linewidth}
    \includegraphics[width=\linewidth]{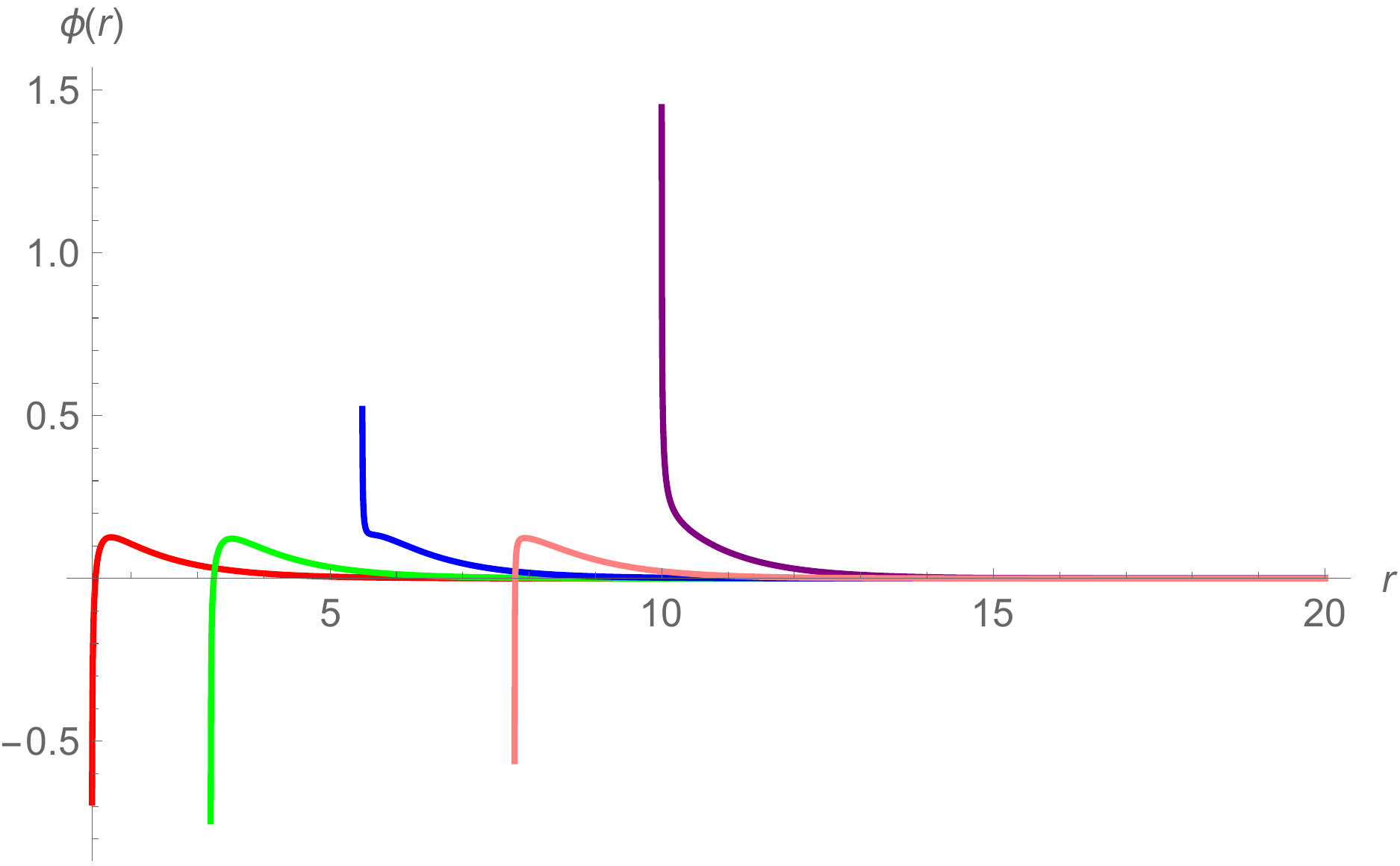}
  \caption{$\phi(r)$ solution}
  \end{subfigure}
  \begin{subfigure}[b]{0.45\linewidth}
    \includegraphics[width=\linewidth]{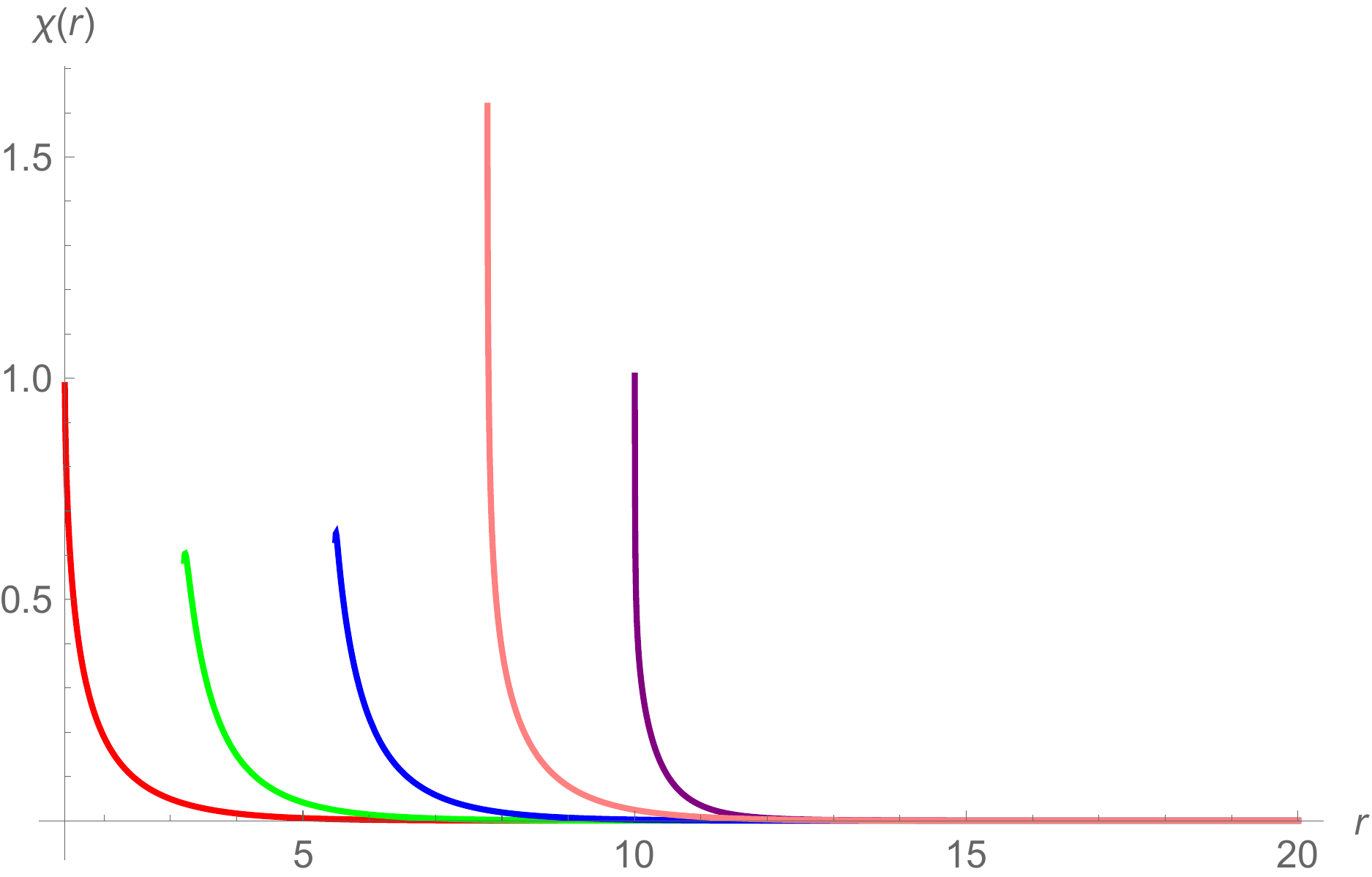}
  \caption{$\chi(r)$ solution}
  \end{subfigure}\\
    \begin{subfigure}[b]{0.45\linewidth}
    \includegraphics[width=\linewidth]{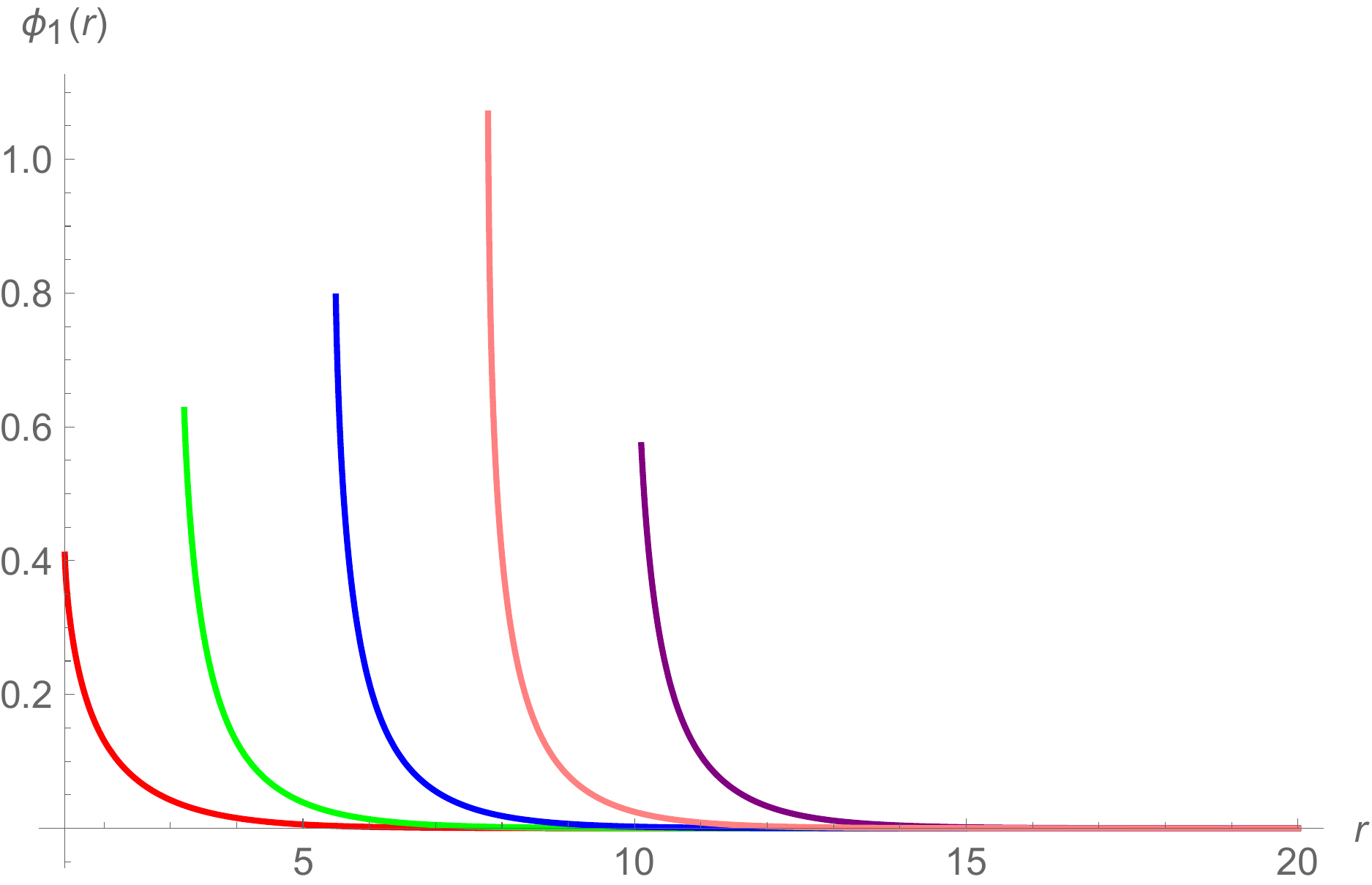}
  \caption{$\phi_1(r)$ solution}
  \end{subfigure}
  \begin{subfigure}[b]{0.45\linewidth}
    \includegraphics[width=\linewidth]{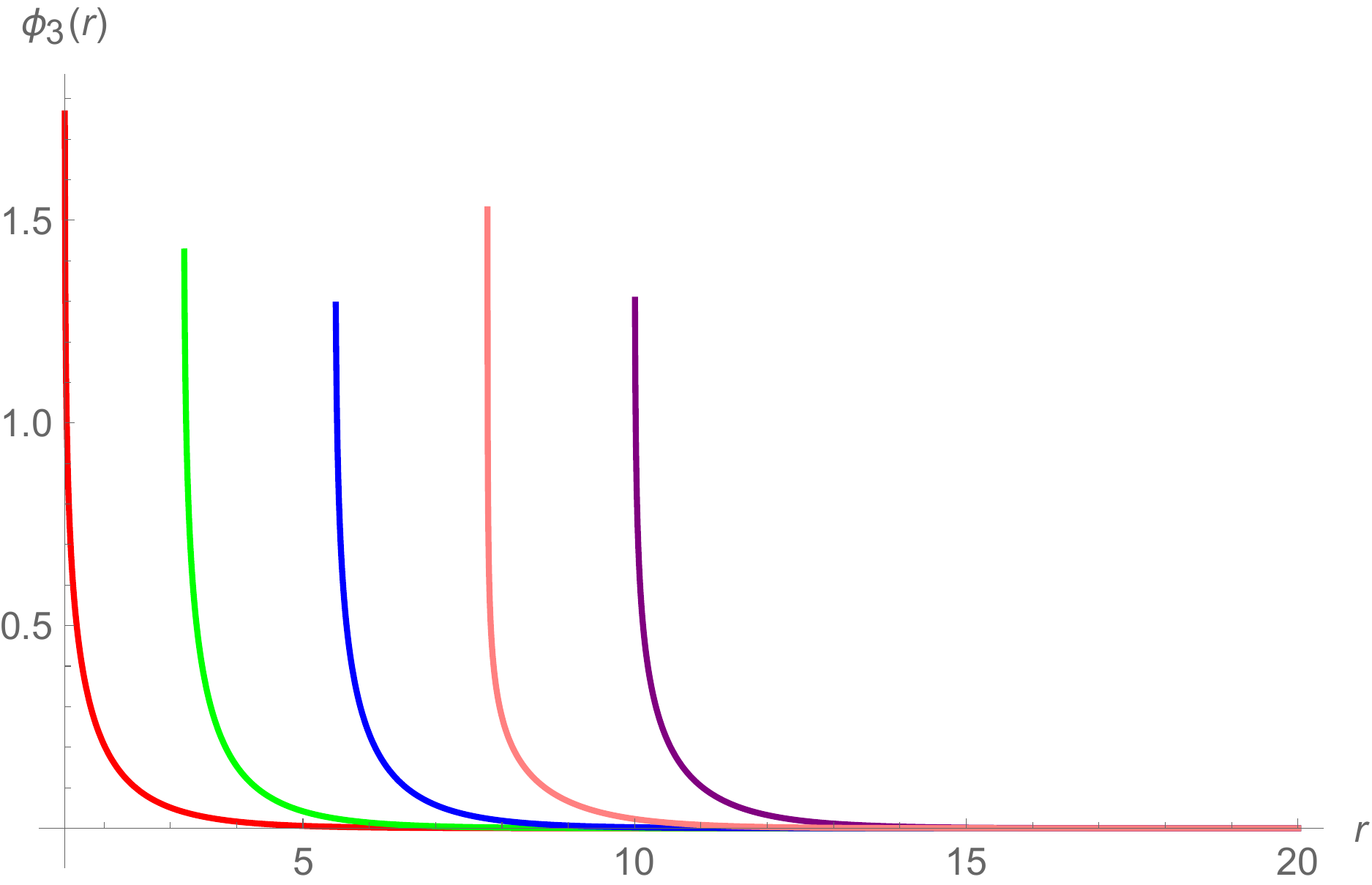}
  \caption{$\phi_3(r)$ solution}
  \end{subfigure}\\
   \begin{subfigure}[b]{0.45\linewidth}
    \includegraphics[width=\linewidth]{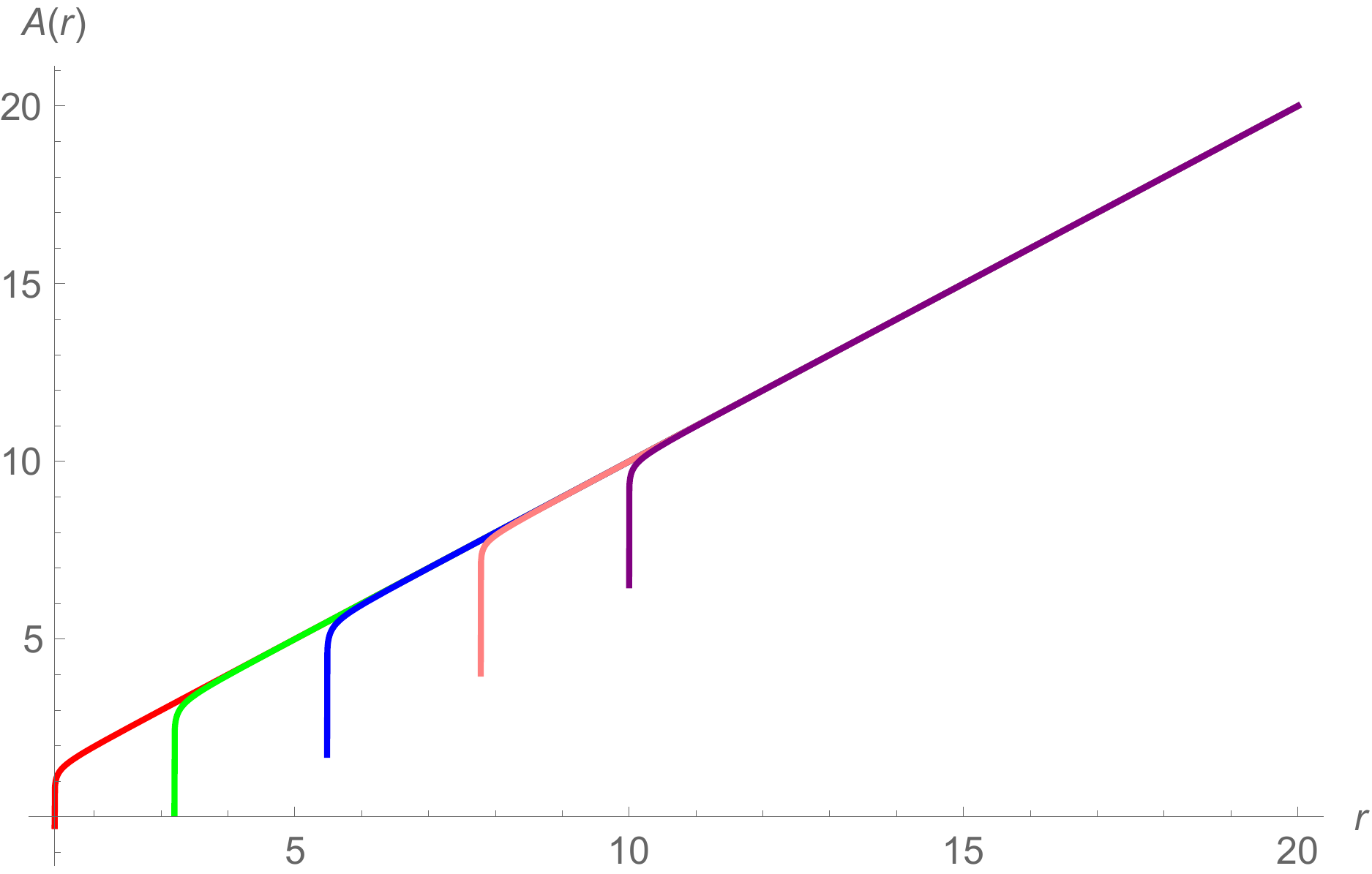}
  \caption{$A(r)$ solution}
   \end{subfigure} 
 \begin{subfigure}[b]{0.45\linewidth}
    \includegraphics[width=\linewidth]{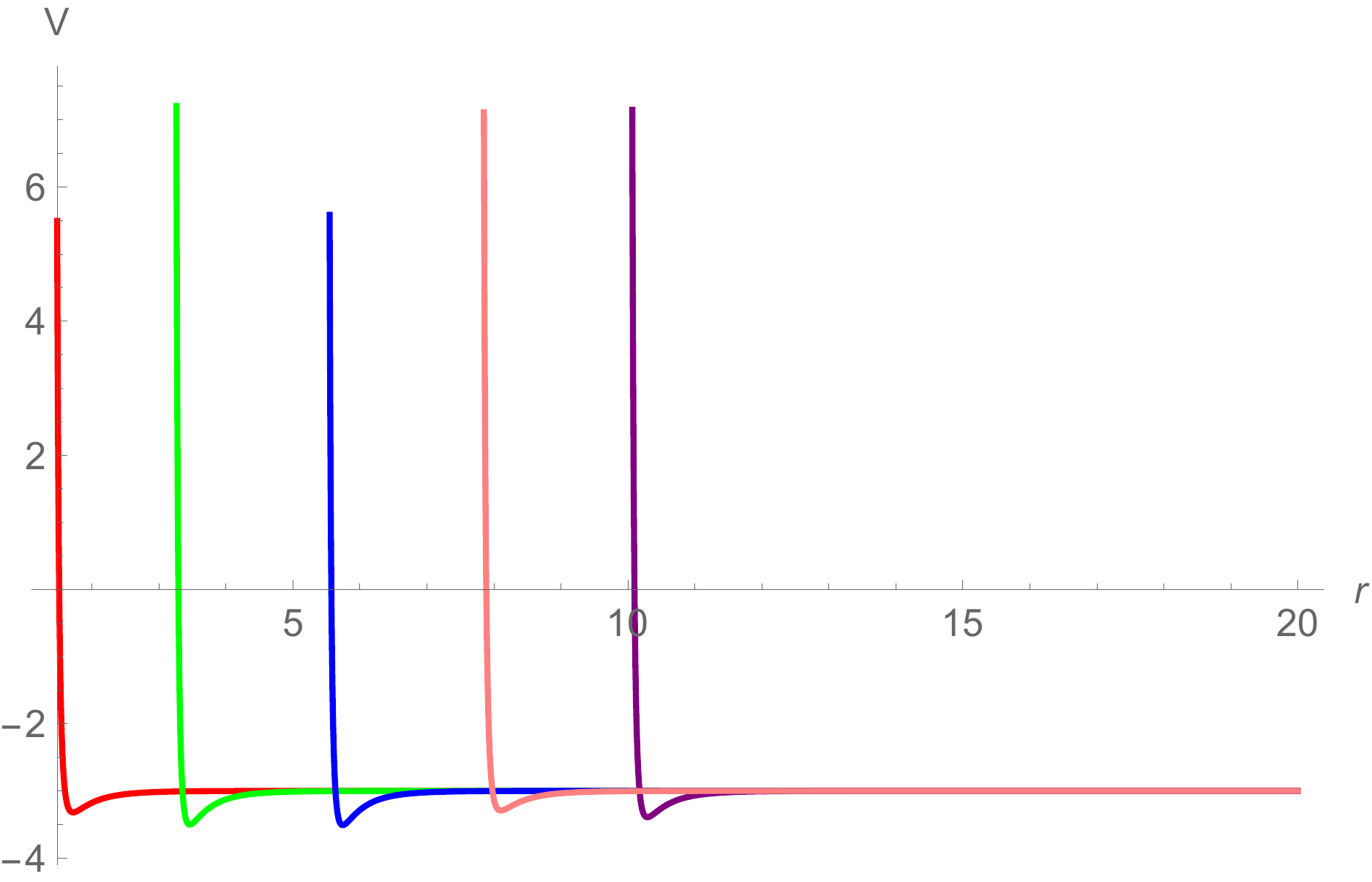}
  \caption{Scalar potential}
   \end{subfigure} 
  \caption{Examples of $N=1$ RG flows from the $N=4$ SCFT with $SO(4)\times SO(4)$ symmetry in the UV to non-conformal phases in the IR for different values of the electric-magnetic phase $\beta_1=0$ (red), $\frac{\pi}{6}$ (green), $\frac{\pi}{4}$ (blue), $\frac{\pi}{3}$ (purple),$\frac{\pi}{2}$ (pink) with $g=1$ and $h_1=2$.}
  \label{Fig4}
\end{figure}

\begin{figure}
  \centering
  \begin{subfigure}[b]{0.45\linewidth}
    \includegraphics[width=\linewidth]{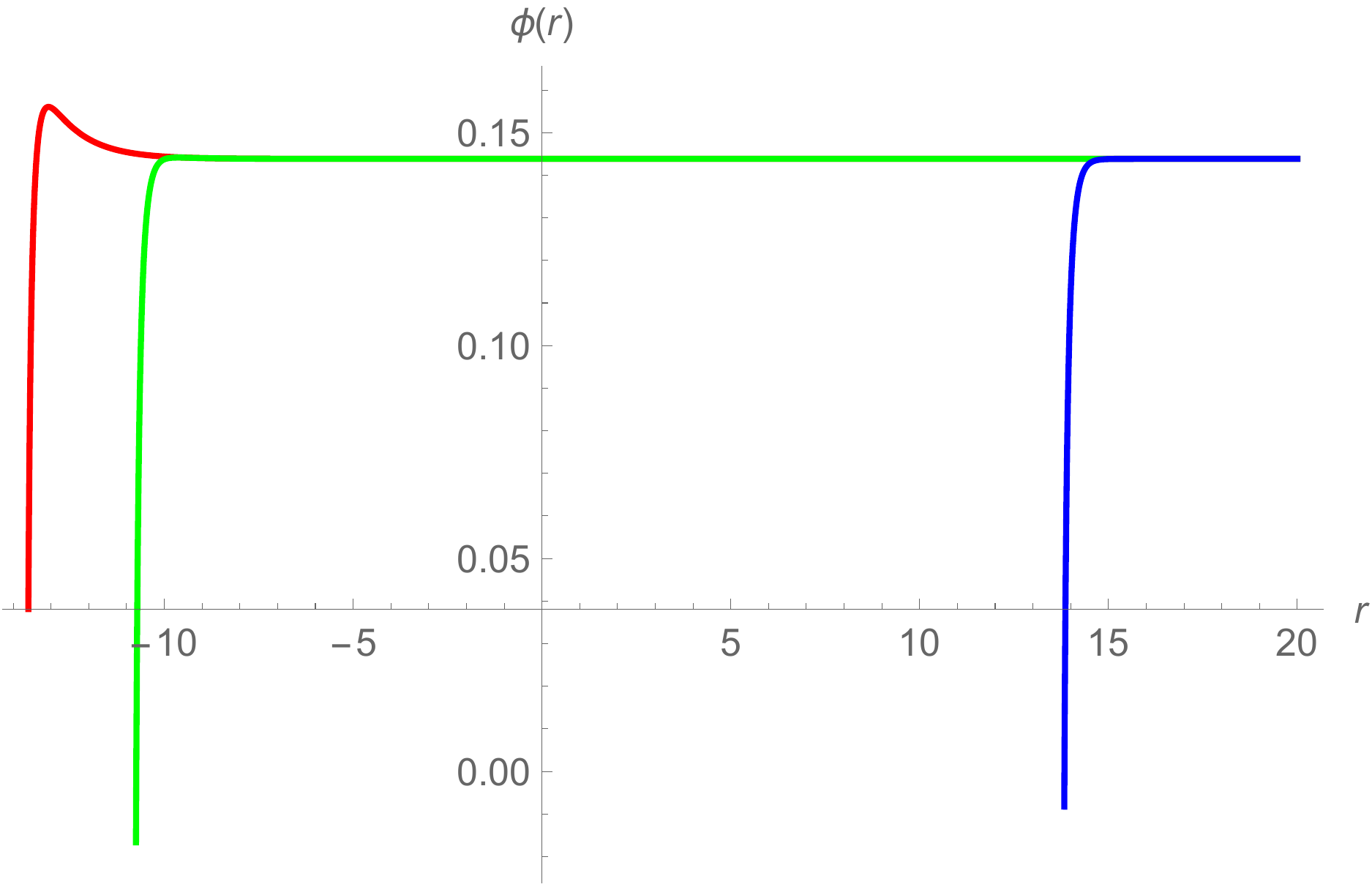}
  \caption{$\phi(r)$ solution}
  \end{subfigure}
  \begin{subfigure}[b]{0.45\linewidth}
    \includegraphics[width=\linewidth]{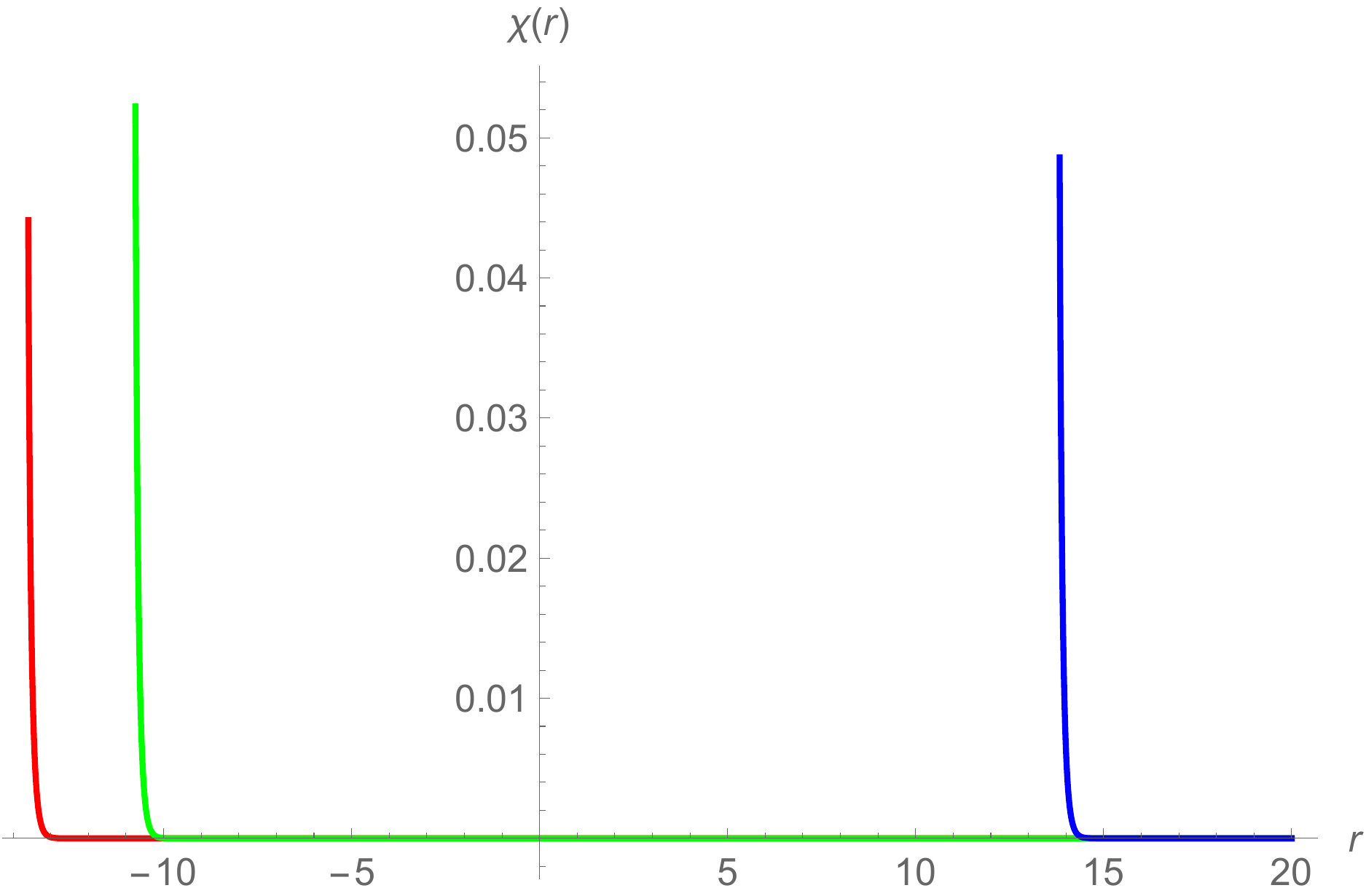}
  \caption{$\chi(r)$ solution}
  \end{subfigure}\\
    \begin{subfigure}[b]{0.45\linewidth}
    \includegraphics[width=\linewidth]{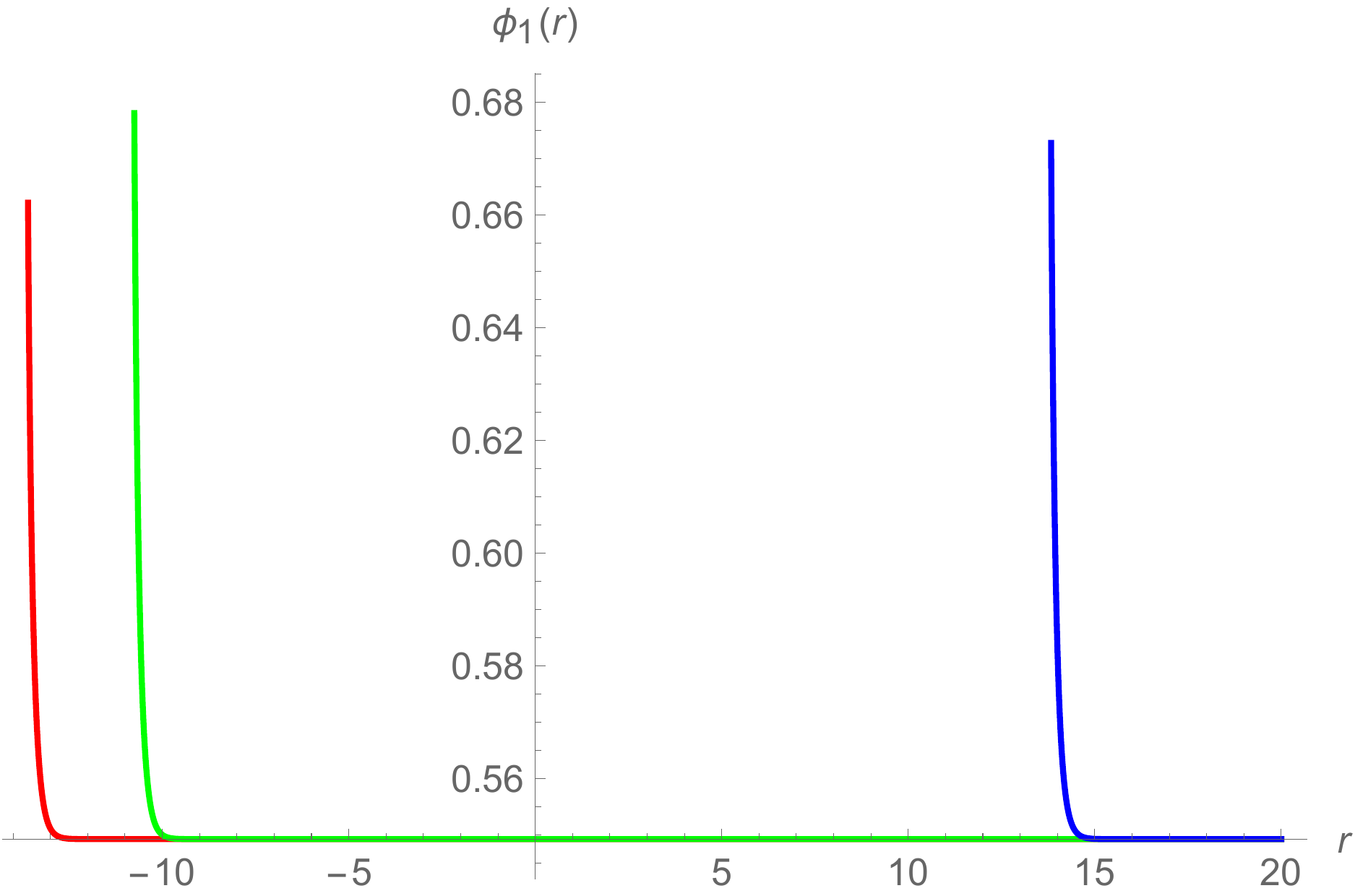}
  \caption{$\phi_1(r)$ solution}
  \end{subfigure}
  \begin{subfigure}[b]{0.45\linewidth}
    \includegraphics[width=\linewidth]{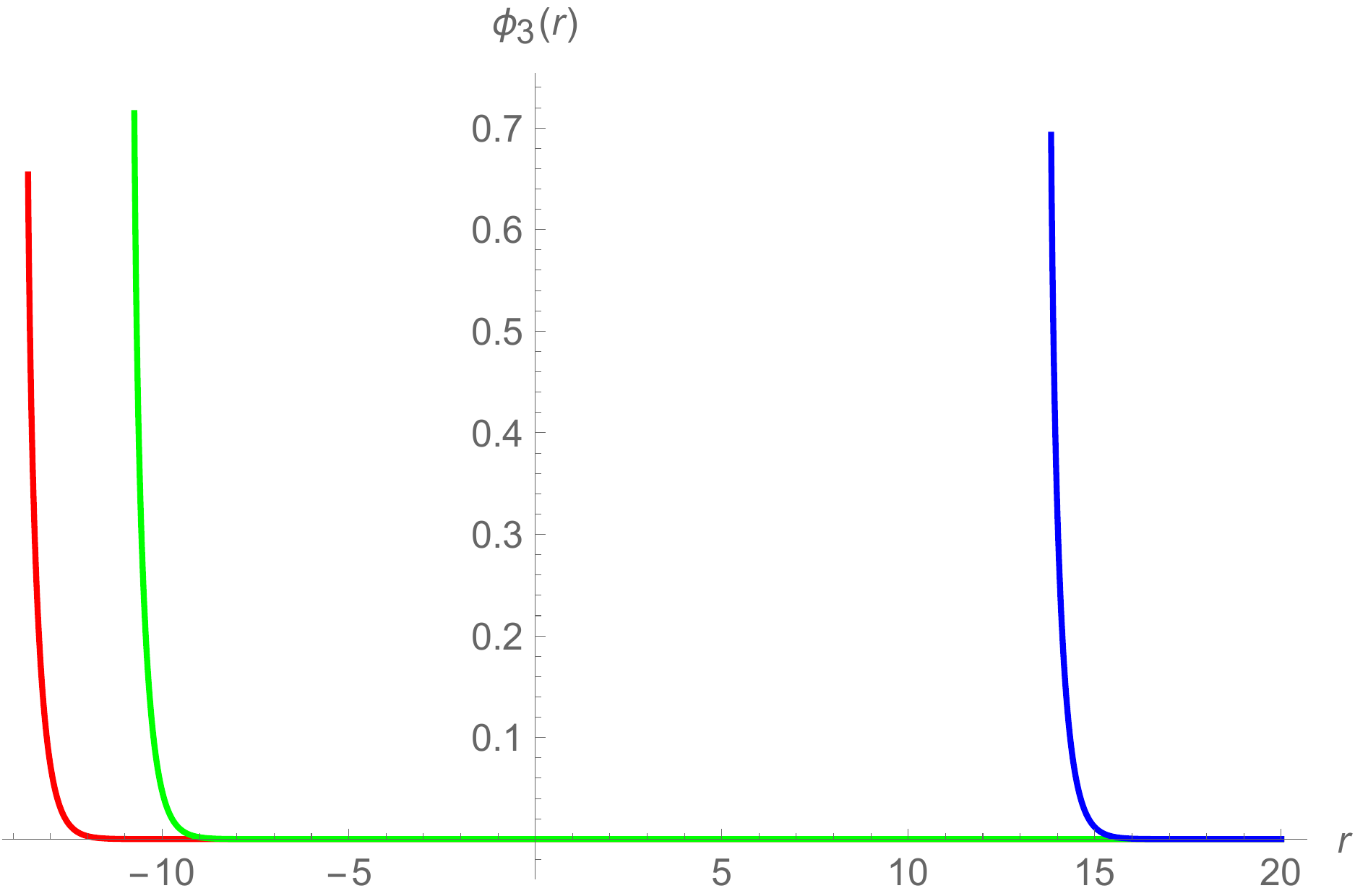}
  \caption{$\phi_3(r)$ solution}
  \end{subfigure}\\
   \begin{subfigure}[b]{0.45\linewidth}
    \includegraphics[width=\linewidth]{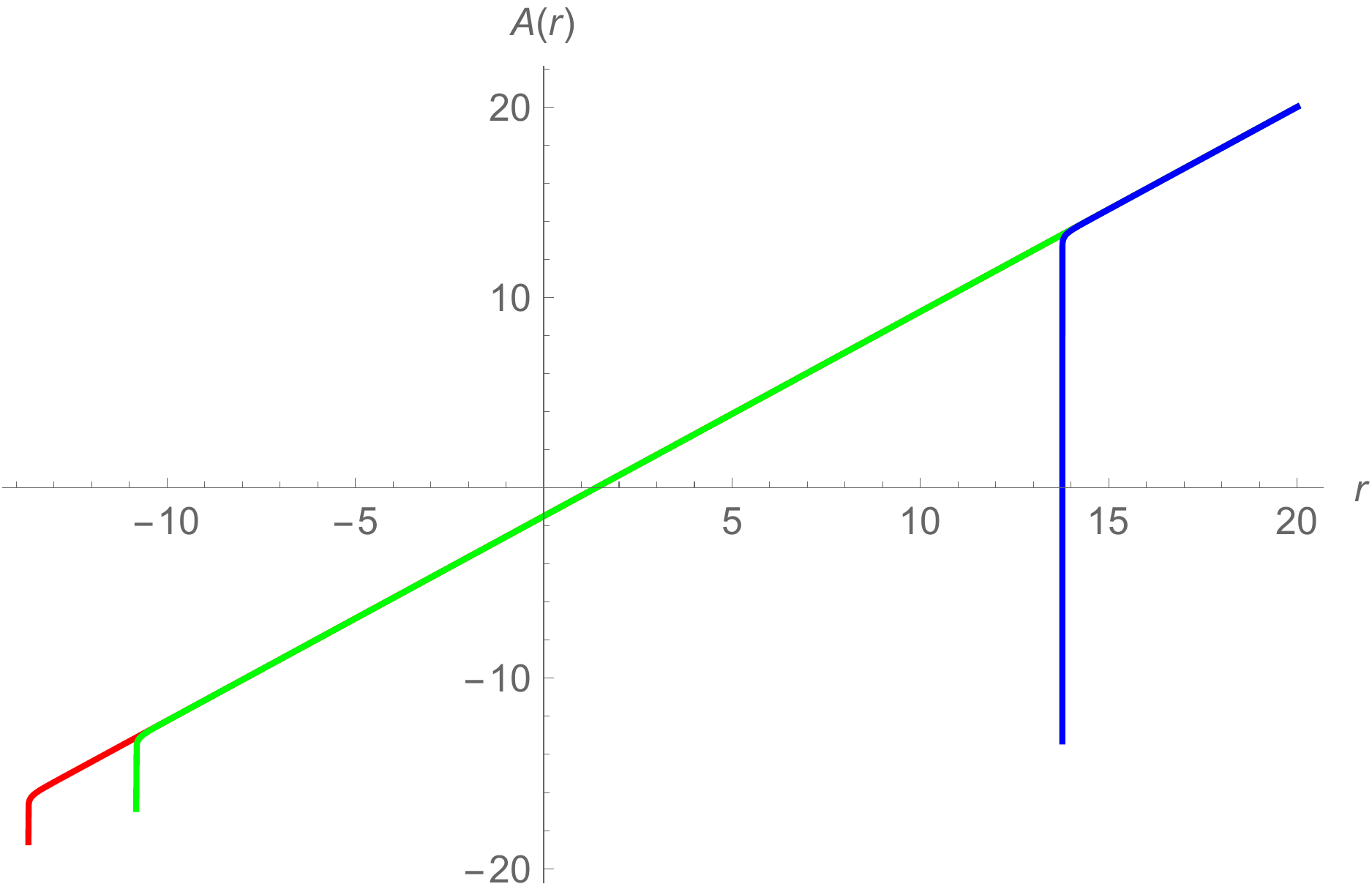}
  \caption{$A(r)$ solution}
   \end{subfigure} 
    \begin{subfigure}[b]{0.45\linewidth}
    \includegraphics[width=\linewidth]{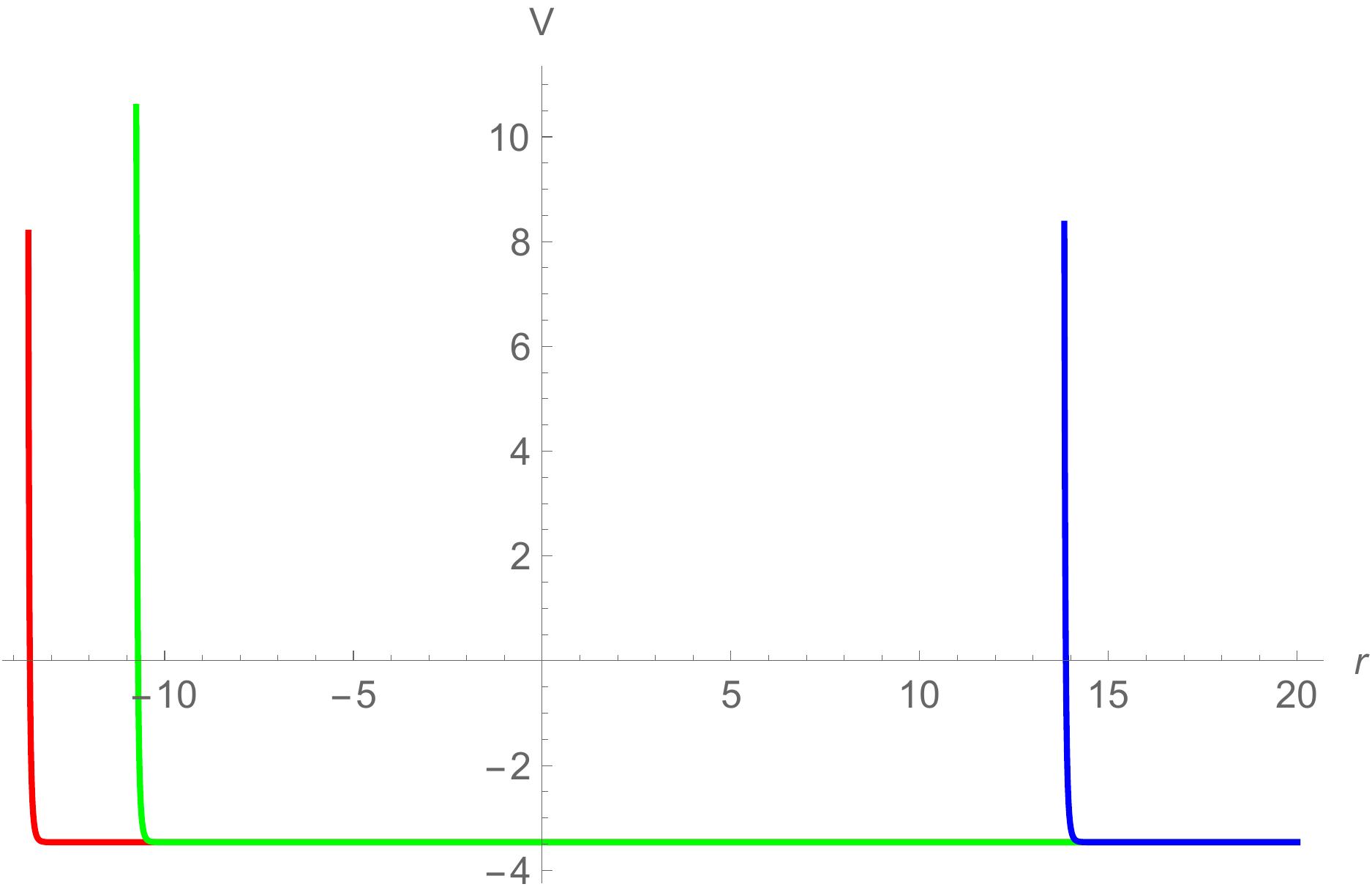}
  \caption{Scalar potential}
   \end{subfigure} 
  \caption{Examples of $N=1$ RG flows from the $N=4$ SCFT with $SO(3)_{\textrm{diag}}\times SO(3)\times SO(3)$ symmetry (critical point $i$) to non-conformal phases in the IR with $\beta_1=0$, $g=1$ and $h_1=2$.}
  \label{Fig5}
\end{figure}

\begin{figure}
  \centering
  \begin{subfigure}[b]{0.45\linewidth}
    \includegraphics[width=\linewidth]{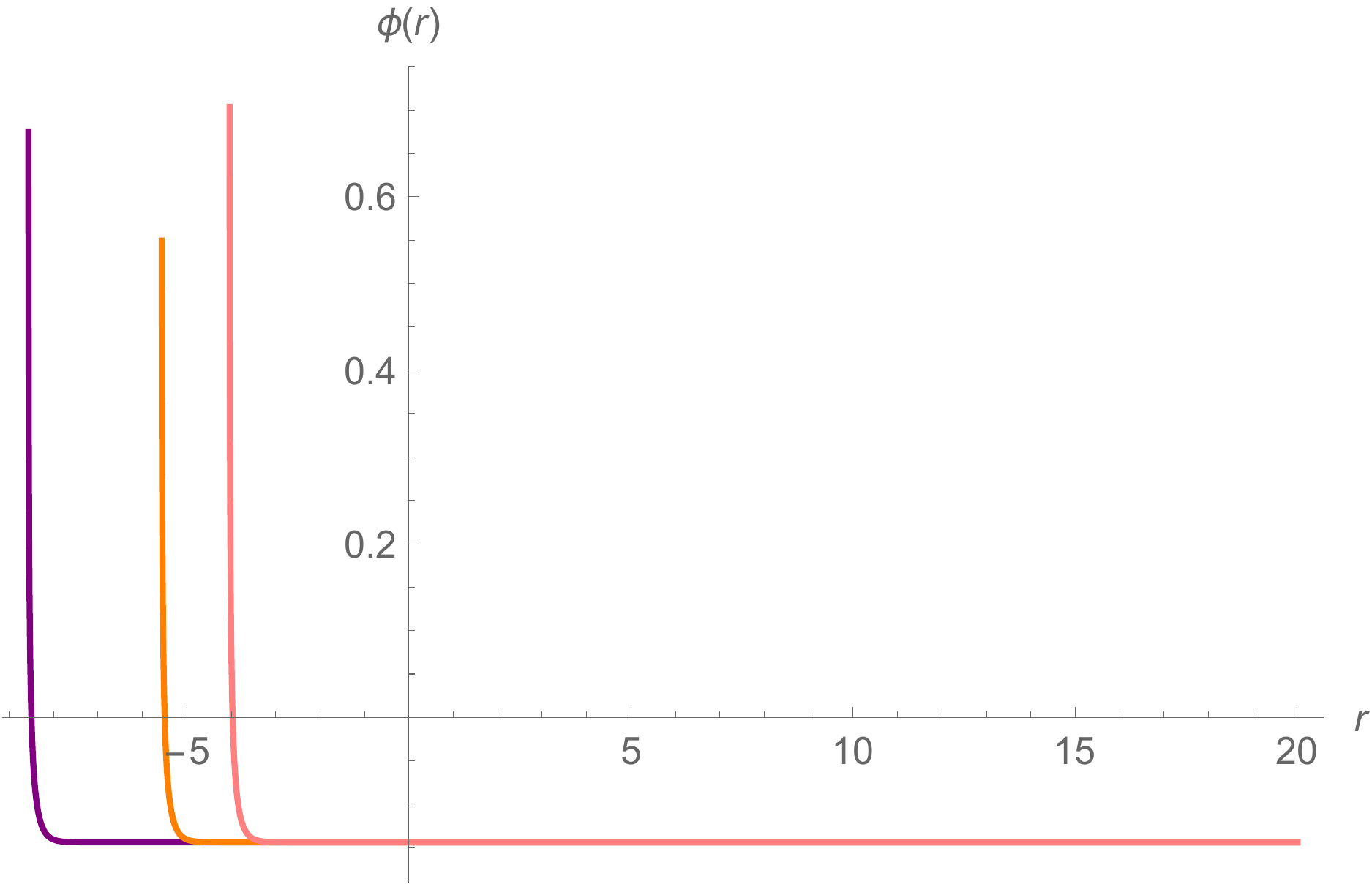}
  \caption{$\phi(r)$ solution}
  \end{subfigure}
  \begin{subfigure}[b]{0.45\linewidth}
    \includegraphics[width=\linewidth]{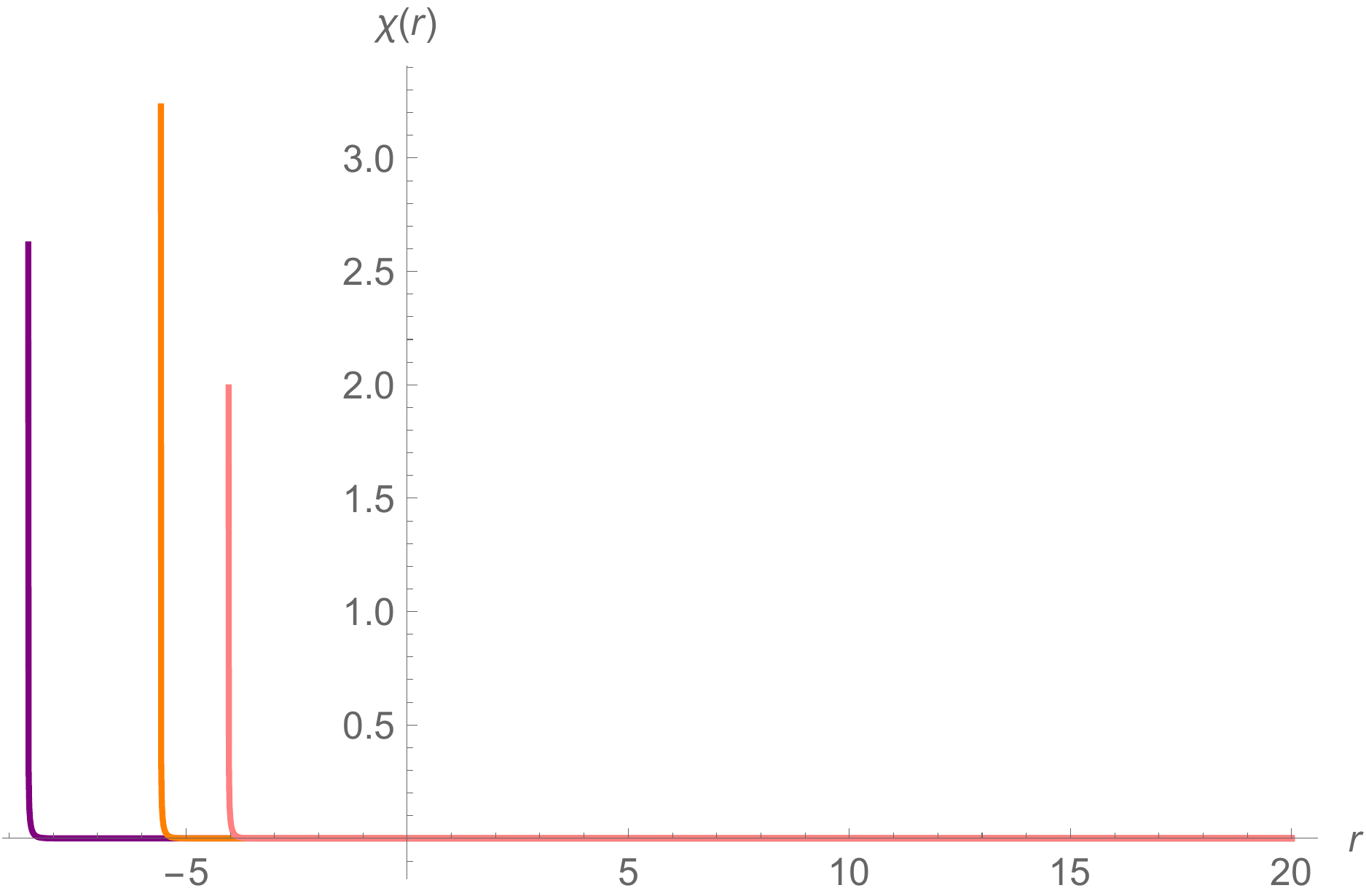}
  \caption{$\chi(r)$ solution}
  \end{subfigure}\\
    \begin{subfigure}[b]{0.45\linewidth}
    \includegraphics[width=\linewidth]{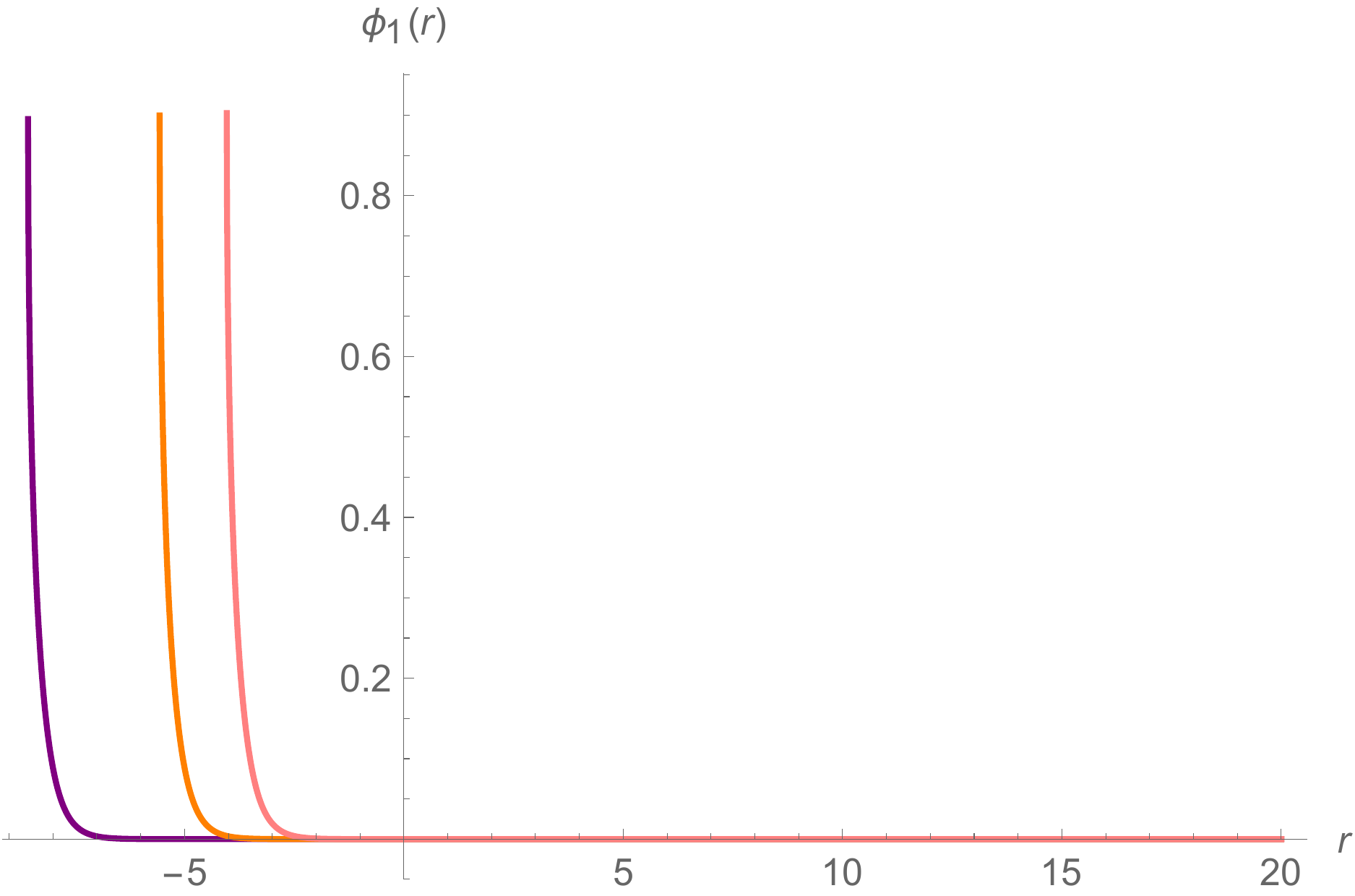}
  \caption{$\phi_1(r)$ solution}
  \end{subfigure}
  \begin{subfigure}[b]{0.45\linewidth}
    \includegraphics[width=\linewidth]{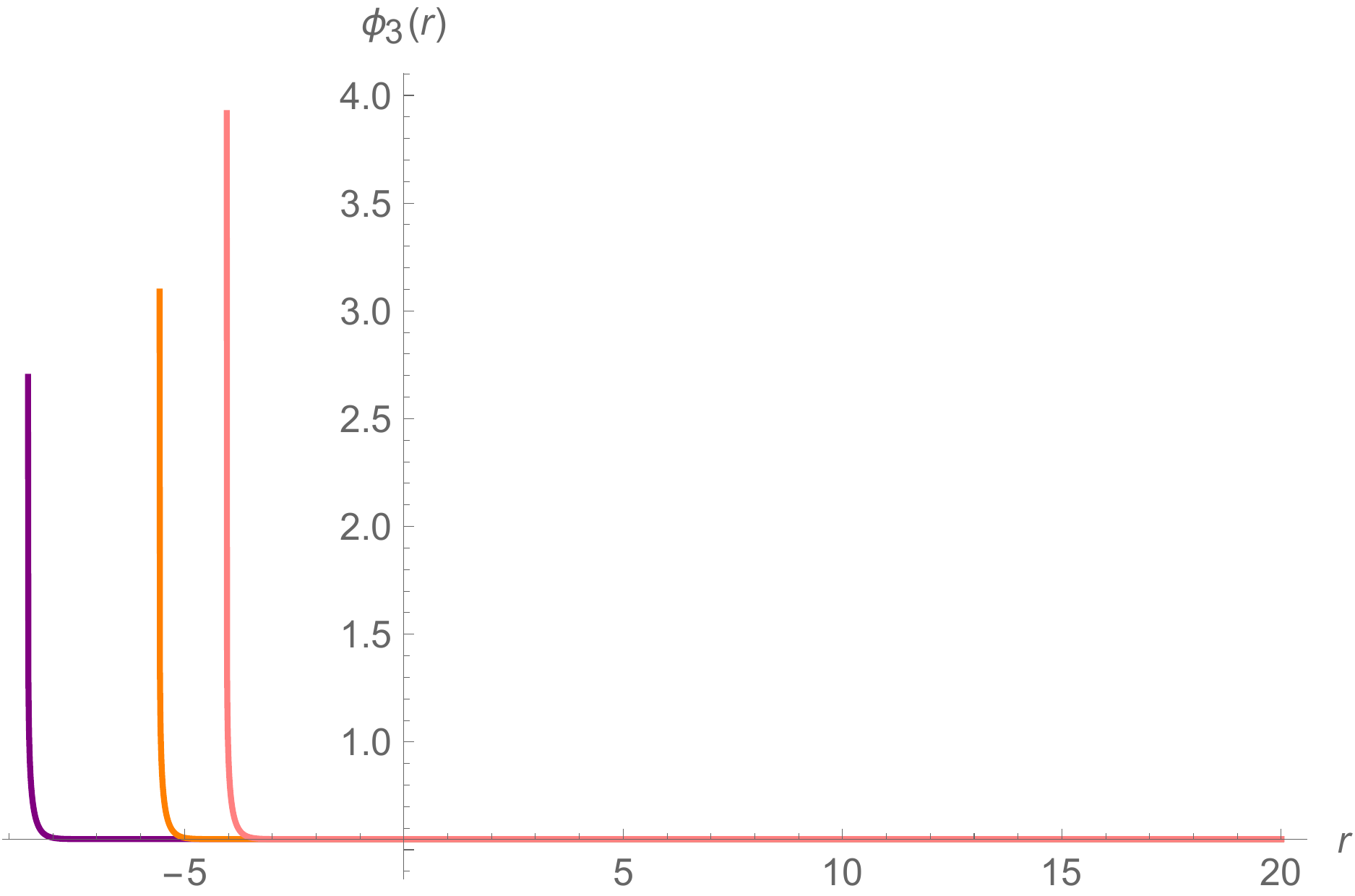}
  \caption{$\phi_3(r)$ solution}
  \end{subfigure}\\
   \begin{subfigure}[b]{0.45\linewidth}
    \includegraphics[width=\linewidth]{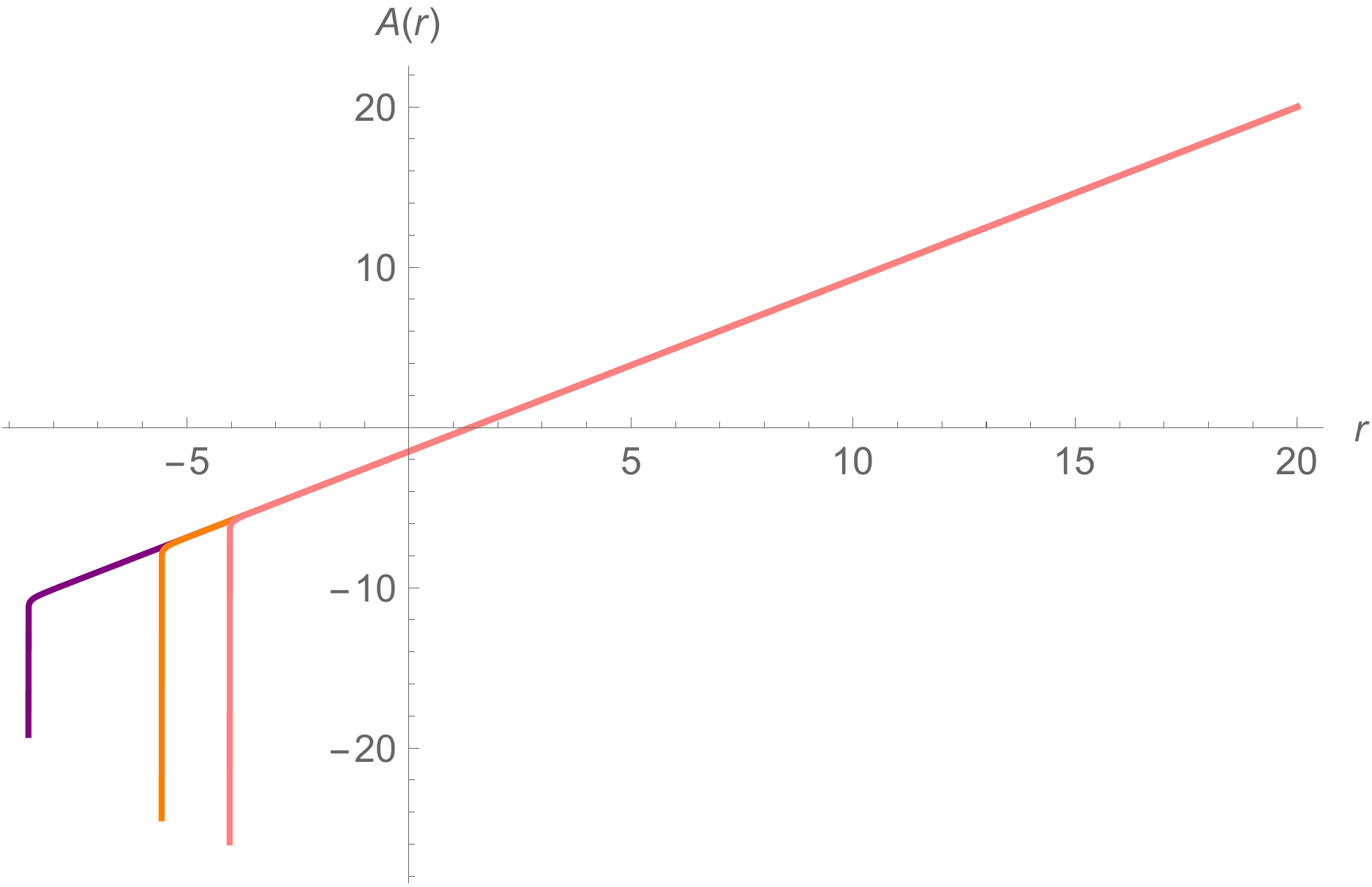}
  \caption{$A(r)$ solution}
   \end{subfigure} 
 \begin{subfigure}[b]{0.45\linewidth}
    \includegraphics[width=\linewidth]{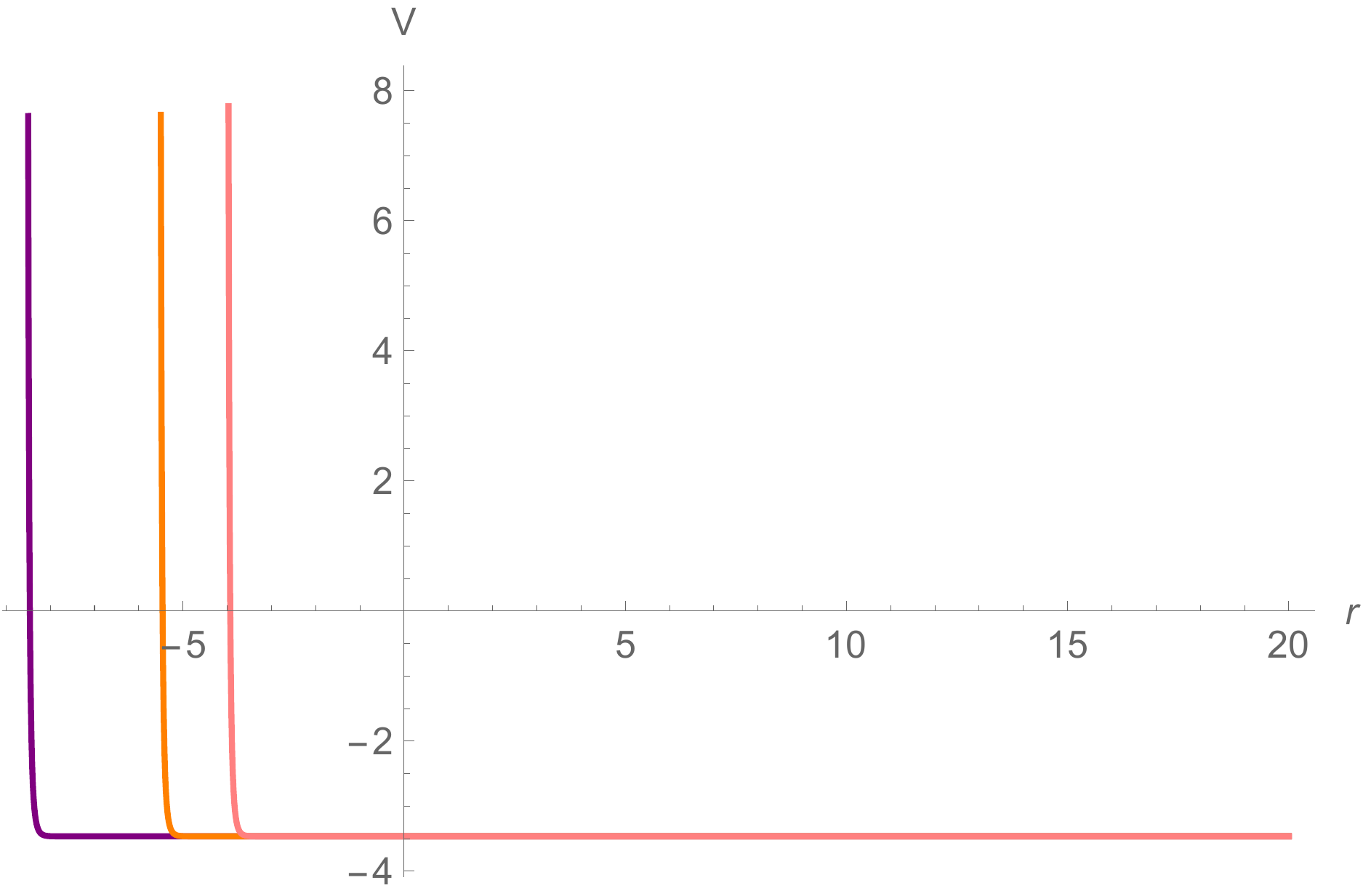}
  \caption{Scalar potential}
   \end{subfigure} 
  \caption{Examples of $N=1$ RG flows from the $N=4$ SCFT with $SO(3)\times SO(3)_{\textrm{diag}}\times SO(3)$ symmetry (critical point $ii$) to non-conformal phases in the IR with $\beta_1=\frac{\pi}{2}$, $g=1$ and $h_1=2$.}
  \label{Fig6}
\end{figure}

A similar $N=1$ flow solution from the $SO(4)\times SO(4)$ fixed point to critical point $ii$ can also be found. This is shown in figure \ref{Fig3}. For other values of the phase $\beta_1$, we have not found any non-trivial $AdS_4$ critical points. Examples of RG flows from the $SO(4)\times SO(4)$ fixed point to non-conformal phases are given in figure \ref{Fig4}. There are also RG flows from $AdS_4$ critical points $i$ and $ii$ to non-conformal phases. Examples of these solutions are given in figures \ref{Fig5} and \ref{Fig6}. Unlike the $N=2$ RG flows given in the previous section, these $N=1$ RG flows turn out to be unphysical according to the criterion of \cite{Gubser_singularity} due to $V\rightarrow \infty$ as seen from the figure. It could be interesting to see whether these singularities are physical in the (if any) uplifted solutions to ten or eleven dimensions.

\section{Conclusions and discussions}\label{conclusion}
In this paper, we have studied symplectically deformed $N=4$ gauged supergravity with $SO(4)\times SO(4)\sim SO(3)\times SO(3)\times SO(3)\times SO(3)$ gauge group with two independent electric-magnetic phases. We have considered three scalar sectors invariant under $SO(3)_{\textrm{diag}}\times SO(3)_{\textrm{diag}}$, $SO(2)\times SO(2)\times SO(2)\times SO(2)$ and $SO(3)_{\textrm{diag}}\times SO(3)$ subgroups of $SO(4)\times SO(4)$. Similar to the $\omega$-deformed $SO(8)$ maximal gauged supergravity, for the trivial supersymmetric $AdS_4$ vacuum at the origin of the scalar manifold, the cosmological constant and scalar masses are independent from the electric-magnetic phases. However, unlike the $\omega$-deformed $SO(8)$ gauged supergravity, it turns out that other $AdS_4$ critical points are the same or related to those identified previously in \cite{4D_N4_flows} for the ``undeformed'' $SO(4)\times SO(4)$ gauge group. Although we have not found any genuinely new supersymmetric $AdS_4$ vacua, we have given a large number of new holographic RG flows preserving $N=2$ and $N=1$ supersymmetries.

The $N=2$ solutions decribe holographic RG flows from the dual $N=4$ CSM theory to non-conformal phases in the IR driven by relevant operators of dimensions $\Delta=1,2$. The $SO(4)\times SO(4)$ symmetry is broken down to $SO(2)\times SO(2)\times SO(2)\times SO(2)$ along the flows through the IR phases. We have found that all these singular solutions describe physical RG flows in the dual field theories since the singularities are physically acceptable by the criterion of \cite{Gubser_singularity}. Moreover, we have also shown that all non-conformal flows within this sector are physical in the sense that all types of singularities lead to the scalar potential that is bounded from above. These solutions also generalize those given recently in \cite{N4_Janus} within an $SO(2)\times SO(2)\times SO(3)\times SO(2)$ subtruncation. However, in this $SO(2)\times SO(2)\times SO(2)\times SO(2)$ sector, no nontrivial $AdS_4$ critical points appear, so there are no RG flows between confornal fixed points. In addition, no non-trivial electric-magnetic phases appear in the analysis. Given that all values of the phase $\alpha>0$ give rise to equivalent gauged supergravities, this sector is essentially the same as the undeformed $SO(4)\times SO(4)$ gauged supergravity considered in \cite{4D_N4_flows} and \cite{N4_Janus}. 

For $N=1$ solutions within $SO(3)_{\textrm{diag}}\times SO(3)$ sector, the phase $\beta_1$ for an $SO(3)$ factor in the vector multiplets appears. We have found three $N=4$ supersymmetric $AdS_4$ critical points with one of them being the trivial $AdS_4$ critical point. The remaining two non-trivial critical points exist for particular values of $\beta_1=0$ and $\beta_1=\frac{\pi}{2}$. The first one preserves $SO(3)_{\textrm{diag}}\times SO(3)\times SO(3)$ symmetry identified in \cite{4D_N4_flows}. The second one with $SO(3)\times SO(3)_{\textrm{diag}}\times SO(3)$ symmetry is very similar to the first critical point and should be related by electric-magnetic duality. We have studied $N=1$ supersymmetric RG flows between the trivial $AdS_4$ vacuum to these two non-trivial critical points similar to the $N=4$ RG flows given previously in \cite{4D_N4_flows}. These flows are driven by relevant operators of dimensions $\Delta=1,2$ and preserve $N=1$ supersymmetry along the flows. At both the UV and IR fixed points, the supersymmetry enhances to $N=4$. For other values of the phase $\beta_1$, we have not found any non-trivial $AdS_4$ critical point. An intensed numerical search suggests that there are no other supersymmetric $AdS_4$ vacua in this sector. We have also given a number of holographic RG flows from $AdS_4$ critical points to various types of non-conformal phases. Unlike the $N=2$ solutions, it turns out that all these flows are unphysical.       
\\
\indent Similar to the $\omega$-deformed $SO(8)$ gauged supergravity, the $N=4$ gauged supergravity considered here currently has no higher dimensional origin. It would be interesting to find the embedding of this gauged supergravity in ten or eleven dimensions. The relevant consistent truncation ansatze could be obtained by using double field theory at $SL(2)$ angles developed in \cite{Dibitetto_SL2_angle} similar to the embedding of half-maximal gauged supergravities in higher dimensions studied in \cite{Dibitetto_non_Geometric,7D_N2_Malek,EFT_Heterotic_Malek,Malek_half_SUSY_EFT,Henning_Malek_AdS7_6_pure,Henning_Malek_AdS7_6}. These could be used to uplift the solutions given here to ten/eleven dimensions resulting in a complete AdS$_4$/CFT$_3$ holography in the framework of string/M-theory. In particular, the unphysical singularities of $N=1$ non-conformal RG flows might be resolved in ten/eleven dimensions and give rise to genuine gravity duals of three-dimensional field theories. It would also be interesting to identify the dual $N=4$ SCFTs and relevant deformations dual to the solutions given in this paper. In addition, the $SO(3)_{\textrm{diag}}$ sector in which both the phases $\beta_1$ and $\beta_2$ appear deserves further study and might lead to new $AdS_4$ vacua. Finally, other types of solutions such as Janus solutions and $AdS_4$ black holes are also worth considering.   
\vspace{0.5cm}\\
{\large{\textbf{Acknowledgement}}} \\
This work is funded by National Research Council of Thailand (NRCT) and Chulalongkorn University under grant N42A650263. The author would like to thank G. Inverso for a useful correspondence and T. Assawasowan for collaboration in a related project.

\appendix
\section{Useful formulae}
In this appendix, we give some formulae used in the main text in particular the convention on 't Hooft matrices and explicit forms of the scalar potential and BPS equations in $SO(3)_{\textrm{diag}}\times SO(3)$ sector.

\subsection{'t Hooft matrices}
We use the following representation of the 't Hooft matrices
\begin{eqnarray}
G_1^{ij}&=&\left[
                     \begin{array}{cccc}
                       0 & 1 & 0 & 0 \\
                       -1 & 0 & 0 & 0 \\
                       0 & 0 & 0  & 1\\
                        0 & 0 & -1  & 0\\
                     \end{array}
                   \right],\qquad
G_2^{ij}=\left[
                     \begin{array}{cccc}
                       0 & 0 & 1 & 0\\
                       0 & 0 & 0 & -1\\
                       -1 & 0 & 0  & 0\\
                        0 & 1 & 0  & 0\\
                     \end{array}
                   \right],\nonumber \\
G_3^{ij}&=&\left[
                     \begin{array}{cccc}
                       0 & 0 & 0 & 1\\
                       0 & 0 & 1 & 0\\
                       0 & -1 & 0  & 0\\
                        -1 & 0 & 0  & 0\\
                     \end{array}
                   \right],\qquad
G_4^{ij}=\left[
                     \begin{array}{cccc}
                       0 & i & 0 & 0\\
                       -i & 0 & 0 & 0\\
                       0 & 0 & 0  & -i\\
                        0 & 0 & i  & 0\\
                     \end{array}
                   \right],\nonumber \\
G_5^{ij}&=&\left[
                     \begin{array}{cccc}
                       0 & 0 & i & 0\\
                       0 & 0 & 0 & i\\
                       -i & 0 & 0  & 0\\
                        0 & -i & 0  & 0\\
                     \end{array}
                   \right],\qquad
G_6^{ij}=\left[
                     \begin{array}{cccc}
                       0 & 0 & 0 & i\\
                       0 & 0 & -i & 0\\
                       0 & i & 0  & 0\\
                        -i & 0 & 0  & 0\\
                     \end{array}
                   \right].  
\end{eqnarray}
These matrices satisfy the relations
\begin{equation}
G_{mij}=(G^{ij}_m)^*=\frac{1}{2}\epsilon_{ijkl}G^{kl}_m\, .
\end{equation}

\subsection{Scalar potential in $SO(3)_{\textrm{diag}}\times SO(3)$ sector}
The scalar potential in $SO(3)_{\textrm{diag}}\times SO(3)$ sector is explicitly given by
\begin{eqnarray}
V&=&\frac{1}{64}g^2e^{-\phi}\left[(3+\cosh2\phi_1+2\cosh2\phi_3\sinh^2\phi_1)^2(2\cosh2\phi_3\sinh^2\phi_1 \right.\nonumber \\
& &\left.+\cosh2\phi_2-3) -16\sinh^3\phi_1\sinh^32\phi_3\chi+32\cosh^4\phi_3\chi^2(\cosh2\phi_3-2)\right]\nonumber \\
& &+\frac{1}{64}e^{-\phi}h_1^2(2\cosh^2\phi_1\cosh2\phi_3+\cosh2\phi_1-3)^2(3+2\cosh^2\phi_1\cosh2\phi_3\nonumber \\
& &+\cosh2\phi_1)(\cos^2\beta_1+\sin^2\beta_1(e^{2\phi}+\chi^2)-\sin2\beta_1\chi)-2g^2\cosh^3\phi_1\cosh^3\phi_3\nonumber \\
& &-\frac{1}{4}e^{-\phi}gh_1\left[8\cos\beta_1\cosh^3\phi_1\cosh^6\phi_3\sinh^3\phi_1-8e^\phi\sin\beta_1\sinh^3\phi_3 \right.\nonumber \\
& &+8e^\phi\cosh^3\phi_3(e^\phi\sin\beta_1\cosh^3\phi_1\sinh^3\phi_3-\cos\beta_1\sinh^3\phi_1)\nonumber \\
& &-\chi(\sin\beta_1\cosh^6\phi_3\sinh^32\phi_1+\cos\beta_1\cosh^3\phi_1\sinh^32\phi_3)\nonumber \\
& &\left.+\sin\beta_1\cosh^3\phi_1\sinh^32\phi_3\chi^2 \right]+\frac{1}{2}g^2e^\phi\cosh^4\phi_3(\cosh2\phi_3-2).
\end{eqnarray}

\subsection{BPS equations in $SO(3)_{\textrm{diag}}\times SO(3)$ sector}
In this section, we collect all the BPS equations in $SO(3)_{\textrm{diag}}\times SO(3)$ sector. These are given by
\begin{eqnarray}
W \phi'&=&\frac{1}{2}g^2e^{-\phi}\left[(\cosh^2\phi_1+\sinh^2\phi_1\sinh^2\phi_3)^3+\cosh^6\chi^2-e^{2\phi}\cosh^6\phi_3 \phantom{\frac{1}{4}}\right.\nonumber \\
& &\left. +\left(6\cosh^2\phi_1\cosh^3\phi_3\sinh\phi_1\sinh\phi_3-\frac{1}{4}\sinh^3\phi_1\sinh^32\phi_3\right)\chi \right]\nonumber \\
& &+\frac{1}{256}h_1^2e^{-\phi}\left[1+\cos2\beta_1-2e^{2\phi}\sin^2\beta_1+2\sin\beta_1\chi(\sin\beta_1\chi-2\cos\beta_1)\right]\times \nonumber \\
& &(2\cosh^2\phi_1\cosh2\phi_3+\cosh2\phi_1-3)^3+\frac{1}{64}g h_1e^{-\phi}\left[-2\cos\beta_1\cosh^6\phi_3\times \right.\nonumber \\
& &\sinh6\phi_1+8(\cosh^2\phi_1\sinh^32\phi_3-24\cosh^3\phi_3\sinh^2\phi_1\sinh\phi_3)\times\nonumber \\
& &e^{2\phi}\sin\beta_1\cosh\phi_1+\chi\left[-6\sin\beta_1\sinh2\phi_1\sinh^22\phi_3(3+\cosh4\phi_1) \right.\nonumber \\
& &+8\cos\beta_1\cosh^3\phi_1\sinh^32\phi_3+\cosh^2\phi_3\left\{\sin\beta_1\sinh^32\phi_1(\cosh4\phi_3\right.\nonumber \\
& &\left.+28\cosh2\phi_3-21)-96\cos\beta_1\cosh\phi_1\sinh^2\phi_1\sinh2\phi_3 \right\}\nonumber \\
& &\left.-64\sin\beta_1\cosh\phi_1\cosh^3\phi_3\sinh\phi_3\chi(\cosh^2\phi_1\sinh^2\phi_3-3\sinh^2\phi_1)\right]\nonumber \\
& &\left. +\frac{3}{4}\cos\beta_1\cosh^2\phi_3\sinh2\phi_1(\cosh4\phi_3+68\cosh2\phi_3-61)\right],
\end{eqnarray}
\begin{eqnarray}
W\chi'&=&h_1^2e^\phi\sin\beta_1(\cos\beta_1-\sin\beta_1\chi)\left(\sinh^2\phi_1+\cosh^2\phi_1\sinh^2\phi_3\right)^3\nonumber \\
& &-g^2e^\phi\cosh^3\phi_3\left(3\cosh^2\phi_1\sinh\phi_1\sinh\phi_3-\sinh^3\phi_1\sinh^3\phi_3+\cosh^3\chi\right)\nonumber\\
& &+\frac{1}{8}gh_1e^\phi\cosh\phi_1\left[16\sin\beta_1\sinh\phi_3\cosh^3\phi_3\chi(\cosh^2\phi_1\sinh^2\phi_3-3\sinh^2\phi_1) \right. \nonumber \\ 
\end{eqnarray}
\begin{eqnarray}
& &-\cosh^2\phi_1\cosh^2\phi_3\left\{\sin\beta_1\sinh^3\phi_1(\cosh4\phi_3+28\cosh2\phi_3-21)\right.\nonumber \\
& &\left.+8\cos\beta_1\cosh\phi_3\sinh^3\phi_3 \right\}+\frac{3}{2}\sinh\phi_1\sinh2\phi_3\times
\nonumber \\
& &\left.
\{8\cos\beta_1\cosh^2\phi_3\sinh\phi_1+(3+\cosh4\phi_1)\sin\beta_1\sinh2\phi_3\}\right],
\end{eqnarray}
\begin{eqnarray}
W\phi_1'&=&\frac{1}{32}g^2e^{-\phi}\left[\left\{\sinh\phi_1\sinh2\phi_1\sinh4\phi_3-(\cosh\phi_1+7\cosh3\phi_1)\sinh2\phi_3\right\}\chi\right. \nonumber \\
& &+e^\phi\left\{(3\cosh3\phi_3-7\cosh\phi_3)\sinh3\phi_1-4\cosh^3\phi_3\sinh\phi_1\right\}\nonumber \\
& &\left. -16\cosh^5\phi_1\sinh\phi_1-4\sinh^32\phi_1\sinh^2\phi_3-16\cosh\phi_1\sinh^5\phi_1\sinh^4\phi_3 \right]\nonumber \\
& & -\frac{1}{128}h_1^2e^{-\phi}\left\{1+\cos2\beta_1+2e^{2\phi}\sin^2\beta_1+2\sin\beta_1\chi(\sin\beta_1\chi-2\cos\beta_1)\right\}\times\nonumber \\
& &\sinh2\phi_1(2\cosh^2\phi_1\cosh2\phi_3+\cosh2\phi_1-3)^2\nonumber \\
& &+\frac{1}{16}gh_1e^{-\phi}\left[\cos\beta_1\left\{\cosh^2\phi_1\cosh2\phi_1\sinh^2\phi_1(\cosh4\phi_3+8\cosh2\phi_3-1)\right.\right.\nonumber \\
& &\left.
-8\sinh^2\phi_3
(\cosh^6\phi_1+\sinh^6\phi_1)\sinh^2\phi_3 \right\} -8e^{\phi}\cosh\phi_1\cosh\phi_3\times \nonumber \\
& & \left\{\cos\beta_1\sinh^2\phi_1(\cosh2\phi_3-2)+\cos\beta_1\cosh^2\phi_1\sinh^2\phi_3\right.\nonumber \\
& &\left. +\sin\beta_1\cosh2\phi_1\sinh\phi_1\sinh2\phi_3 \right\}-e^{2\phi}\sin\beta_1\sinh\phi_1\times\nonumber \\
& &\{3+7\cosh2\phi_1\sinh2\phi_3-\cosh^2\phi_1\sinh4\phi_3\}-\frac{1}{16}\sin\beta_1\chi\cosh2\phi_1\times \nonumber \\
& &\left\{58+6\cosh4\phi_1+8\cosh2\phi_3(\cosh4\phi_1-9)+4\cosh4\phi_3\sinh^22\phi_1 \right\}\nonumber \\
& &+\sin\beta_1\sinh\phi_1\chi^2\left\{\cosh^2\phi_1\sinh4\phi_4-\sinh2\phi_3(3+7\cosh2\phi_1) \right\}\nonumber \\
& &\left.+\cos\beta_1\sinh\phi_1\sinh2\phi_3\chi(3+7\cosh2\phi_1-2\cosh^2\phi_1\cosh2\phi_3)\right],
\end{eqnarray}
\begin{eqnarray}
W\phi_3'&=&-\frac{1}{4}e^{-\phi}\cosh\phi_3\left[\sinh\phi_3\left\{2e^{2\phi}\cosh^4\phi_3+e^\phi\cosh\phi_1\cosh\phi_3[2\cosh^2\phi_1 \right. \right.\nonumber \\
& &\left.+\sinh^2\phi_1
(1-5\cosh2\phi_3)]+2\sinh^2\phi_1(\cosh^2\phi_1+\sinh^2\phi_1\sinh^2\phi_3)^2 \right\}\nonumber \\
& &+2\cosh\phi_3\chi\left\{\cosh^3\phi_3\sinh\phi_3\chi -\cosh2\phi_3\sinh^3\phi_1\sinh^2\phi_3\right.\nonumber \\
& &\left.\left.-\cosh^2\phi_1\sinh\phi_1(1-2\cosh2\phi_3) \right\}\right]-\frac{1}{2}h_1^2e^{-\phi}\cosh^2\phi_1\cosh\phi_3\sinh\phi_3\times\nonumber \\
& &(\sinh^2\phi_1+\cosh^2\phi_1\sinh^2\phi_3)^2\left[e^{2\phi}\sin^2\beta_1
+(\cos\beta_1-\sin\beta_1\chi)^2\right]\nonumber \\
& &+\frac{1}{256}gh_1e^{-\phi}\cosh\phi_3\left[32e^{2\phi}\sin\beta_1\cosh\phi_1\left\{4\cosh\phi_3\sinh^2\phi_1\times\right.\right.\nonumber \\
& &\left.(1-2\cosh2\phi_3)+\cosh^2\phi_1\sinh\phi_3\sinh4\phi_3 \right\}-2e^\phi\left\{33\sin\beta_1\right.\nonumber \\
& &\left.+\sin\beta_1[40\cosh^2\phi_1\cosh4\phi_3\sinh^2\phi_1-\cosh4\phi_1+4\cosh2\phi_3(\cosh4\phi_1-9)]\right.\nonumber \\
& &\left. +8\cos\beta_1\sinh\phi_1[(1-3\cosh2\phi_1)\sinh2\phi_3+5\cosh^2\phi_1\sinh4\phi_3]\right\}\nonumber \\
& &+\cos\beta_1\sinh2\phi_1\left\{2\sinh\phi_3(31+8\cosh4\phi_1\cosh^4\phi_3)-67\sinh3\phi_3\right. \nonumber \\
& &\left.-\sinh5\phi_3 \right\}+\chi\left\{\sin\beta_1\sinh\phi_3[-8\cosh^4\phi_3\sinh6\phi_1 \right.\nonumber \\
& &+(9+140\cosh2\phi_3+3\cosh4\phi_3)\sinh2\phi_1]+64\cos\beta_1\cosh\phi_1\times \nonumber 
\end{eqnarray}
\begin{eqnarray}
& & \left\{2\cosh3\phi_3\sinh^2\phi_1-\cosh^2\phi_1(\cosh\phi_3+\cosh3\phi_3)\sinh^2\phi_3 \right\}\nonumber \\ 
& &\left.\left.+4\sin\beta_1\chi\left\{4\cosh^3\phi_1\cosh5\phi_3+(5\cosh\phi_1-9\cosh3\phi_1)\cosh3\phi_3\right\} \right\}\right]\nonumber \\
& &
\end{eqnarray} 
together with an equation for the metric function $A'=W$.



\begin{thebibliography}{99}
\bibitem{omega_N8_1}  G. Dall’Agata, G. Inverso and M. Trigiante, ``Evidence for a family of $SO(8)$ gauged
supergravity theories'', Phys. Rev. Lett. \textbf{109} (2012) 201301, arXiv:1209.0760.
\bibitem{omega_N8_2} G. Dall’Agata, G. Inverso and A. Marrani, ``Symplectic Deformations of Gauged Maximal Supergravity'', JHEP 07 (2014) \textbf{133}, arXiv:1405.2437.
\bibitem{omega_Range1} A. Borghese, A. Guarino, and D. Roest, ``Triality, Periodicity and Stability of $SO(8)$ Gauged Supergravity'', JHEP 05 (2013) \textbf{107}, arXiv: 1302.6057. 
\bibitem{omega_deWit} B. de Wit and H. Nicolai, ``Deformations of gauged $SO(8)$ supergravity and supergravity in eleven dimensions'', JHEP 05 (2013) \textbf{077}, arXiv: 1302.6219.
\bibitem{SO8_deWit} B. de Wit and H. Nicolai, ``$N=8$ Supergravity'', Nucl. Phys. \textbf{B208} (1982) 323.
\bibitem{maldacena} J. M. Maldacena, ``The large $N$ limit of superconformal field theories and supergravity'', Adv. Theor. Math. Phys. \textbf{2} (1998) 231-252, arXiv: hep-th/9711200.
\bibitem{Gubser_AdS_CFT} S. S. Gubser, I. R. Klebanov and A. M. Polyakov, ``Gauge Theory Correlators from Non-Critical String Theory'', Phys. Lett. \textbf{B428} (1998) 105-114, arXiv: hep-th/9802109.
\bibitem{Witten_AdS_CFT} E. Witten, ``Anti De Sitter Space and holography'', Adv. Theor. Math. Phys. \textbf{2} (1998) 253-291, arXiv: hep-th/9802150.
\bibitem{omega_vacua1} A. Borghese, A. Guarino and D. Roest, ``All $G_2$ invariant critical points of maximal supergravity'', JHEP 12 (2012) \textbf{108}, arXiv: 1209.3003.
\bibitem{omega_vacua2} K. Kodama and M. Nozawa, ``Classification and stability of vacua in maximal gauged supergravity'', JHEP 01 (2013) \textbf{045}, arXiv: 1210.4238.
\bibitem{omega_vacua}  A. Borghese, G. Dibitetto, A. Guarino, D. Roest, and O. Varela, ``The $SU(3)$-invariant sector of new maximal supergravity,” JHEP 03 (2013) \textbf{082}, arXiv:1211.5335.
\bibitem{Guarino_BPS_DW} A. Guarino, ``On new maximal supergravity and its BPS domain-walls'',  JHEP 02 (2014) \textbf{026}, arXiv: 1311.0785.
\bibitem{Elec_mag_flows} J. Tarrio and O. Varela, ``Electric/magnetic duality and RG flows in $AdS_4/CFT_3$'', JHEP 01 (2014) \textbf{071}, arXiv: 1311.2933.
\bibitem{Yi_4D_flow} Y. Pang, C. N. Pope and J. Rong, ``Holographic RG Flow in a New $SO(3)\times SO(3)$ Sector of $\omega$-Deformed $SO(8)$ Gauged $N=8$ Supergravity'', JHEP 08 (2015) \textbf{122}, arXiv: 1506.04270.
\bibitem{N8_omega_Janus} P. Karndumri and C. Maneerat, ``Supersymmetric Janus solutions in $\omega$-deformed $N=8$ gauged supergravity'', Eur. Phys. J. \textbf{C81} (2021) 801, arXiv: 2012.15763.
\bibitem{Inverso_symplectic} G. Inverso, ``Electric-magnetic deformations of $D=4$ gauged supergravities'', JHEP 03 (2016) \textbf{138}, arXiv: 1512.04500.
\bibitem{de_Roo_N4_4D} M. de Roo and P. Wagemans, ``Gauged matter coupling in $N=4$ supergravity'', Nucl. Phys. \textbf{B262} (1985) 644-660.
\bibitem{N4_Wagemans} P. Wagemans, ``Breaking of $N=4$ supergravity to $N=1$, $N=2$ at $\Lambda=0$, Phys. Lett. \textbf{B206} (1988) 241.
\bibitem{dS_Roest} D. Roest and J. Rosseel, ``De Sitter in Extended Supergravity'', Phys. Lett. \textbf{B685} (2010) 201-207, arXiv: 0912.4440.
\bibitem{4D_N4_flows} P. Karndumri and K. Upathambhakul, "Holographic RG flows in N = 4 SCFTs from half-maximal gauged supergravity", Eur. Phys. J. \textbf{C78} (2018) 626, arXiv:hep-th/1806.01819.
\bibitem{N4_Janus} P. Karndumri, ``Holographic RG flows and Janus solutions from matter-coupled $N=4$ gauged supergravity'', Eur. Phys. J. \textbf{C81} (2021) 520, arXiv: 2102.05532.
\bibitem{N3_4_AdS4_BH} P. Karndumri, ``Supersymmetric $AdS_4$ black holes from matter-coupled $N=3,4$ gauged supergravities'', arXiv: 2106.11275.
\bibitem{ABJM} O. Aharony, O. Bergman, D. L. Jafferis and J. Maldacena, ``$N=6$ superconformal Chern-Simons-matter theories, M2-branes and their gravity duals'', JHEP 10 (2008) \textbf{091}, arXiv: 0806.1218.
\bibitem{Bena} I. Bena, ``The M-theory dual of a 3 dimensional theory with reduced supersymmetry'', Phys. Rev. \textbf{D62} (2000) 126006, arXiv: hep-th/0004142.
\bibitem{BL1} J. Bagger and N. Lambert, ``Modeling multiple M2’s'', Phys. Rev. \textbf{D75} (2007) 045020, arXiv: hep-th/0611108.
\bibitem{BL2} J. Bagger and N. Lambert, ``Gauge Symmetry and Supersymmetry of
Multiple M2-Branes'', Phys. Rev. \textbf{D77} (2008) 065008, arXiv: 0711.0955.
\bibitem{BL3} J. Bagger and N. Lambert, ``Comments On Multiple M2-branes'', JHEP 01 (2008) \textbf{105}, arXiv: 0712.3738.
\bibitem{Gustavsson} A. Gustavsson, ``Algebraic structures on parallel M2-branes'', Nucl.
Phys. \textbf{B811} (2009) 66, arXiv: 0709.1260.
\bibitem{Basu_Harvey} A. Basu and J. A. Harvey, ``The M2-M5 brane system and a generalized Nahm’s equation'', Nucl. Phys. \textbf{B713} (2005) 136, arXiv: hep-th/0412310.
\bibitem{Schwarz_3D_SCFT} J. H. Schwarz, ``Superconformal Chern-Simons theories'', JHEP 11 (2004) \textbf{078}, arXiv: hep-th/0411077.
\bibitem{ABJ} O. Aharony, O. Bergman and D. L. Jafferis, ``Fractional M2-branes'',
JHEP 11 (2008) \textbf{043}, arXiv: 0807.4924.
\bibitem{tri-sasakian-flow} P. Karndumri, ``Supersymmetric deformations of 3D SCFTs from tri-sasakian truncation'', Eur. Phys. J. C (2017) \textbf{77}, 130, arXiv: 1610.07983.
\bibitem{Warner_membrane_flow} R. Corrado, K. Pilch and N. P. Warner, ``An $N=2$ supersymmetric membrane flow'', Nucl. Phys. \textbf{B629} (2002) 74-96, arXiv: hep-th/0107220.
\bibitem{Warner_M_F_theory_flow} C. N. Gowdigere and N. P. Warner, ``Flowing with Eight Supersymmetries in M-Theory and F-theory'', JHEP 12 (2003) \textbf{048}, arXiv: hep-th/0212190.
\bibitem{Warner_higher_Dflow} K. Pilch, A. Tyukov and N. P. Warner, ``Flowing to Higher Dimensions: A New Strongly-Coupled Phase on M2 Branes'', JHEP 11 (2015) \textbf{170}, arXiv: 1506.01045.
\bibitem{Flow_in_N8_4D} C. Ahn and K. Woo, ``Supersymmetric Domain Wall and RG Flow from 4-Dimensional Gauged $N=8$ Supergravity'', Nucl. Phys. \textbf{B599} (2001) 83-118, arXiv: hep-th/0011121.
\bibitem{4D_G2_flow} C. Ahn and T. Itoh, ``An $N=1$ Supersymmetric $G_2$-invariant Flow in M-theory'', Nucl. Phys. \textbf{B627} (2002) 45-65, arXiv: hep-th/0112010.
\bibitem{Warner_M2_flow} N. Bobev, N. Halmagyi, K. Pilch and N. P. Warner, ``Holographic, $N=1$ Supersymmetric RG Flows on M2 Branes'', JHEP 09 (2009) \textbf{043}, arXiv: 0901.2376.
\bibitem{N3_SU2_SU3} P. Karndumri, ``Holographic RG flows in $N=3$ Chern-Simons-Matter theory from $N=3$ 4D gauged supergravity'', Phys. Rev. \textbf{D94} (2016) 045006, arXiv: 1601.05703.
\bibitem{N3_4D_gauging} P. Karndumri and K. Upathambhakul, ``Gaugings of four-dimensional $N=3$ supergravity and AdS$_4$/CFT$_3$ holography'', Phys. Rev. \textbf{D93} (2016) 125017 arXiv: 1602.02254.
\bibitem{orbifold_flow} P. Karndumri and K. Upathambhakul, ``Supersymmetric RG flows and Janus from type II orbifold compactification'', Eur. Phys. J. C (2017) \textbf{77}, 455, arXiv: 1704.00538.
\bibitem{N4_from_ISO7} A. Guarino, J. Tarrio and O. Varela, ``Halving $ISO(7)$ supergravity'', JHEP 11 (2019) \textbf{143}, arXiv: 1907.11681.
\bibitem{ISO7_N3_flow} A. Guarino, J. Tarrio and O. Varela, ``Flowing to $N=3$ Chern-Simons-matter theory'', JHEP 03 (2020) \textbf{100}, arXiv: 1910.06866. 
\bibitem{N5_flow} P. Karndumri and C. Maneerat, ``Supersymmetric solutions from $N=5$ gauged supergravity'', Phys. Rev. \textbf{D101} (2020) 126015, arXiv: 2003.05889.
\bibitem{N6_flow} P. Karndumri and J. Seeyangnok, ``Supersymmetric solutions from $N=6$ gauged supergravity'', Phys.Rev. \textbf{D103} (2021) 6, 066023, arXiv: 2012.10978.
\bibitem{Dibitetto_SL2_angle} F. Ciceri, G. Dibitetto, J. J. Fernandez-Melgarejo, A. Guarino and G. Inverso, ``Double Field Theory at $SL(2)$ angles'', JHEP 05 (2017) \textbf{028}, arXiv: 1612.05230.
\bibitem{N4_gauged_SUGRA} J. Schon and M. Weidner, ``Gauged $N=4$ supergravities'', JHEP 05 (2006) \textbf{034}, arXiv: hep-th/0602024.
\bibitem{Eric_N4_4D} E. Bergshoeff, I. G. Koh and E. Sezgin, ``Coupling of Yang-Mills to $N=4$, $d=4$ supergravity'', Phys. Lett. \textbf{B155} (1985) 71-75.
\bibitem{AdS4_N4_Jan} J. Louis and H. Triendl, ``Maximally supersymmetric $AdS_4$ vacua in $N=4$ supergravity'', JHEP 10 (2014) \textbf{007}, arXiv:1406.3363.
\bibitem{Gubser_singularity} S. S. Gubser, ``Curvature singularities: the good, the bad and the naked'', Adv.
Theor. Math. Phys. \textbf{4} (2000) 679-745, arXiv: hep-th/0002160.
\bibitem{Dibitetto_non_Geometric} G. Dibitetto, J. J. Fernandez-Melgarejo, D. Marques and D. Roest, ``Duality orbits of non-geometric fluxes'', Fortschr. Phys. \textbf{60} (2012) 11, arXiv: 1203.6562.
\bibitem{7D_N2_Malek} E. Malek, ``7-dimensional $N=2$ Consistent Truncations using $SL(5)$ Exceptional Field Theory'', JHEP 06 (2017) \textbf{026}, arXiv: 1612.01692.
\bibitem{EFT_Heterotic_Malek} E. Malek, ``From Exceptional Field Theory to Heterotic Double Field Theory via K3'', JHEP 03 (2017) \textbf{057}, arXiv: 1612.01990.
\bibitem{Malek_half_SUSY_EFT} E. Malek, ``Half-maximal supersymmetry from exceptional field theory'', Fortsch. Phys. \textbf{65} (2017) 1700061, arXiv: 1707.00714.
\bibitem{Henning_Malek_AdS7_6_pure} E. Malek, H. Samtleben and V. V. Camell, ``Supersymmetric $AdS_7$ and $AdS_6$ vacua and their minimal consistent truncations from exceptional field theory'', Phys. Lett. \textbf{B786} (2018) 171-179, arXiv: 1808.05597.
\bibitem{Henning_Malek_AdS7_6} E. Malek, H. Samtleben and V. V. Camell, ``Supersymmetric $AdS_7$ and $AdS_6$ vacua and their consistent truncations with vector multiplets'', JHEP 04 (2019) \textbf{088}, arXiv: 1901.11039.
\end{thebibliography}
\end{document}